\RequirePackage{rotating}
\documentclass[usenatbib]{mnras}
\usepackage{epsfig}
\usepackage{amsmath}
\usepackage{graphicx}
\usepackage{array}
\usepackage{textcomp}
\usepackage{amssymb}
\usepackage{rotating}
\usepackage{pdflscape}
\usepackage{caption}
\usepackage{subcaption}
\usepackage{longtable}
\usepackage{xtab}
\def\mnras {{\sc MNRAS}}
\def\apj {ApJ}
\def\apjs {ApJS}
\def\apjl {ApJL}
\def\aj {AJ}

\def\pasp {PASP}
\def\pasj {PASJ}
\def\aap {A\&A} 
\def\nat {Nature}

\title[Passive Spiral Galaxies]
      {Multiple Mechanisms Quench Passive Spiral Galaxies}
\author[A.\ Fraser-McKelvie et al.]
       {Amelia Fraser-McKelvie$^{1,2,3}$\thanks{Amelia.Fraser-Mckelvie@nottingham.ac.uk}, 
       Michael J. I. Brown$^{2,3}$, Kevin Pimbblet$^{4}$, \and Tim Dolley$^{2,3}$, Nicolas J. Bonne$^{5}$
.       
        \vspace*{1mm}\\
        $^{1}$ School of Physics \& Astronomy, University of Nottingham, University Park, Nottingham, NG7 2RD, U.K.\\
        $^{2}$ School of Physics and Astronomy, Monash University, Clayton, Victoria 3800, Australia\\
        $^{3}$ Monash Centre for Astrophysics (MoCA), Monash University, Clayton, Victoria 3800, Australia\\
	$^{4}$ E. A. Milne Centre for Astrophysics, University of Hull, Cottingham Road, Kingston-upon-Hull, HU6 7RX, U.K.\\
	$^{5}$ Institute of Cosmology and Gravitation, University of Portsmouth, Dennis Sciama Building, Burnaby Road, Portsmouth PO1 3FX, U.K.\\
	}

\begin{document}
\maketitle
\begin{abstract}
We examine the properties of a sample of 35 nearby passive spiral galaxies in order to determine their dominant quenching mechanism(s).
All five low mass ($\textrm{M}_{\star}<1~\times10^{10}~\textrm{M}_{\odot}$) passive spiral galaxies are located in the rich Virgo cluster. This is in contrast to low mass spiral galaxies with star formation, which inhabit a range of environments. We postulate that cluster-scale gas stripping and heating mechanisms operating only in rich clusters are required to quench low mass passive spirals, and ram-pressure stripping and strangulation are obvious candidates.

For higher mass passive spirals, while trends are present, the story is less clear. 
The passive spiral bar fraction is high: $74\pm15\%$, compared with $36\pm5\%$ for a mass, redshift, and T-type matched comparison sample of star forming spiral galaxies. 
The high mass passive spirals occur mostly, but not exclusively, in groups, and can be central or satellite galaxies. The passive spiral group fraction of $74\pm15\%$ is similar to that of the comparison sample of star forming galaxies at $61\pm7\%$. 
We find evidence for both quenching via internal structure and environment in our passive spiral sample, though some galaxies have evidence of neither. From this, we conclude no one mechanism is responsible for quenching star formation in passive spiral galaxies - rather, a mixture of mechanisms are required to produce the passive spiral distribution we see today.

\end{abstract}
 

\section{Introduction}
In the established picture of galaxy evolution, a galaxy is likely to be quenched if it is massive \citep[e.g.][]{Kauffmann03}, or located in a dense environment \citep[e.g.][]{Peng10}. Low mass quenched galaxies are preferentially satellites \citep[e.g.][]{Geha12, Davies16}, and the vast majority of quenched galaxies possess early type morphology \citep[e.g.][]{Strateva01,Bell12}. This implies that the mechanism(s) responsible for quenching star formation in most galaxies also result in morphological transformation, or vice versa.

Quenching mechanisms that alter morphology include processes that strip a galaxy of its gas upon entry into a denser environment, such as galaxy harassment \citep{Lake98, Moore96}, galaxy-galaxy mergers \citep{Toomre72, White78, Kormendy13}, and tidal stripping. 
There do exist environmental mechanisms that can quench a galaxy without impacting its morphology, however. Ram-pressure stripping \citep{Gunn72, vanGorkom04, Bekki09} occurs in large galaxy clusters and strips the halo and disk of cold gas used as fuel for star formation without destroying the disk \citep[e.g.][]{Weinmann06}. Strangulation also acts to cut off the gas supply from the galaxy's sub-halo, causing star formation to cease when its gas reservoir is consumed \citep[e.g.][]{Larson80, Balogh00}. Mass quenching mechanisms such as AGN heating also act to cease star formation without destroying a galaxy's disk \citep[e.g.][]{Tabor93, Fabian94}.

There exist galaxies that do not conform to the above quenching paradigm, such as massive, star forming disks \citep[e.g.][]{Ogle16}, and spiral galaxies that show no signs of star formation \citep{Fraser16}. Passive spiral galaxies are rare but intriguing objects, as their existence asserts that morphological transformation is not always required to quench star formation. 
While red spiral galaxies have been discussed in the literature for over 40 years, \citep[e.g.][]{vandenBergh76, Goto03,Ishigaki07, Bamford09, Skibba09, Wolf09, Bundy10, Masters10, Rowlands12}, these earlier samples often showed evidence of nebular line emission, ultraviolet (UV) light from young stellar populations, or infrared (IR) excess from warm dust \citep[e.g.][]{Cortese12}. For this reason, we define spiral galaxies that are optically red as red spirals, and those spirals without any signs of star formation as passive spiral galaxies.

In \citet{Fraser16}, we presented a photometrically and spectroscopically confirmed sample of passive spiral galaxies, selected using a mid-IR colour cut to ensure quiescence. These galaxies spanned a range of stellar masses, yet were uniformly passive and contained undisturbed spiral arms.
Given that the mechanism(s) that cease star formation in passive spiral galaxies must do so without disrupting spiral structure, we may question whether the traditional quenching mechanisms that often destroy internal structure are occurring within these galaxies. The alternative hypothesis is that unique quenching pathways may be invoked to quench passive spiral galaxies, and this is the topic of this paper.

Alternate quenching mechanisms that do not require high stellar mass nor dense environmental regions have been characterised in the literature: for example morphological quenching \citep[e.g.][]{Martig09}, or extra heating provided by the winds of dying low mass stars \citep{Conroy15}. These mechanisms have only been described in early type galaxies, however, and it is unclear whether they are also effective at quenching disk galaxies. In low mass galaxies, supernovae winds can expel a large fraction of interstellar medium on short timescales, also quenching star formation \citep[e.g.][]{Dekel86, Yepes97, Scannapieco08, Bower17}.

The role of bars in galaxy evolution and quenching is well studied \citep[e.g.][]{Kormendy79, Kormendy04, Ellison11, Cheung13}.
By channelling cold gas into the central regions of galaxies forcing a short lived starburst, bars are one of the most efficient redistributors of gas in the disks of galaxies \citep[e.g.][]{Combes81, Weinberg85, Friedli95,  Athanassoula02, Knapen02, Masters11, Athanassoula13, Holmes15}. 

 Simulations show strong bars are difficult to destroy once created \citep[e.g.][]{Shen04,Debattista06}, and are capable of driving gas into the nuclear regions of galaxies \citep[e.g.][]{Shlosman89, Martinet97,Jogee05, Spinoso17, Khoperskov17}. 
The resultant quenched galaxy retains its spiral structure \citep[e.g.][]{Cheung13, Gavazzi15}.
 Therefore, naturally we may suspect bars (or the mechanisms that create them) as being involved in passive spiral quenching. 

 Given the above quenching pathways, we wish to determine whether passive spiral galaxies have particular morphologies or environments that clearly distinguish them from other spiral galaxies, and thus identify or constrain their quenching mechanisms.
 To achieve this, we define a sample of passive spiral galaxies, along with a mass, redshift ($z$), and T-type-matched comparison sample.
 
 This paper is organised as follows: in Section~\ref{sample} we describe the sample of passive spiral galaxies used for this work, and in Section~\ref{comp_sample}, we detail the control sample of spiral galaxies used for comparison. In Section~\ref{quenching_mech} we examine the quenching mechanisms responsible for the formation of passive spirals by splitting our sample into high and low mass bins. Throughout this paper we use AB magnitudes and a flat $\Lambda$CDM cosmology, with $\Omega_{\textrm{m}}=0.3$, $\Omega_{\Lambda}=0.7$ and $H_{0}=70~\textrm{km}~\textrm{s}^{-1}~\textrm{Mpc}^{2}$.
  
  \section{Passive Spiral Sample}
  \label{sample}
   To create our passive spiral sample, we use a similar method to \citet{Fraser16}, with some added refinements.
  We begin with the catalogue of \citet{Bonne15}, which is an all-sky sample of 13,325 local Universe galaxies drawn from the 2-Micron All-Sky Survey (2MASS) Extended Source Catalogue \citep{Jarrett00}. This catalogue has a redshift and morphological completeness  of 99\% to $K=12.59$, with the majority of morphologies (in the form of T-types) coming from the PGC catalogue \citep{Paturel03}. 
  
  We limit our analysis to the Sloan Digital Sky Survey (SDSS) imaging regions, to aid in accurate morphological classification using a large sample of uniform imaging. We select galaxies with $-3<\textrm{T-type}<8$, which allows for the misclassification of spirals as lenticulars, and perform the mid-IR colour cut of $M_{K}-M_{W3}<-2.73$, 
 where $M_{K}$ and $M_{W3}$ are the 2MASS $K$-band and Wide-Field Survey Explorer (WISE) 12 $\mu$m k-corrected absolute magnitudes respectively. \citet{Fraser16} showed this mid-IR colour cut is effective at separating passive spiral galaxies from optically red galaxies suffering from dust-obscured star formation.
 We confirm spiral morphology by visually inspecting each passive spiral candidate using SDSS colour images. Edge on galaxies, shell galaxies, merger remnants, and elliptical and lenticular galaxies are rejected from our sample, leaving 35 bona fide spiral galaxies with passive mid-IR colour. 
 
 Our passive spiral sample spans the SDSS DR13 coverage region, and:
 \begin{itemize}
  \item $0.0024 < z < 0.033$ 
 \item $3.9\times10^{9}~\textrm{M}_{\odot} < \textrm{M}_{\star} < 8.5\times10^{10}~\textrm{M}_{\odot}$
 \item $1<\textrm{T-type}<8$,

 \end{itemize}
\noindent{where stellar masses are sourced from the NASA Sloan Atlas\footnote{\url{www.nsatlas.org}}.}

\begin{figure}
\centering
\includegraphics[width=0.48\textwidth]{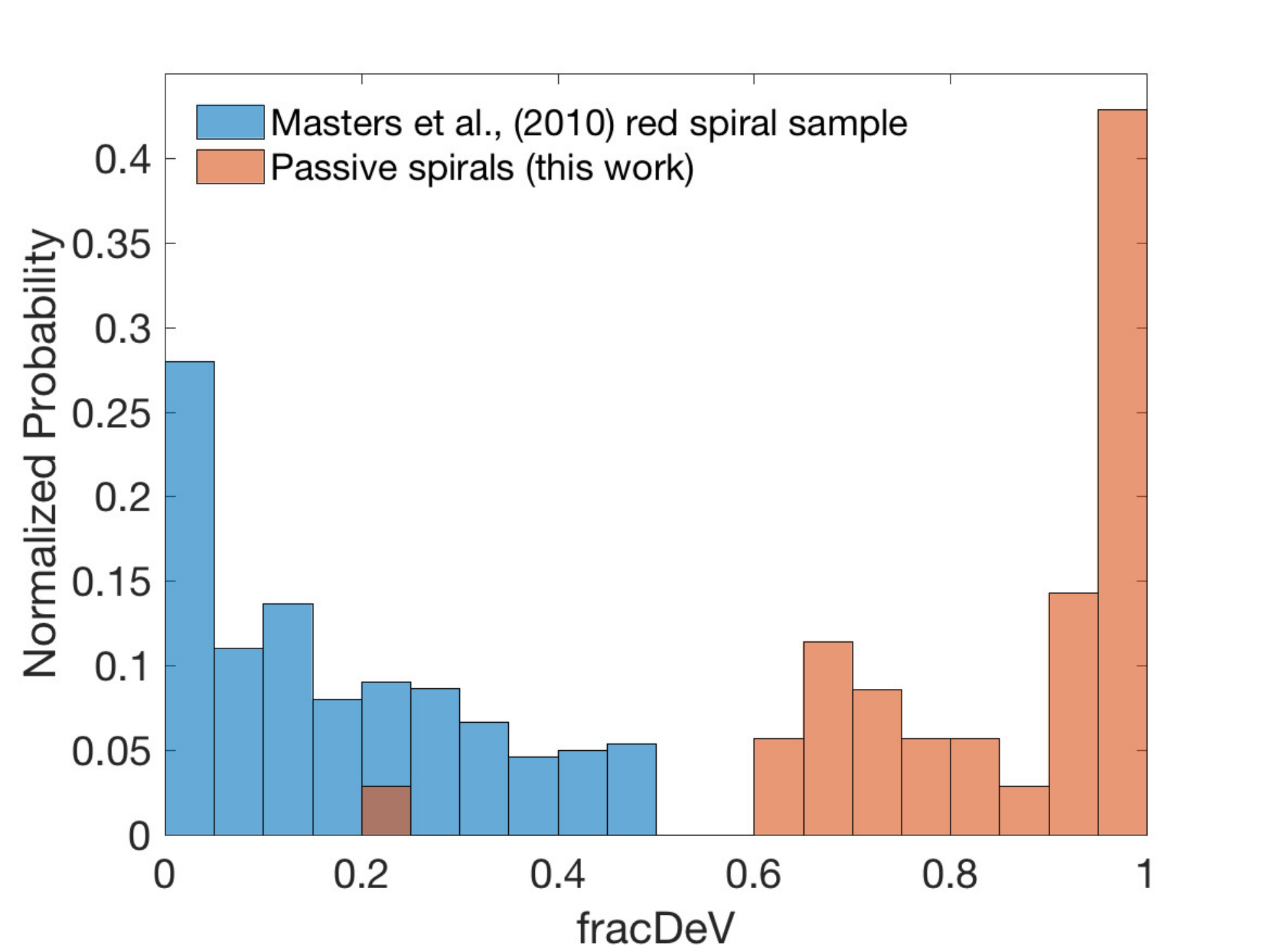} 
\caption{Comparison of the \texttt{fracdeV} quantity, a proxy for bulge size between the red spirals sample of \citet{Masters10} in blue, and the passive spirals selected in this work (orange). Just one galaxy from our sample has \texttt{fracdeV}$<$0.5, and this is NGC 4880. We expect the almost bimodal distribution in bulge size between this work and that of \citet{Masters10} is the reason that our passive spiral sample is more passive.}

\label{fracdeV}
\end{figure}

We note there is no crossover between our passive spiral sample and that of the optically-identified red spirals of \citet{Masters10}. 
In an effort to produce a sample dominated by disky spirals, \citet{Masters10} selected only red spirals with a small bulge size using the SDSS quantity \texttt{fracdeV}$<0.5$, where \texttt{fracdeV} measures the fraction of the galaxy light fit by a de Vaucouleurs profile.
In Figure~\ref{fracdeV} we plot a normalised probability histogram of the SDSS value \texttt{fracdeV} of both our passive spiral sample, and the \citet{Masters10} red spirals. Just one galaxy in our sample has \texttt{fracdeV} $<0.5$, highlighting the dichotomy between the \citet{Masters10} sample and our own. 
We suggest that the reason our passive spiral sample is more passive than the red spirals of \citet{Masters10} is simply because by making a cut in \texttt{fracdeV}, they excluded the most passive spiral galaxies.

\section{Comparison Sample Selection}
\label{comp_sample}
In order to analyse the trends seen in our passive spiral sample, we create a control sample of spiral galaxies. As bar fraction is correlated with stellar mass \citep[e.g.][]{Cameron10,Melvin14} and T-type \citep[e.g.][]{Nair10,Martinez07}, we elect to match our control sample in stellar mass, T-type, and $z$.
To create the control sample, we take all galaxies from the \citet{Bonne15} catalogue, the parent catalogue of the passive spiral sample,  and select 
the four galaxies that are closest in $z$ and mass to each passive spiral galaxy. We impose the constraint that the T-type of the comparison galaxy must match that of the passive spiral galaxy it is being compared to. If the T-type of the passive spiral galaxy is listed as $<1$, we re-classify the galaxy, and select comparison galaxies according to the new T-type given. We note that to ensure a meaningful comparison can be made between passive and non-passive spirals, we include the restriction that a galaxy already designated as a passive spiral galaxy cannot be used as a comparison for any other passive spiral galaxy and each comparison galaxy may only be used once. We select the four galaxies nearest in mass and $z$ range to each passive spiral galaxy with the same T-type for a sample of 140, spanning:

\begin{itemize}
\item $0.0027<z<0.043$ 
\item $7.1\times10^{8}~\textrm{M}_{\odot}<\textrm{M}_{\star}<9.0\times10^{10}~\textrm{M}_{\odot}$,
\item $1<\textrm{T-type}<8$.
\end{itemize}

\noindent{We additionally require that all galaxies are within the SDSS DR13 imaging regions to ensure ease of morphological classification, and with axis ratio greater than 0.4, to enable easy feature identification. We clean the sample to remove any lenticulars or merging galaxies that have been misclassified.
  
In Figure~\ref{ps_examples} we provide some examples of SDSS images of the passive spiral galaxies in our sample in the left column, and the mass-matched comparison galaxies are shown in the four right columns. All galaxies not already shown in the body of this paper are included in Figure~\ref{ps_highmass}. 
We use this comparison sample to compare the trends seen in the passive spiral sample.

\begin{figure*}  
\centering
\begin{subfigure}{0.19\textwidth}
\includegraphics[width=\textwidth]{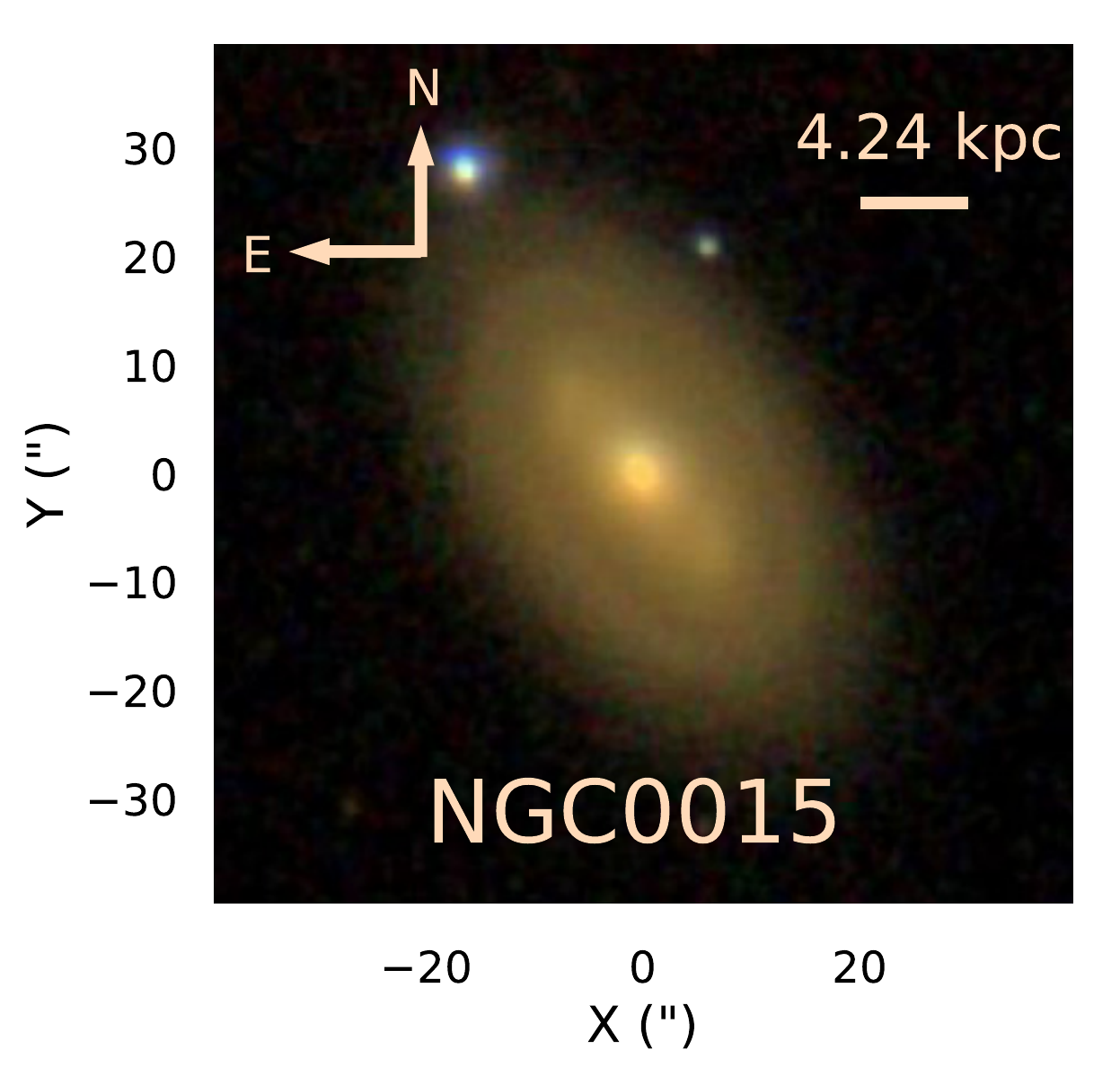}
\end{subfigure}
\hfill
\begin{subfigure}{0.19\textwidth}
\includegraphics[width=\textwidth]{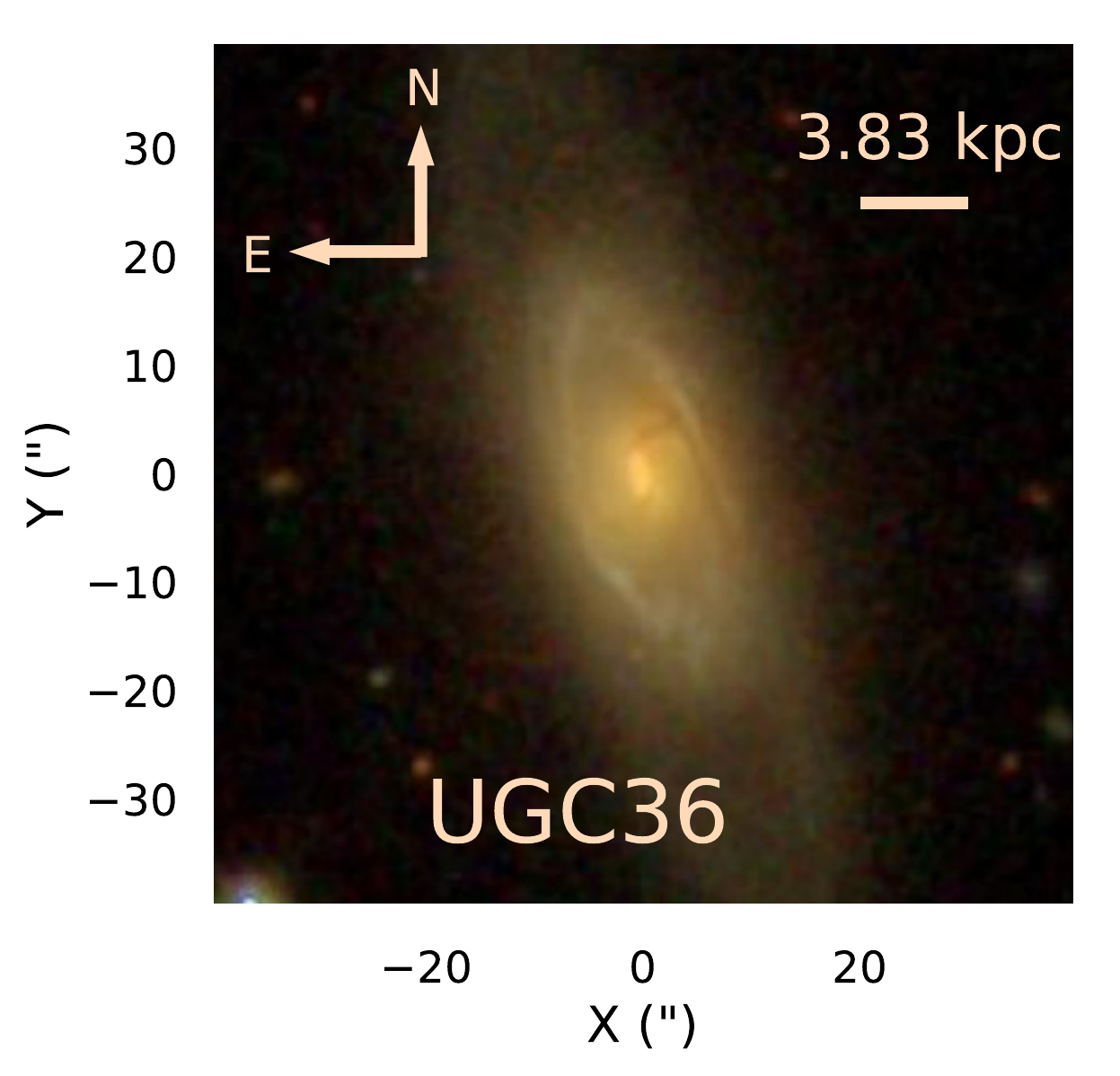}
\end{subfigure}
\hfill
\begin{subfigure}{0.19\textwidth}
\includegraphics[width=\textwidth]{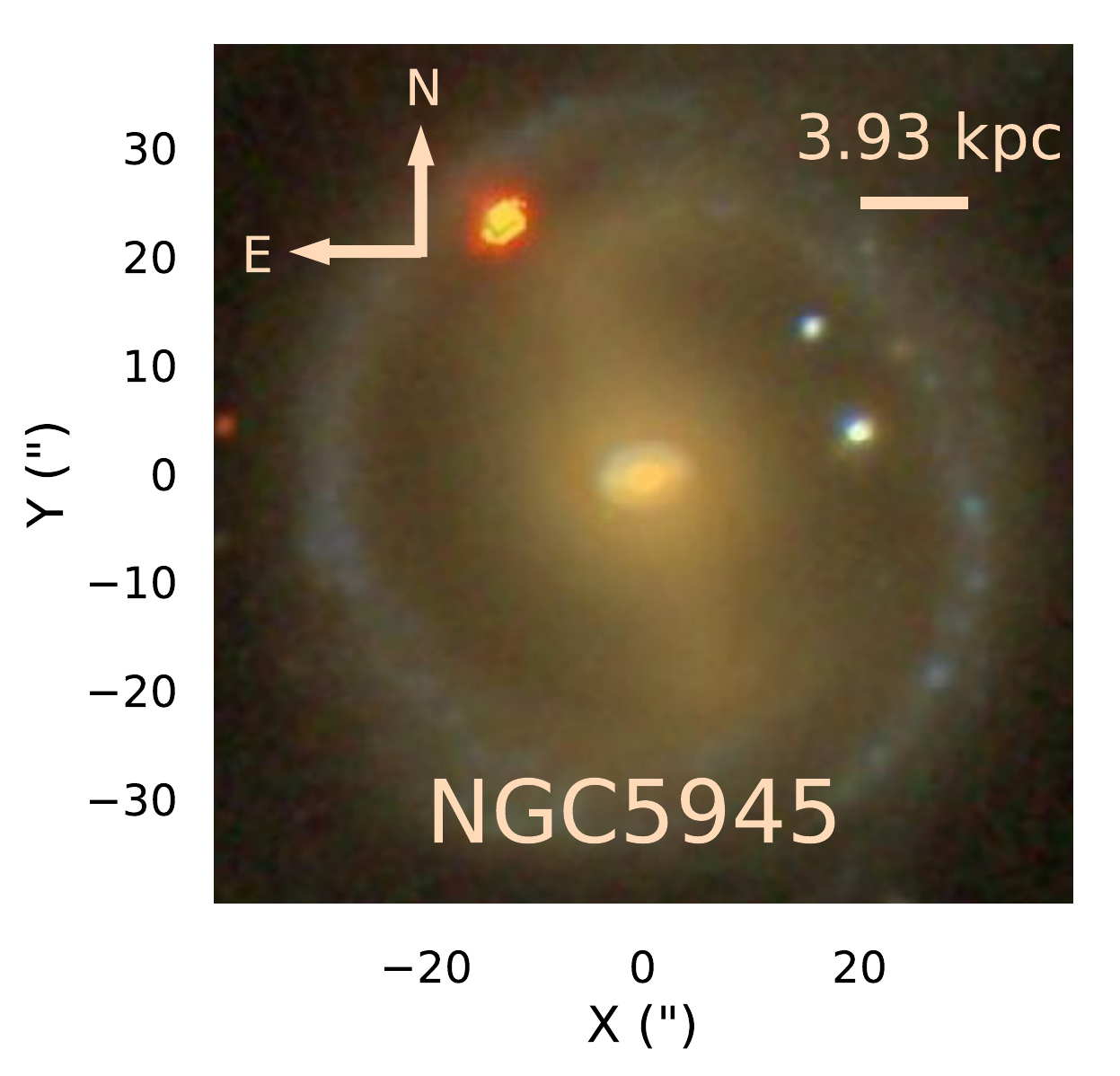}
\end{subfigure}
\hfill
\begin{subfigure}{0.19\textwidth}
\includegraphics[width=\textwidth]{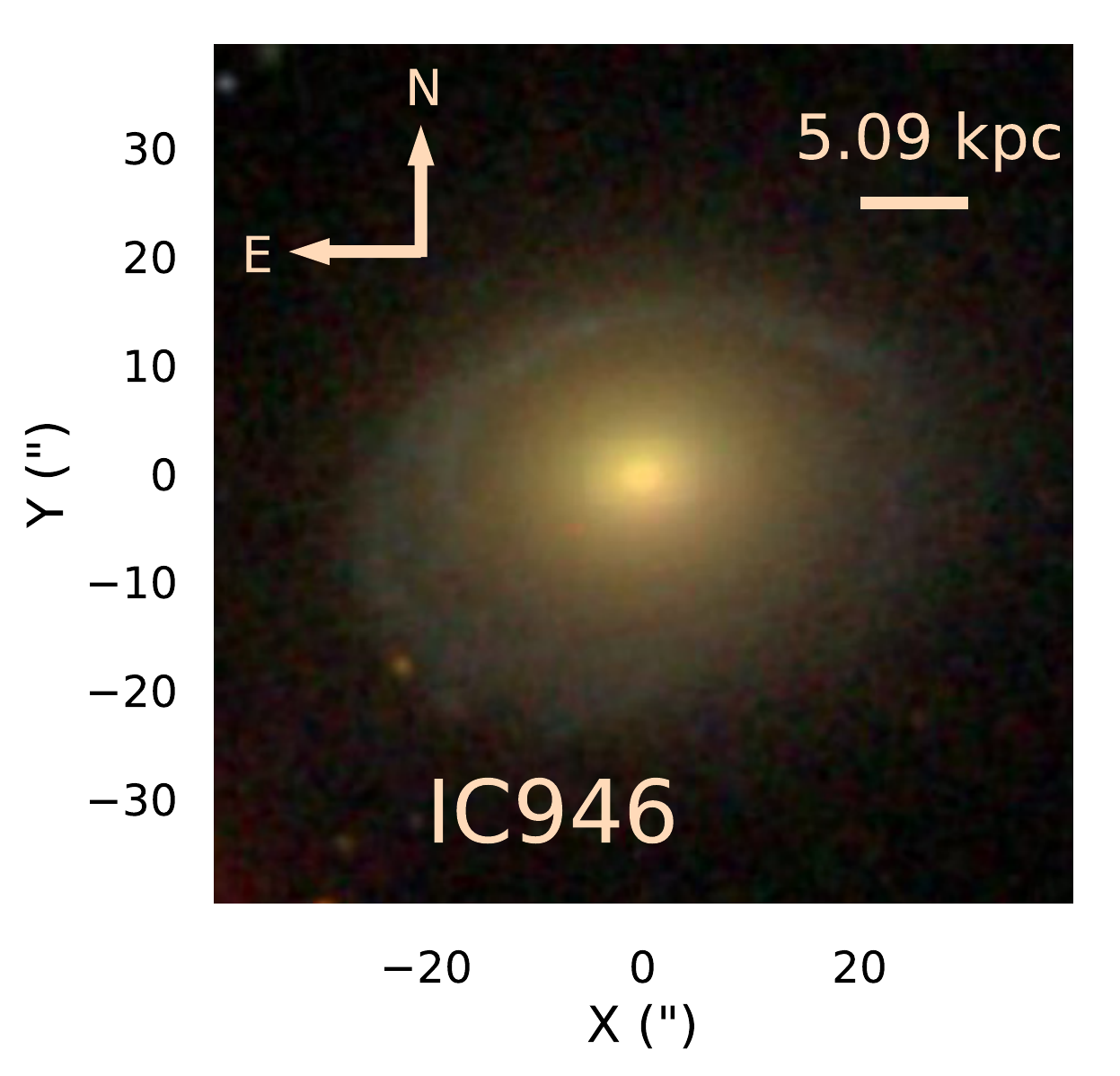}
\end{subfigure}
\hfill
\begin{subfigure}{0.19\textwidth}
\includegraphics[width=\textwidth]{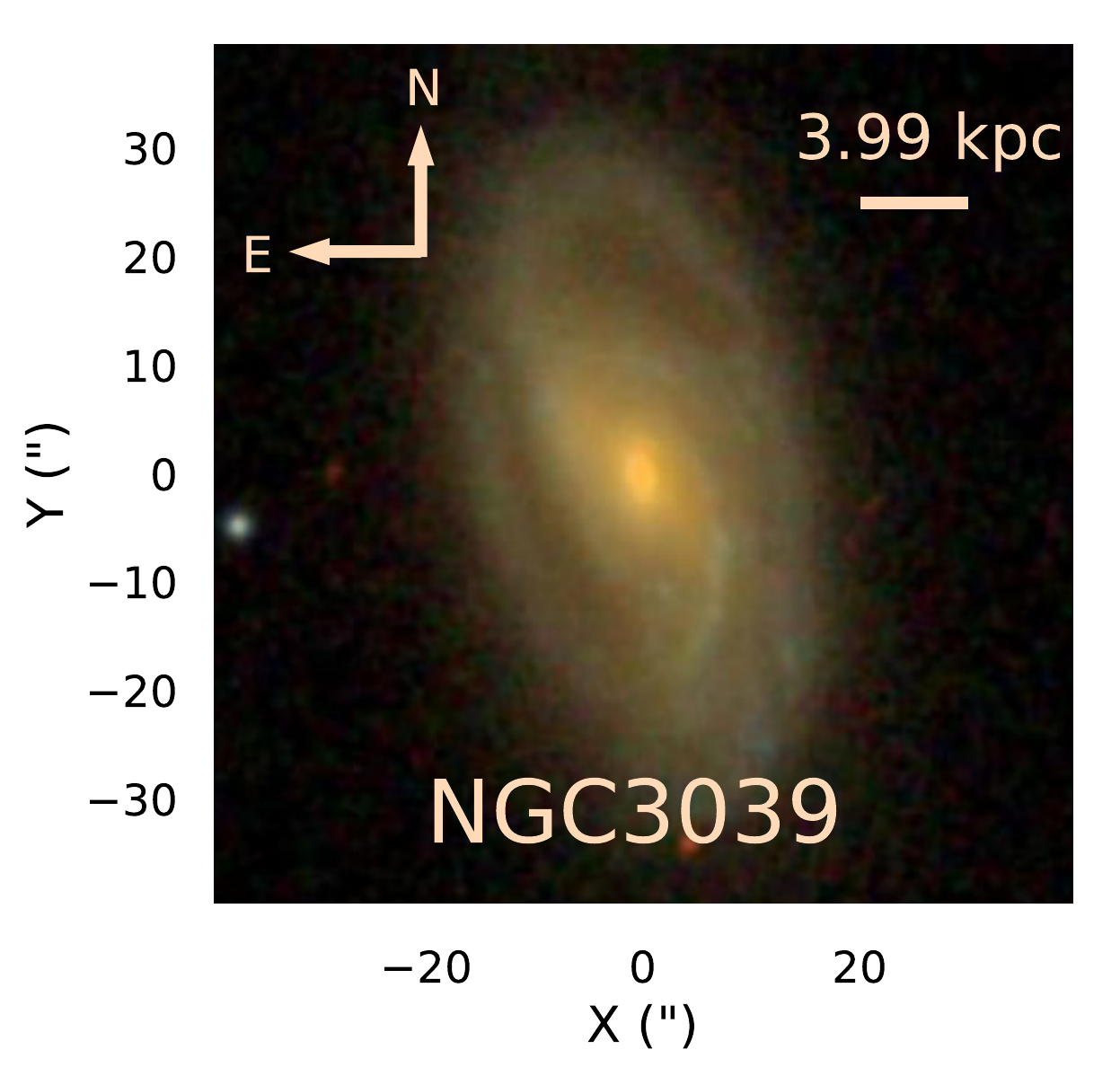}
\end{subfigure}

\vskip\baselineskip

\begin{subfigure}{0.19\textwidth}
\includegraphics[width=\textwidth]{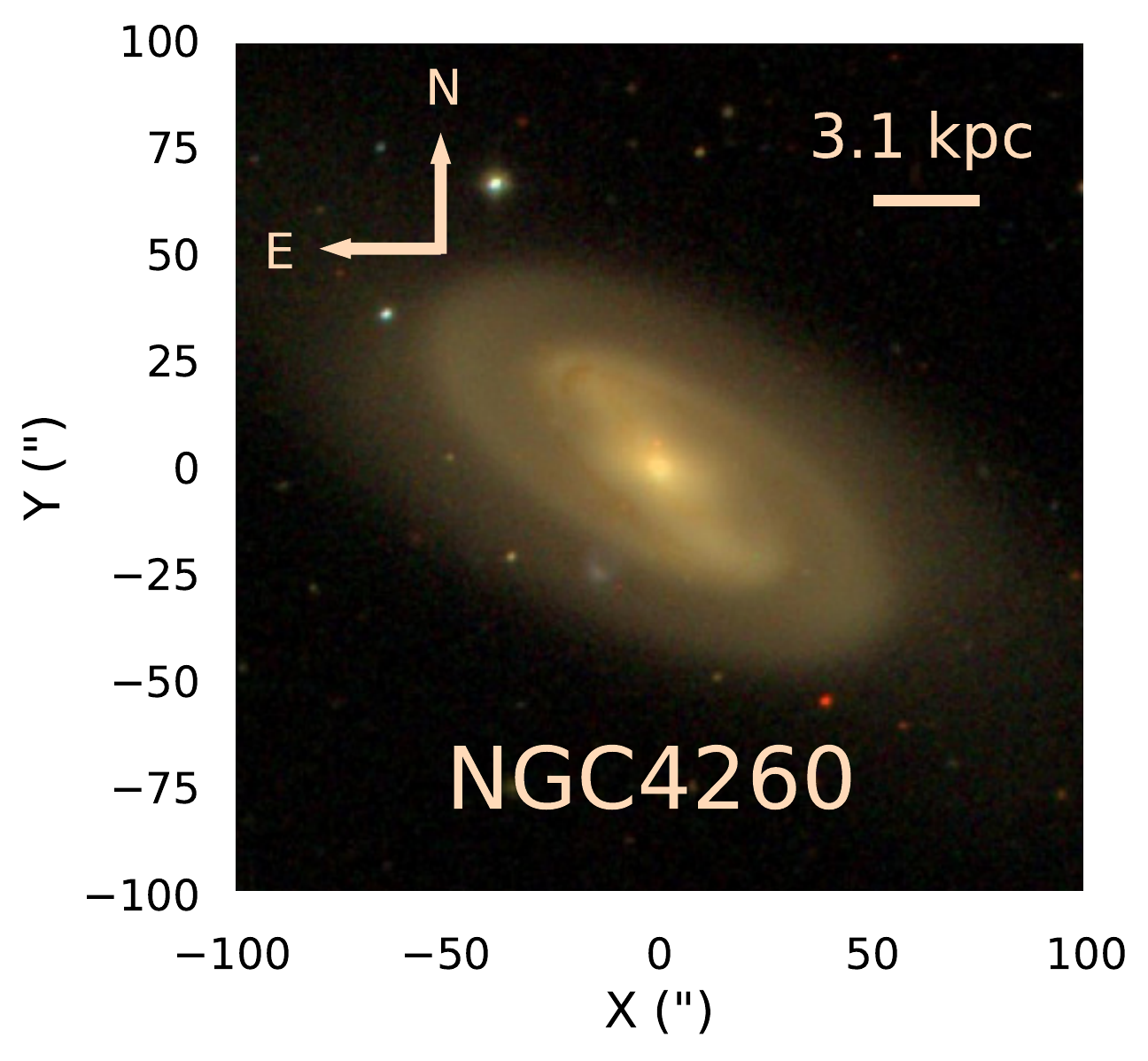}
\end{subfigure}
\hfill
\begin{subfigure}{0.19\textwidth}
\includegraphics[width=\textwidth]{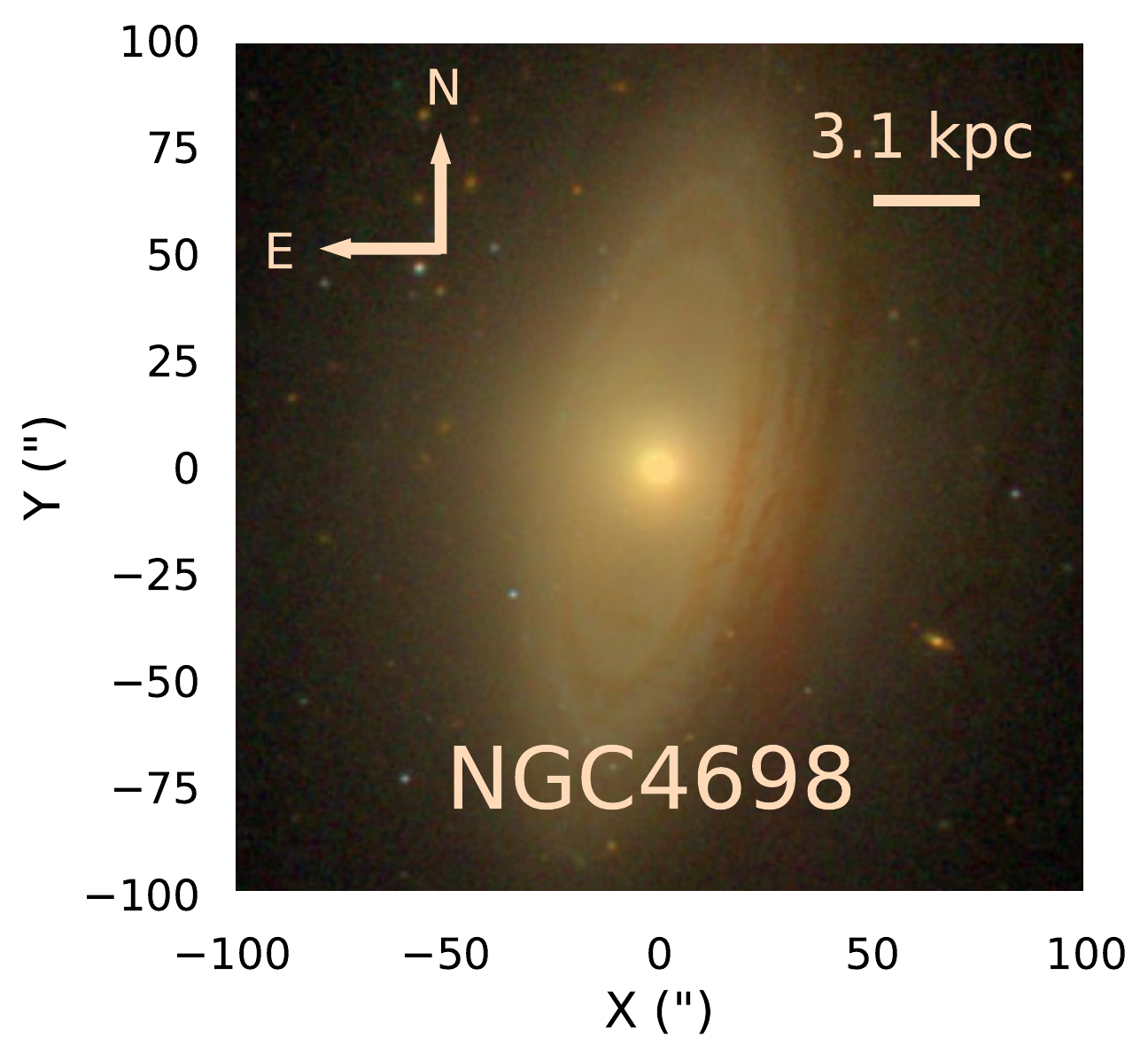}
\end{subfigure}
\hfill
\begin{subfigure}{0.19\textwidth}
\includegraphics[width=\textwidth]{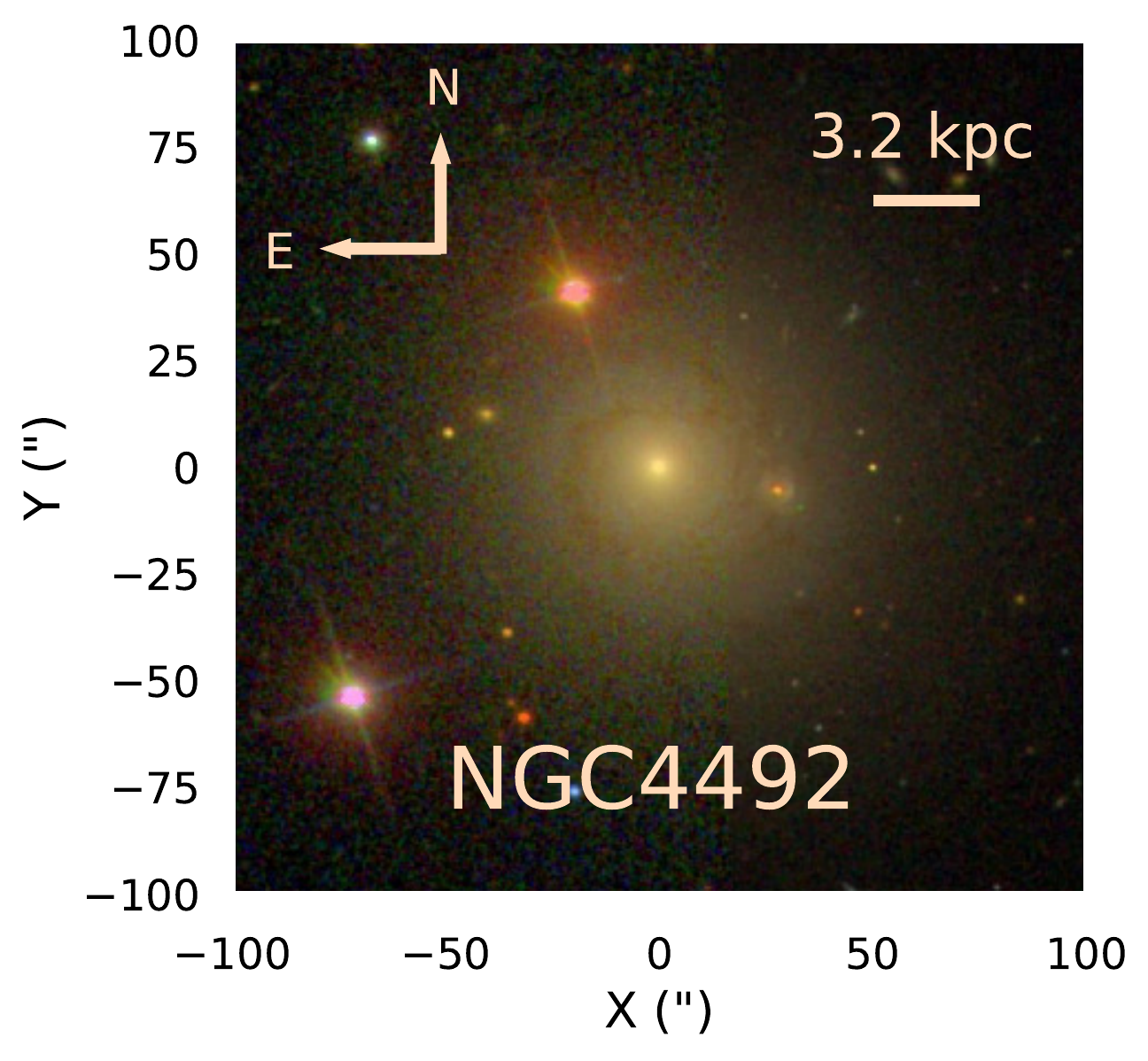}
\end{subfigure}
\hfill
\begin{subfigure}{0.19\textwidth}
\includegraphics[width=\textwidth]{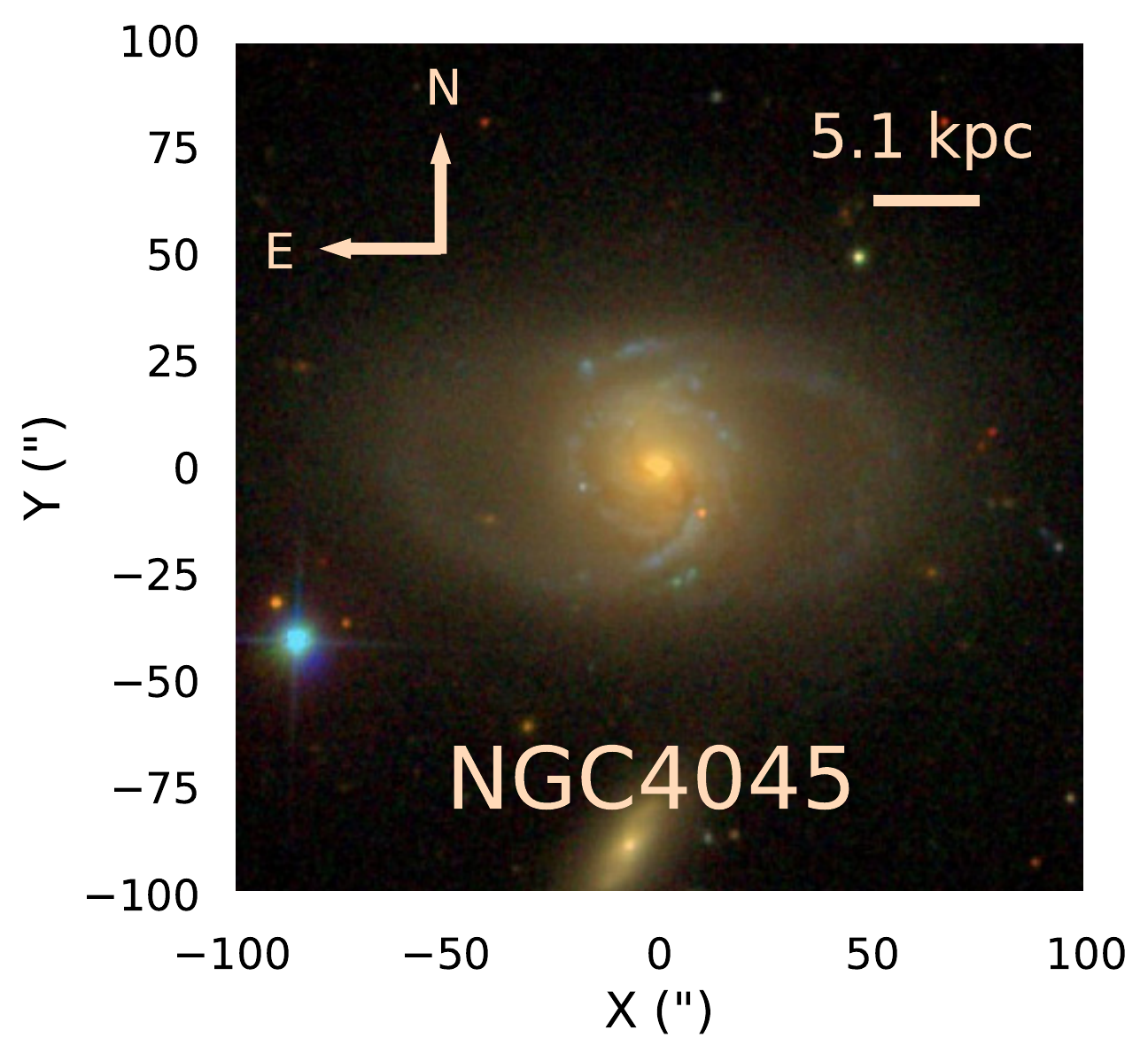}
\end{subfigure}
\hfill
\begin{subfigure}{0.19\textwidth}
\includegraphics[width=\textwidth]{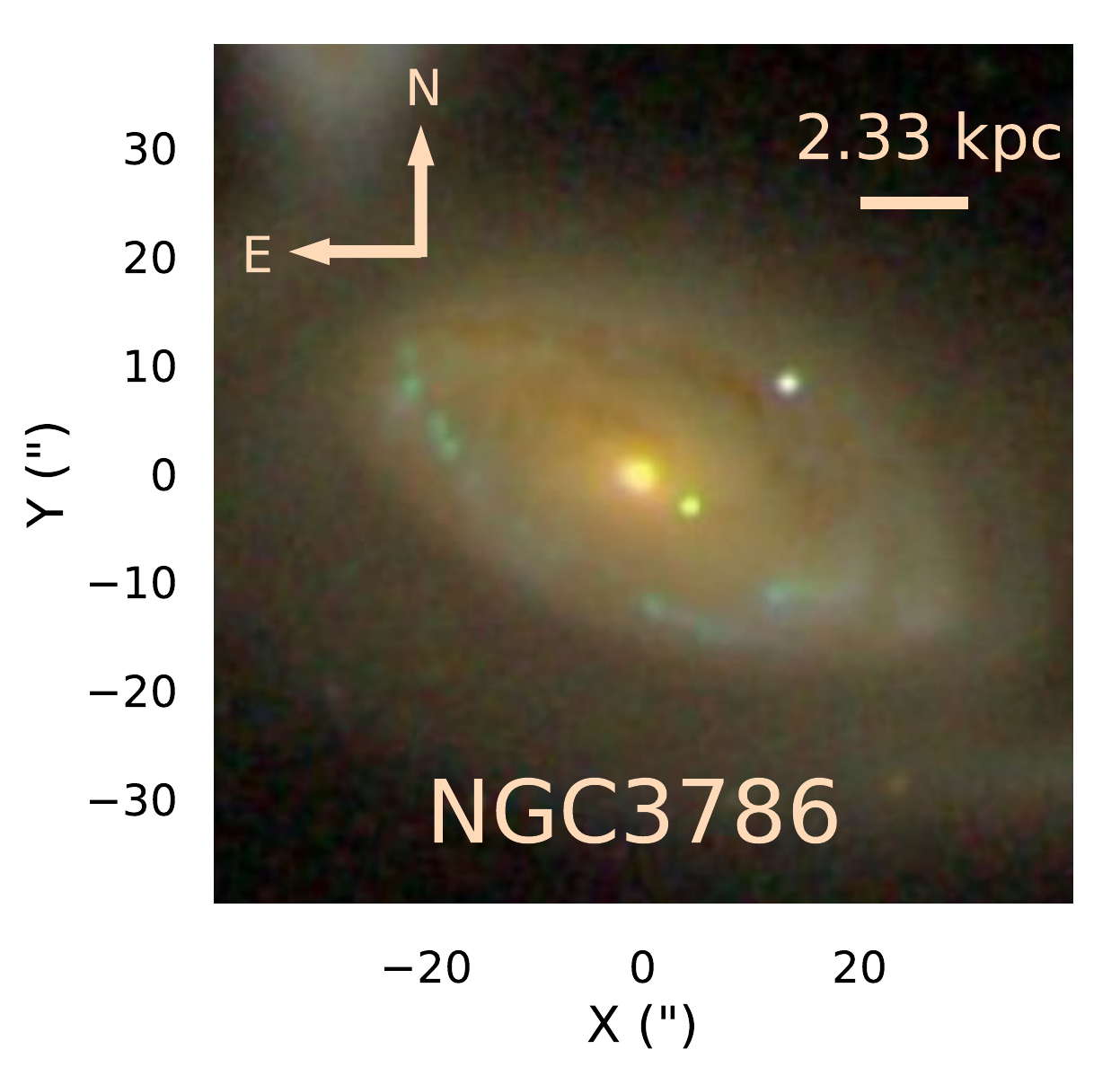}
\end{subfigure}

\vskip\baselineskip

\begin{subfigure}{0.19\textwidth}
\includegraphics[width=\textwidth]{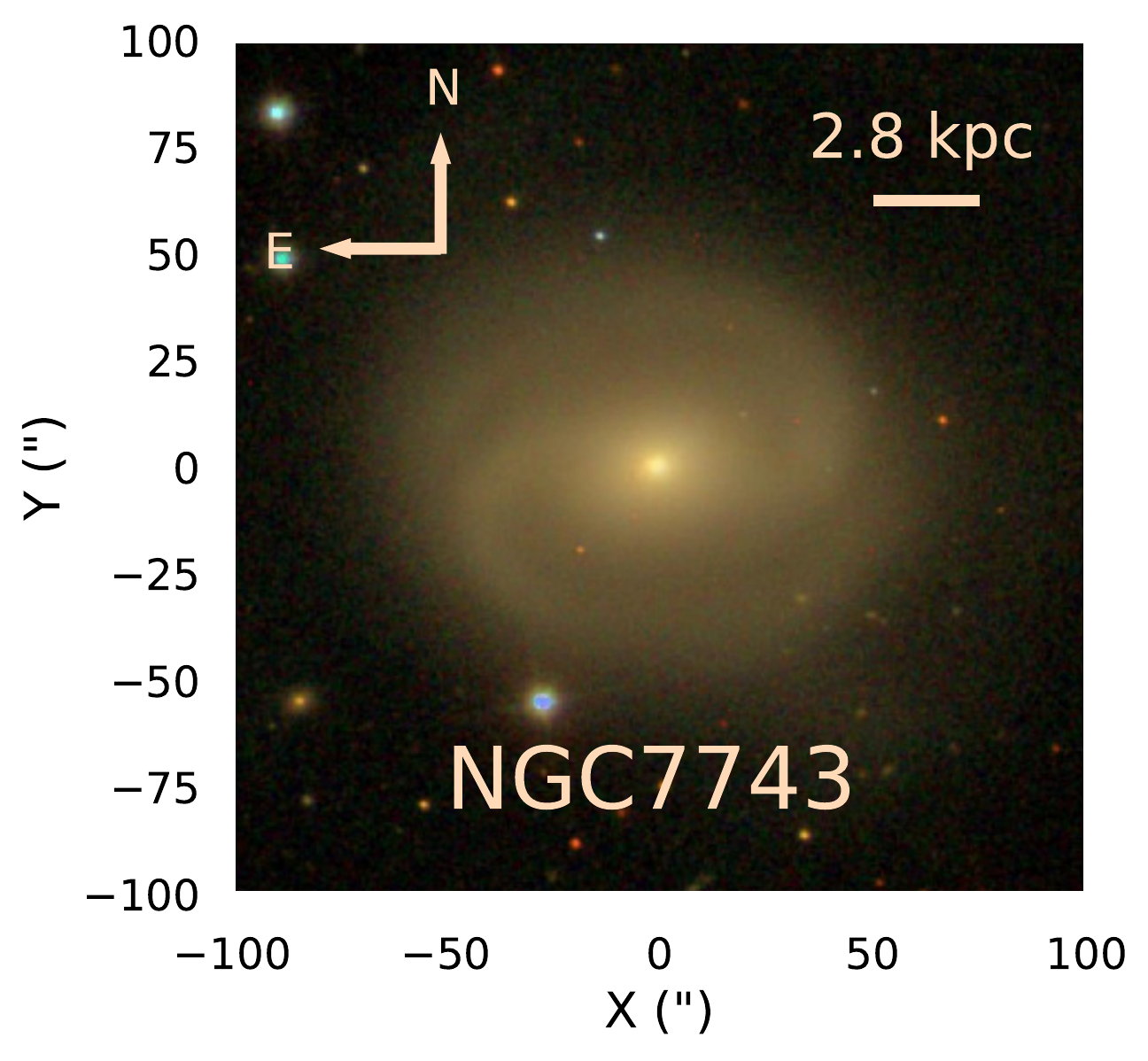}
\end{subfigure}
\hfill
\begin{subfigure}{0.19\textwidth}
\includegraphics[width=\textwidth]{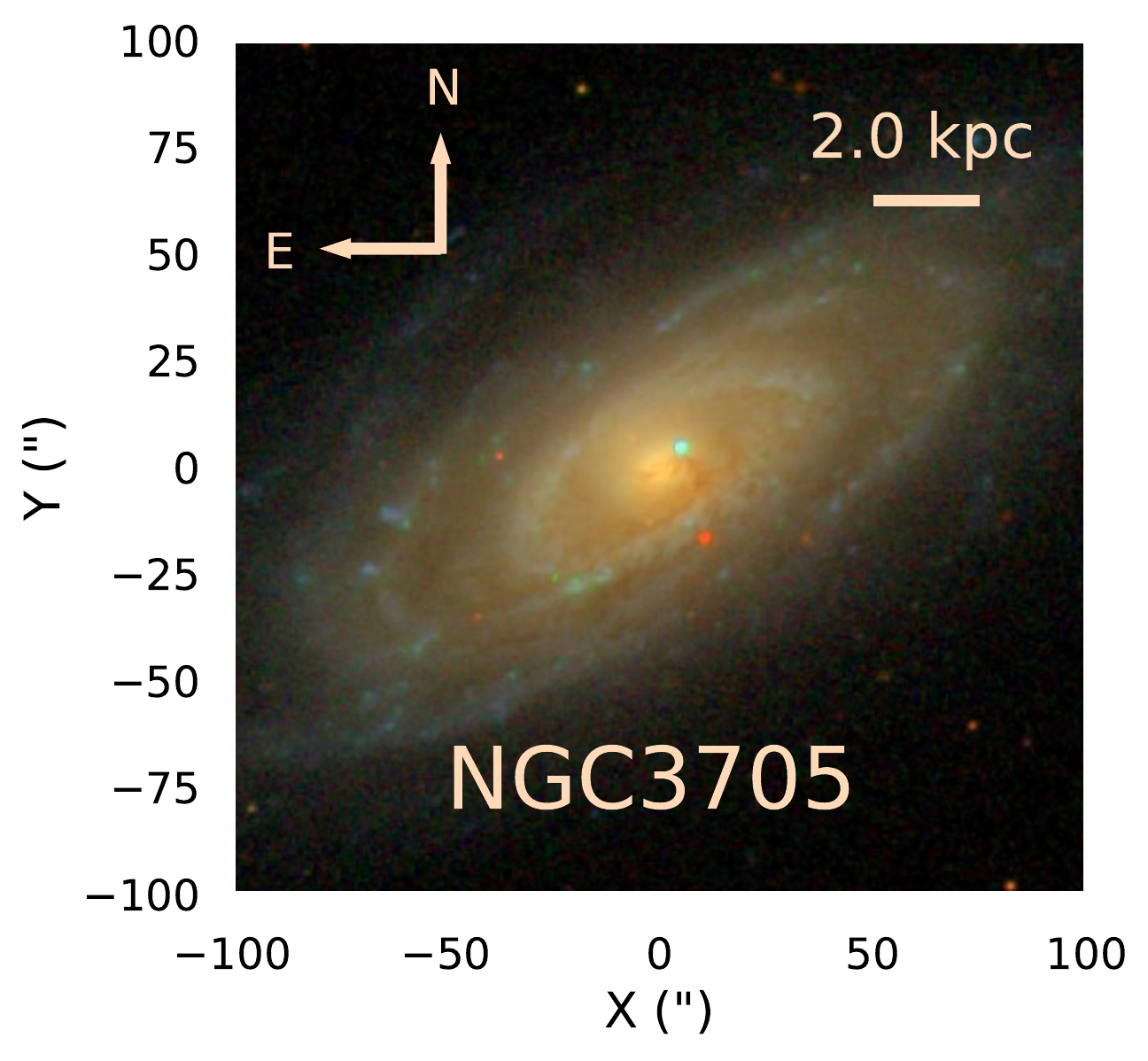}
\end{subfigure}
\hfill
\begin{subfigure}{0.19\textwidth}
\includegraphics[width=\textwidth]{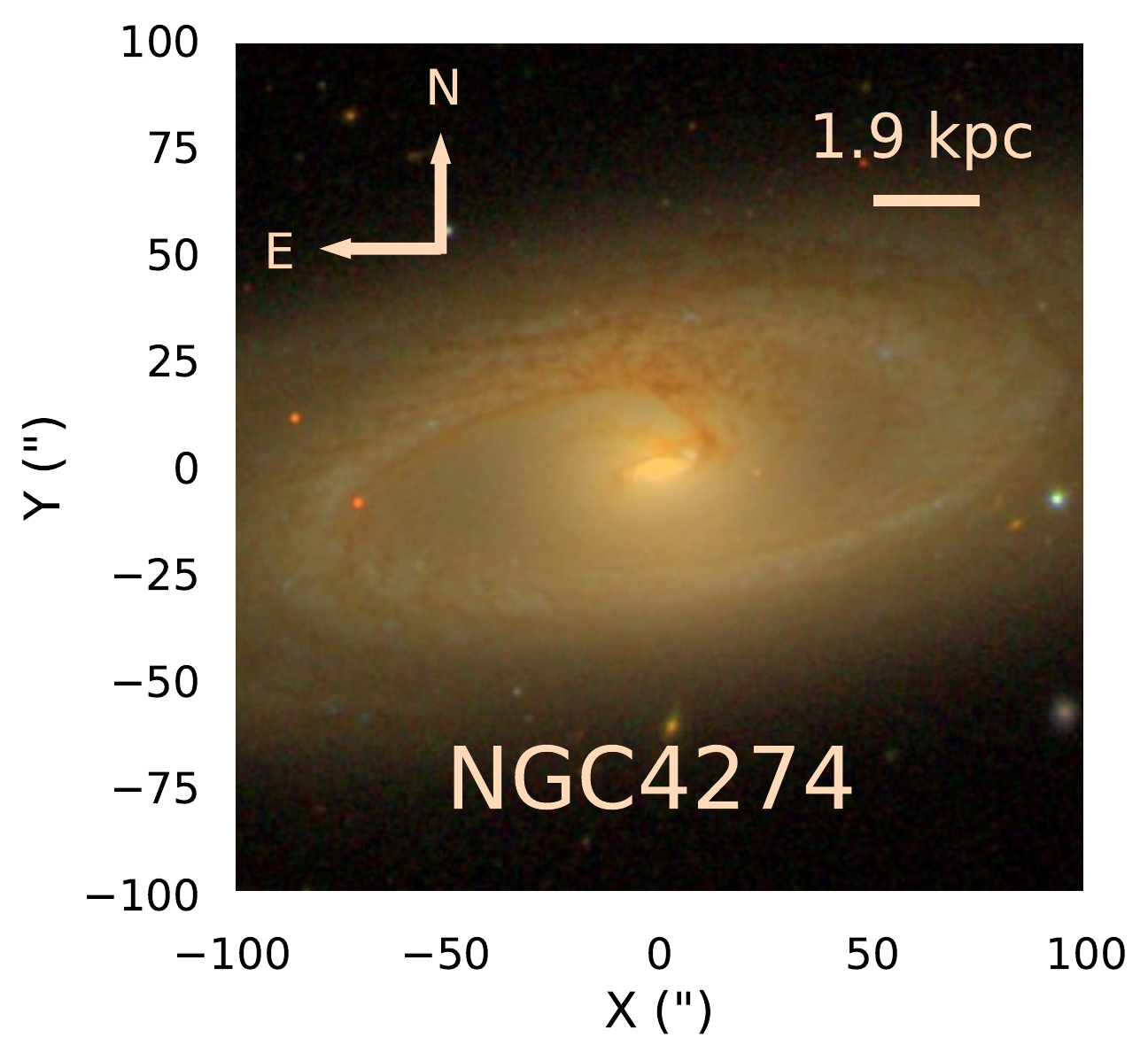}
\end{subfigure}
\hfill
\begin{subfigure}{0.19\textwidth}
\includegraphics[width=\textwidth]{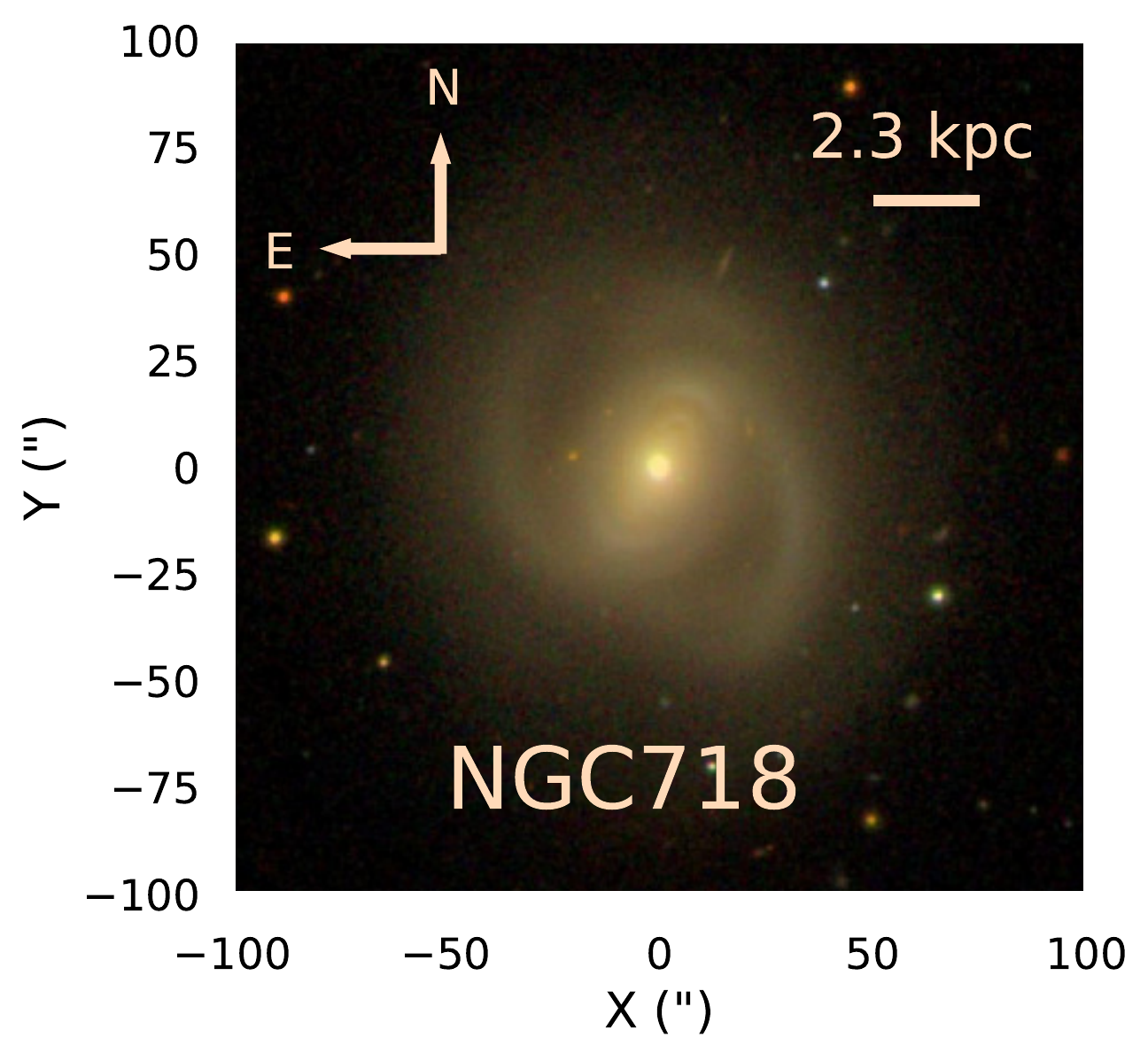}
\end{subfigure}
\hfill
\begin{subfigure}{0.19\textwidth}
\includegraphics[width=\textwidth]{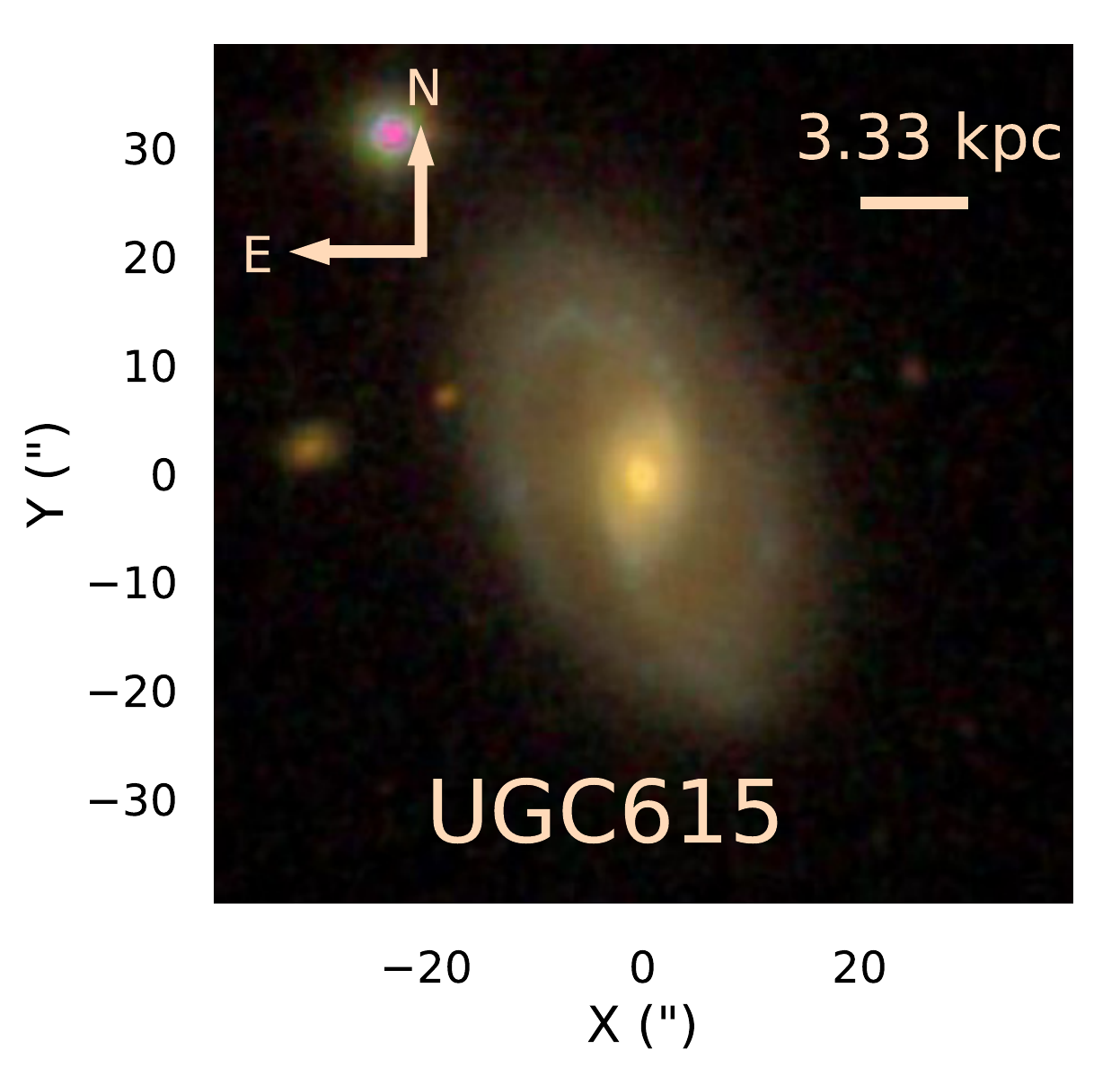}
\end{subfigure}
\caption{SDSS cutout $gri$ images of some example passive spiral galaxies (left) and their four galaxies from the comparison sample closest in both $z$ and stellar mass with the same T-type in the right four columns. While some of the galaxies in the comparison sample are red (a consequence of matching by T-type), the comparison galaxies generally show evidence for star formation including blue stellar populations and dust lanes.}
\label{ps_examples}
\end{figure*}

 \section{Quenching Mechanisms}
 \label{quenching_mech}
We search for viable quenching pathways for our sample of passive spiral galaxies by determining their mass, environmental, and internal structure properties. Given the dichotomy in galaxy properties and traditional quenching pathways present in mass-selected samples of galaxies \citep[e.g.][]{Geha12}, we split our analysis into low mass passive spirals in Section~\ref{low_mass} and high mass in Section~\ref{high_mass}.

\begin{table*}
\centering
\caption{Passive spiral galaxies in our sample and their properties. A horizontal line separates the five galaxies in the low mass subsample discussed in Section~\ref{low_mass} from the higher mass passive spiral galaxies in Section~\ref{high_mass}.}
\label{Table:PS}
\resizebox{\textwidth}{!}{
\begin{tabular}{l c c c c c c c c c c}
\hline
\textbf{Name}        & \textbf{RA}         & \textbf{Dec}       & \textbf{z}$^{1}$    & \textbf{$\textrm{D}$} $^{2}$     & \textbf{Stellar Mass}$^{3}$          & \textbf{T-type}$^{4}$ &  \textbf{$\textrm{N}_{\textrm{group}}^{5}$} & \textbf{Galaxy}$^{5}$  & \textbf{Bar?}$^{6}$ & \textbf{Ansa Bar?}$^{6}$  \\
                              & \textbf{(J2000)}  & \textbf{(J2000)}  &  &      \textbf{(Mpc)}             &   \textbf{($\textrm{M}_{\odot}$)}       &                                      &    & \textbf{Environment} & \\
\hline
 NGC 4440     & 186.9732  & 12.2932 & 0.0024 & 25.69 & 3.89$\times10^{9}$    & 1        & 197  & Satellite &   yes    & yes\\
  NGC 4277     & 185.0154 & 5.3414  & 0.0083 & 27.60 & 5.36$\times10^{9}$  & 1         & 39   & Satellite    &  yes  & no\\ 
  NGC 4880     & 195.0439   & 12.4833 & 0.0051 & 25.55 & 6.85$\times10^{9}$  & 1       & 1    & Isolated$^{7}$ &     no  &  --  \\
  NGC 4305     & 185.5150  & 12.7408 & 0.0064 & 26.59 & 8.84$\times10^{9}$  & 1         & 197    & Satellite  &  no   & --\\
  NGC 4264     & 184.8991 & 5.8468  & 0.0084 & 27.61 & 9.28$\times10^{9}$  & 1        & 39   & Satellite &   yes   & yes \\
  \hline
    NGC 4260    & 184.8427 & 6.0988  & 0.0060 & 38.58$^{*}$ & 1.97$\times10^{10}$  & 1         & 39   & Satellite  &   yes & no \\
      NGC 2692    & 134.2418  & 52.0660 & 0.0126 & 30.03$^{*}$ & 2.12$\times10^{10}$  & 1         & 3    & BGG &      yes & yes\\
  NGC 0357     & 15.8412  & -6.3392 & 0.0078 & 27.22 & 2.13$\times10^{10}$   & 1         & 1    & Isolated      &   yes & yes \\
  NGC 7743     & 356.0881 & 9.9341  & 0.0055 & 20.70$^{*}$ & 2.27$\times10^{10}$  & 1        & 2    & BGG  &    yes   & no \\
    NGC 2648    & 130.6658 & 14.2855 & 0.0069  & 34.01$^{*}$ & 2.45$\times10^{10}$   & 1         & 1    & Isolated$^{8}$ & no   &  --  \\
  NGC 656      & 25.6135  & 26.1431 & 0.0131 & 48.77 & 2.51$\times10^{10}$  & 1         & 1    & Isolated  &    yes & yes \\
  NGC 4608     & 190.3053 & 10.1558 & 0.0062 & 20.18$^{*}$ & 2.95$\times10^{10}$  & 1         & 197  & Satellite & yes  & yes  \\
  UGC 12800    & 357.5797 & 10.7574 & 0.0180 & 68.52 & 3.05$\times10^{10}$  & 1         & 1    & Isolated   &   yes  & yes\\
  NGC 4643     & 190.8339 & 1.9784  & 0.0044 & 26.44 & 3.22$\times10^{10}$  & 1        & 1   & Isolated    &    yes  & yes\\
  NGC 7563     & 348.9831 & 13.1962 & 0.0144  & 58.76$^{*}$ & 3.28$\times10^{10}$ & 1         & 7    & BGG  &  yes & yes \\
  NGC 2878     & 141.4477 & 2.0896  & 0.0243  & 117.63 & 3.30$\times10^{10}$ & 2         & 2    & BGG  & yes & no\\
  NGC 109     & 6.5610   & 21.8074 & 0.0182 & 60.16$^{*}$ & 3.34$\times10^{10}$  & 1         & 18    & Satellite &   yes  & yes  \\
    UGC 01271    & 27.2502  & 13.2112 & 0.0170 & 65.76 & 3.46$\times10^{10}$  & 1         & 10    & Satellite  &  yes  & yes \\
  NGC 538      & 21.3585  & -1.5506 & 0.0182  & 66.26$^{*}$ & 3.59$\times10^{10}$  & 2        & 43    & Satellite  &  yes   &  no\\
    NGC 345      & 15.3421  & -6.8843 & 0.0174  & 67.93 & 4.07$\times10^{10}$  & 1         & 9    & Satellite &    no  &--\\
  NGC 4596     & 189.9831 & 10.1761 & 0.0062 & 26.38 & 4.19$\times10^{10}$ & 1         & 197  & Satellite  &  yes & yes\\
  PGC 047732   & 203.3438 & 54.9491 & 0.0250  & 114.77 & 4.30$\times10^{10}$ & 2         & 1    & Isolated   &     yes & no \\
  UGC 8484   & 202.4019 & 32.4007 & 0.0247 & 117.32 & 4.47$\times10^{10}$  & 3         & 7    & Satellite  &   yes  & yes\\
  NGC 0015     & 2.2603   & 21.6245 & 0.0209 & 122.30$^{*}$ & 4.50$\times10^{10}$  & 1         & 1    & Isolated &    yes  & no \\
  PGC 070141   & 344.5540 & 25.2209 & 0.0251 & 99.78 & 4.81$\times10^{10}$  & 1        & 9    & Satellite &   yes   & yes  \\
  UGC 06163    & 166.7132 & 23.01627 & 0.0214 & 104.45 & 4.94$\times10^{10}$  & 1         & 4   & BGG  &    no &--\\
  NGC 3943     & 178.2358 & 20.4791 & 0.0220 & 107.65 & 4.95$\times10^{10}$  & 2         & 18   & Satellite   &   yes & yes \\
  PGC 67858 & 330.4222 & -2.0983 & 0.0269 & 109.82 & 5.07$\times10^{10}$   & 3         & 7    & Satellite  &    no &-- \\
  NGC 7383     & 342.3986  & 11.5564 & 0.0270 & 108.39 & 5.14$\times10^{10}$  & 1         & 10    & Satellite  &  yes &no   \\
  NGC 7389     & 342.5670 & 11.5662 & 0.0264  & 105.51 & 5.70$\times10^{10}$ & 3         & 10    & Satellite   &  yes & yes \\
  PGC 029301   & 151.4473 & 14.3387 & 0.0312  & 148.71 & 6.27$\times10^{10}$ & 5         & 6   & Satellite   &   yes & yes\\
  UGC 12897    & 0.1581   & 28.3845 & 0.0290 & 126.28$^{*}$ & 7.64$\times10^{10}$ & 2         & 6    & Satellite  &  no &-- \\
  NGC 550      & 21.6773  & 2.0224   & 0.0194  & 92.57$^{*}$ & 7.75$\times10^{10}$  & 1         & 8    & Satellite   &   no & -- \\
  NGC 2618     & 128.9731 & 0.7072  & 0.0134 & 61.05$^{*}$ & 8.17$\times10^{10}$  & 2         & 1    & Isolated & no &--\\
  NGC 3527     & 166.8258 & 28.5278  & 0.0333  & 107.39$^{*}$ & 8.52$\times10^{10}$  & 1         & 27   & Satellite &  yes & yes \\

\hline
\hline
\multicolumn{11}{l}{$^{1}$ From \citet{Bonne15}}\\
\multicolumn{11}{l}{$^{2}$ * denotes redshift independent distances from NED, collated by \citet{Bonne15}, otherwise these are flow-corrected distances, calculated by \citet{Bonne15}.}\\
\multicolumn{11}{l}{$^{3}$ From NASA Sloan Atlas.}\\
\multicolumn{11}{l}{$^{4}$ Compiled by \citet{Bonne15}, most of which are from \citet{Paturel03}.}\\
\multicolumn{11}{l}{$^{5}$ Group information from \citet{Tully15}.}\\
\multicolumn{11}{l}{$^{6}$ From visual inspection by the authors.}\\
\multicolumn{11}{l}{$^{7}$ Whilst listed as isolated by \citet{Tully15}, we expect this galaxy to be within the Virgo cluster \citep[e.g.][]{deVauc61,Eastmond78}.} \\
\multicolumn{11}{l}{$^{8}$ Whilst listed as isolated by \citet{Tully15}, this galaxy has a close companion confirmed by SDSS imaging and spectroscopy.}\\

\end{tabular}
}
\end{table*}

\begin{table*}
\centering
\caption{The environments of both the passive spiral and mass, $z$, and T-type-matched comparison samples as matched to the \citet{Tully15} catalogue. There is no significant difference in group fraction between the passive spiral and comparison samples.}
\label{env_table}
\begin{tabular}{l c c c c}
\hline
   & \textbf{\% in groups of $N\geq 2 $}  & \textbf{\% BGGs} & \textbf{\% in clusters of  $N\geq 10 $}& \textbf{\% Isolated}\\
   \hline
\textbf{Passive Spiral Sample}  & $74\pm15\%$ (26/35)    & $14\pm6\%$(5/35)  &  $20\pm8\%$ (7/35)&$26\pm9\% $ (9/35)   \\ 
\textbf{Comparison Sample} & $61\pm7\%$ (85/140) & $20\pm 4\%$ (28/140) & $19\pm4\%$ (26/140)  & $39\pm5 \%$  (55/140)      \\
\hline
\end{tabular}
\end{table*}

\begin{table*}
\centering
\caption{Bar fractions in the passive spiral sample and the mass, $z$, and T-type-matched comparison sample. The bar fraction and ansa bar fraction of the passive spiral sample are much higher than those of the comparison sample, suggesting bars are involved in the quenching of passive spirals.}
\label{bar_frac_table}
\begin{tabular}{l c c}
\hline
   & \textbf{Bar Fraction}  & \textbf{Ansa Bar Fraction}\\
   \hline
\textbf{Passive Spiral Sample}  &     $74 \pm 15\%$  (26/35)     &   $69 \pm 16\%   $ (18/26)     \\ 
\textbf{Control Sample} & $36 \pm 5\%$ (51/140)     &  $29 \pm 8\%$ (15/51)         \\
\hline
\end{tabular}
\end{table*}

\subsection{The Low Mass Regime}
\label{low_mass}
Environmental quenching can account for nearly all quiescent low mass galaxies at low redshift \citep[e.g.][]{Bamford09, Peng10, Geha12, Kawinwanichakij17}.
Motivated by studies such as these, we examine the environmental properties of the low mass passive spirals in our sample.

\citet{Wolf09} found that optically-red low mass spiral galaxies are rare -- indeed, there are only five with $\textrm{M}_{\star}<1\times10^{10}~\textrm{M}_{\odot}$ in our sample.
These five galaxies - NGC 4440, NGC 4277, NGC 4880, NGC 4305, and NGC 4264, are shown as postage stamp images with their four comparison galaxies in Figure~\ref{ps_lowmass}. 

As a first pass, in Figure~\ref{skyplot} we consider the positions of the low mass passive spirals (gold stars) and their mass, $z$, and T-type-matched comparison galaxies (gold squares) on the sky.
Immediately from their right ascension, declination, and distance listed in Table~\ref{Table:PS}, we notice the low mass passive spiral galaxies are all part of the Virgo cluster. This is in line with results such as \citet{Bamford09}, who found that low mass spirals in the densest regions are mostly optically red.
While the low mass passive spiral galaxies are all satellites, as predicted by the \citet{Peng10} model, none are located in groups, and are instead all members of a rich cluster. 
The same is not true of the low mass comparison galaxies, which are spread across all environments. 
Of the comparison sample galaxies that are satellites, some show obvious star formation in their colour images (e.g. NGC 3380 and NGC 4413 in Figure~\ref{ps_lowmass}). It seems that being a low mass spiral galaxy in a group or cluster is a necessary, but not sufficient condition for quenching. Therefore, the fact these low mass passive spiral galaxies are all in Virgo (and not just any sized group), is very significant.

There are many low mass passive satellite galaxies in groups in the \citet{Bonne15} sample, but these are mostly T-type$<$1 galaxies without discernible spiral structure.
The obvious inference is that upon entering a group, spiral galaxies that have their star formation quenched via environmental processes will also transform their morphology from late to early type. 
Given the only place we have found low mass passive spirals is within the rich cluster environment, we question whether the quenching mechanism that preserves spiral morphology occurs exclusively in galaxy clusters.
Alternatively, gas stripping in rich clusters may occur so quickly that we observe passive spirals before their morphology is transformed. Ram-pressure stripping has been shown in simulations to act on timescales as short as $\sim$2 Gyr \citep{Fillingham15}, for low mass galaxies, and could certainly explain our results.
Either way, we speculate that the only way a spiral can avoid the violent group processes that transform its morphology is by instead entering a cluster. Cluster-specific processes such as ram-pressure stripping or strangulation can then act to strip gas gently and quench the galaxy whilst preserving its morphology.
 
 We note here that a significant portion of our nearby galaxies are located within the Virgo cluster. All of our low mass passive spirals and $5/20$ of their comparison sample counterparts are located within Virgo.
 We also note that we have a small sample size. We therefore determine that the chance that five galaxies are randomly located in Virgo given that 25\% of low mass comparison galaxies are within Virgo to be 0.1\% by the binomial theorem. 
 From this we determine there is only a small chance that the passive spirals may be located in Virgo by random chance.

 From the above analysis, we may infer that cluster-scale gas stripping mechanisms such as ram-pressure stripping and/or strangulation may be the mechanism(s) responsible for ceasing star formation in low mass ($\textrm{M}_{\star}<1\times10^{10}~\textrm{M}_{\odot}$) passive spiral galaxies. 

\subsection{Higher Mass Passive Spirals}
\label{high_mass}
 If prior literature on red spiral galaxies are a guide, it is unlikely higher mass ($\textrm{M}_{\star}>1\times10^{10}~\textrm{M}_{\odot}$) passive spiral galaxies will occupy specific environments or display unique morphologies. 
Previous samples of red spiral galaxies defined by optical colour selection criteria place red spirals in denser regions on average than their more star forming counterparts \citep[e.g.][]{Bamford09, Masters10}, and are more likely to be satellites \citep{Skibba09}, but with considerable spread among these trends. Optically red spirals have been found at all environmental densities, and indeed we see a range of environments listed in Table~\ref{Table:PS}. 

To quantify this, we match the passive spiral and comparison sample to the groups catalogue of \citet{Tully15}, which is an all-sky groups catalogue using the 2MASS Redshift Survey as an input catalogue. While the fraction of passive spiral galaxies in groups (defined as two or more members) is high at $74\pm15\%$, it is comparable to the control sample of spiral galaxies with $61\pm7\%$. The fraction of brightest group galaxies (BGGs) and fraction of galaxies located in clusters (ten or more members) are comparable for both the passive spiral and comparison sample galaxies, though passive spirals are slightly less likely to be isolated ($26\pm9\%$ compared to $39\pm5\%$ for the comparison sample). The environment fractions are listed in Table~\ref{env_table}, and the passive spiral and comparison sample group properties listed in Tables~\ref{Table:PS} and~\ref{Table:comp} respectively. 
Given the lack of environmental trends seen in the passive spiral sample when compared to the comparison sample, we turn instead to other quenching mechanisms, and examine the internal structure of the galaxies in our sample.

\begin{figure*}  
\centering
\begin{subfigure}{0.19\textwidth}
\includegraphics[width=\textwidth]{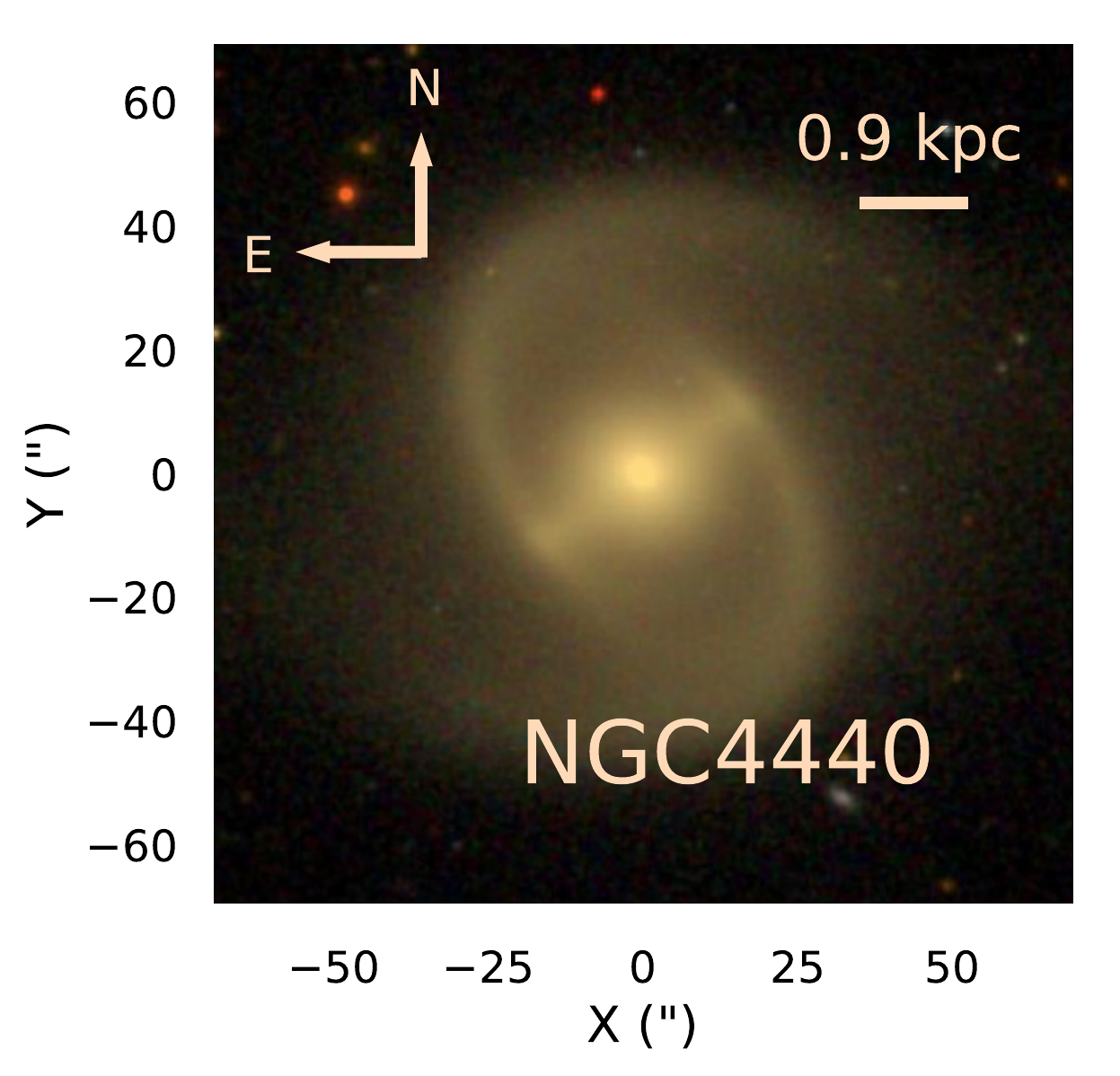}
\end{subfigure}
\hfill
\begin{subfigure}{0.19\textwidth}
\includegraphics[width=\textwidth]{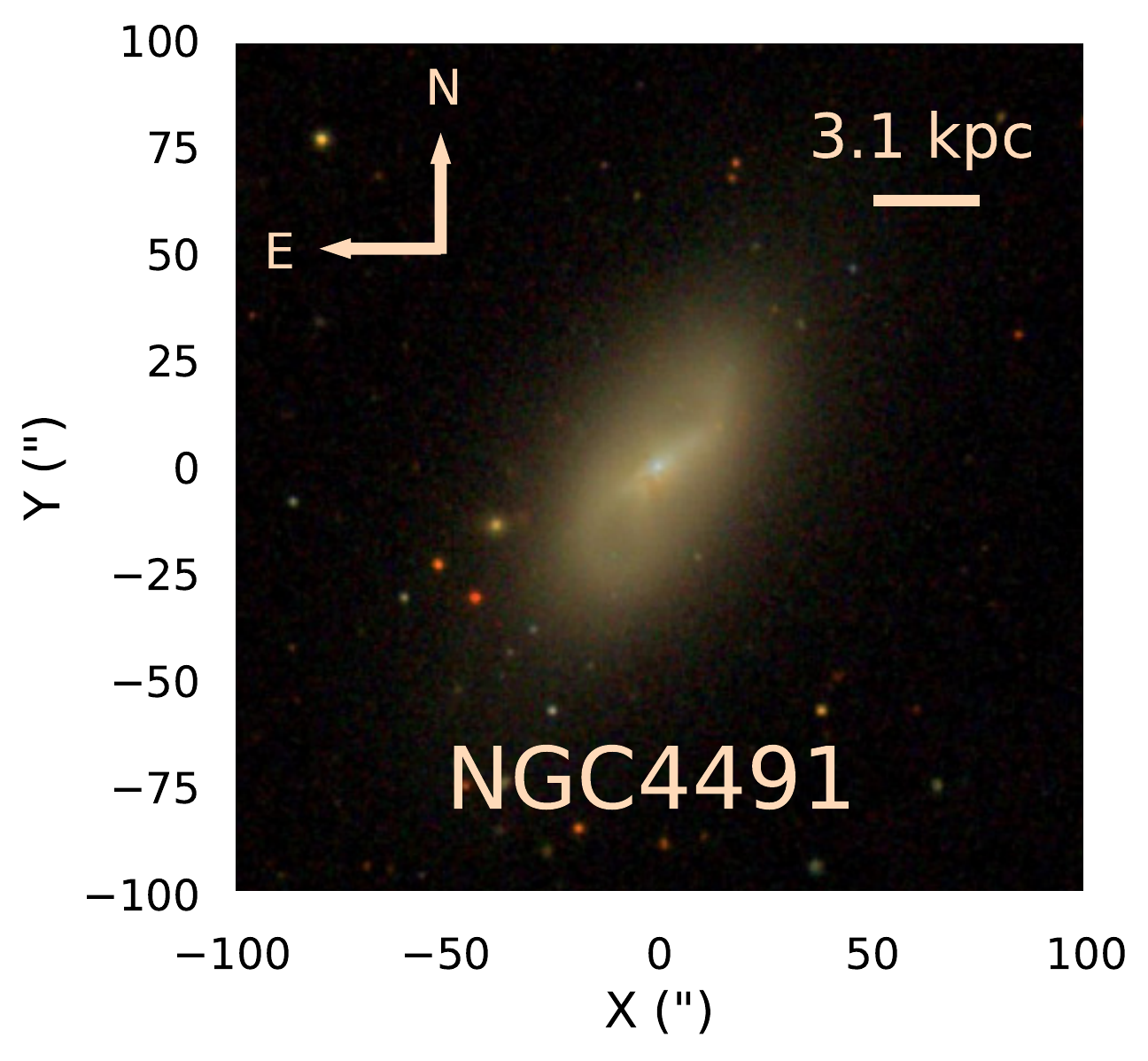}
\end{subfigure}
\hfill
\begin{subfigure}{0.19\textwidth}
\includegraphics[width=\textwidth]{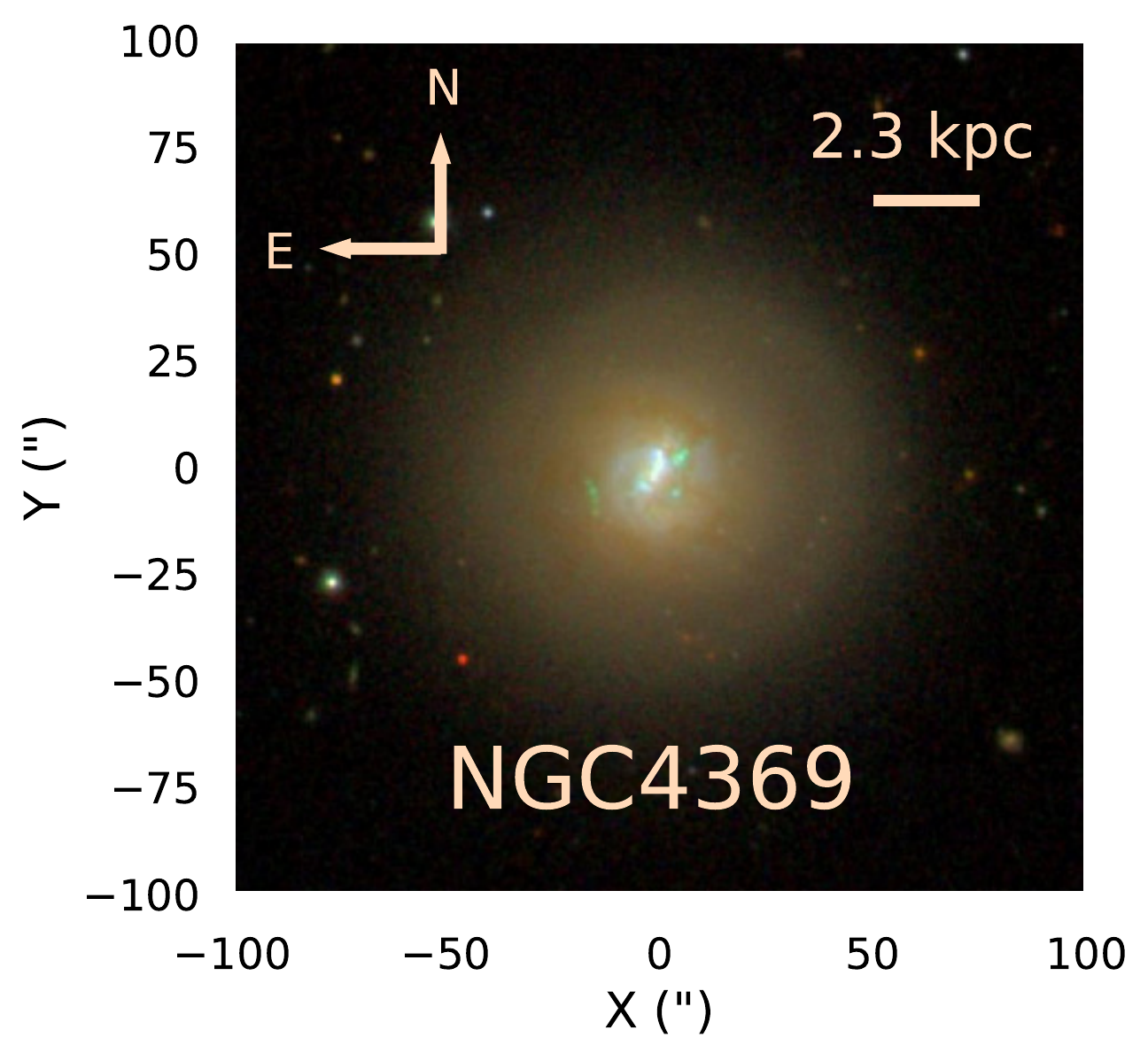}
\end{subfigure}
\hfill
\begin{subfigure}{0.19\textwidth}
\includegraphics[width=\textwidth]{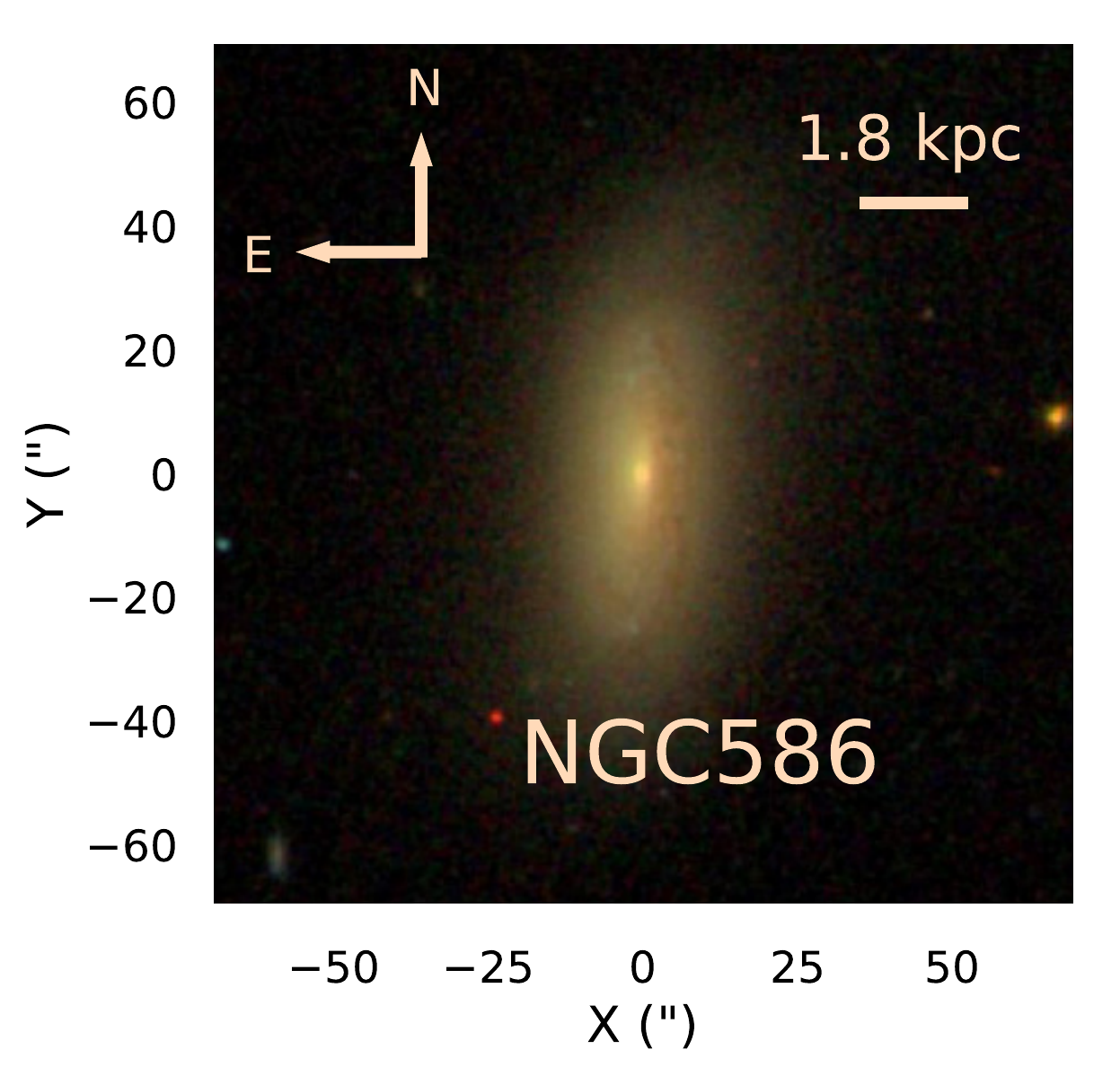}
\end{subfigure}
\hfill
\begin{subfigure}{0.19\textwidth}
\includegraphics[width=\textwidth]{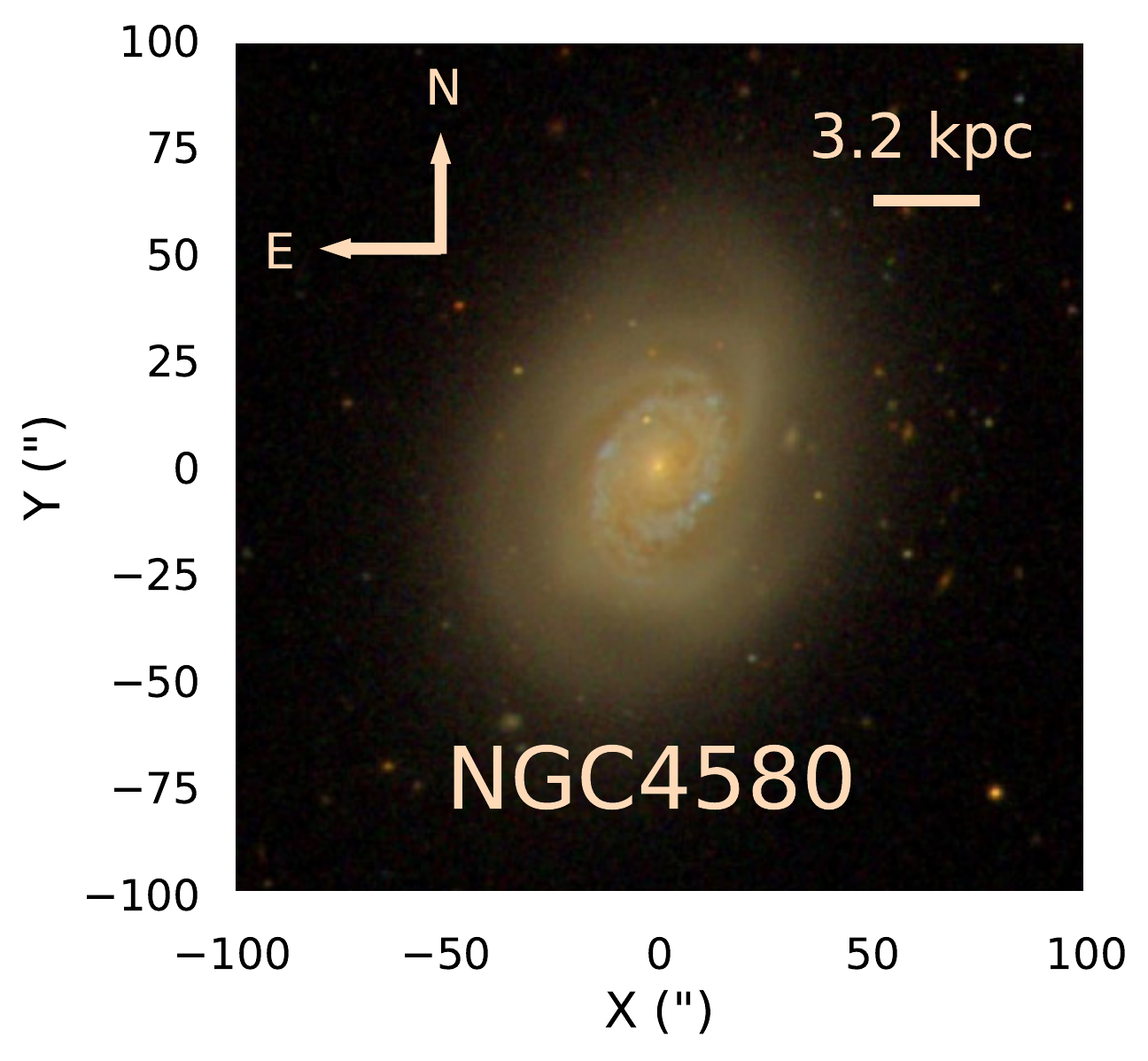}
\end{subfigure}

\vskip\baselineskip

\begin{subfigure}{0.19\textwidth}
\includegraphics[width=\textwidth]{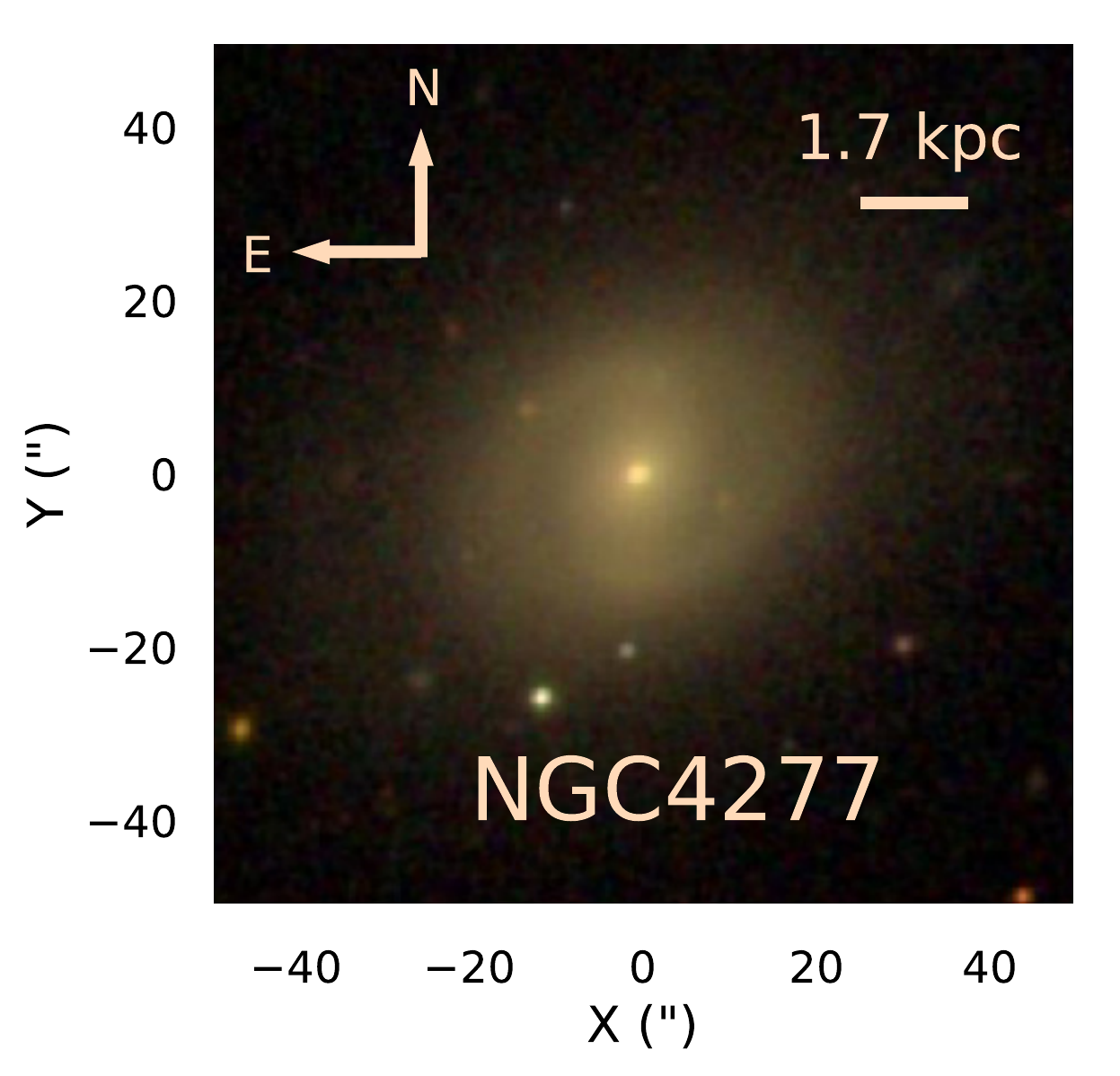}
\end{subfigure}
\hfill
\begin{subfigure}{0.19\textwidth}

\includegraphics[width=\textwidth]{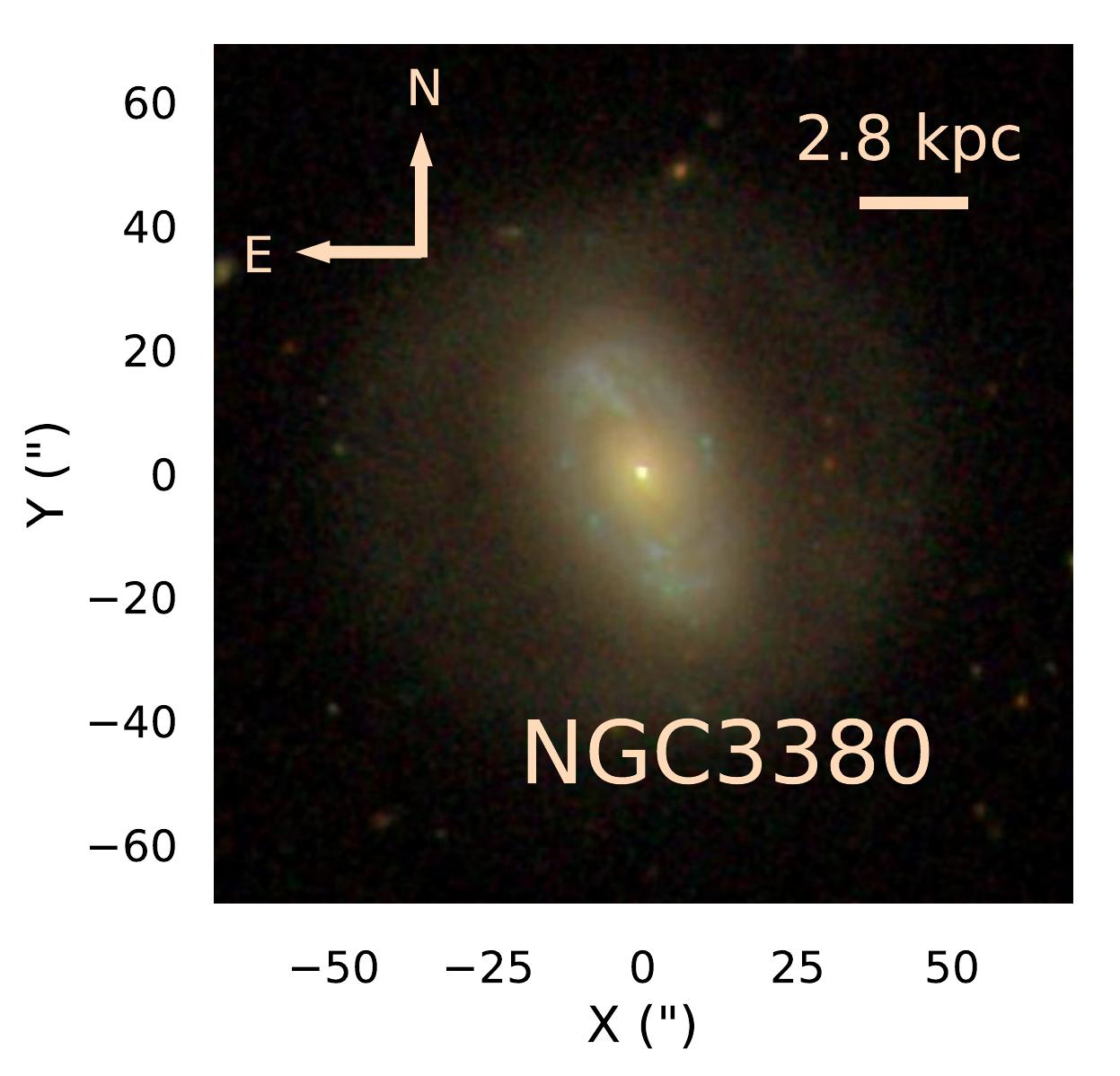}
\end{subfigure}
\hfill
\begin{subfigure}{0.19\textwidth}
\includegraphics[width=\textwidth]{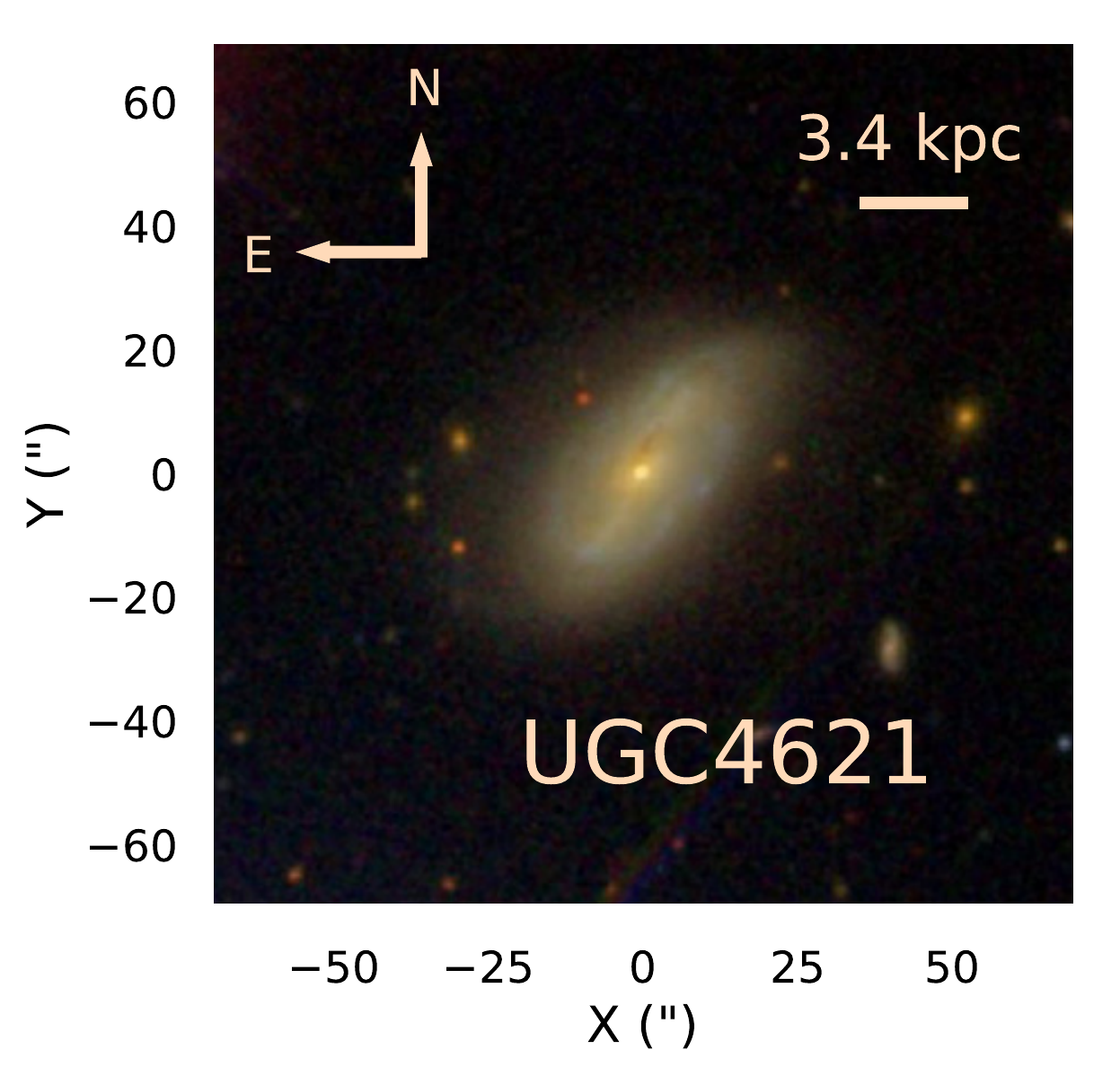}
\end{subfigure}
\hfill
\begin{subfigure}{0.19\textwidth}
\includegraphics[width=\textwidth]{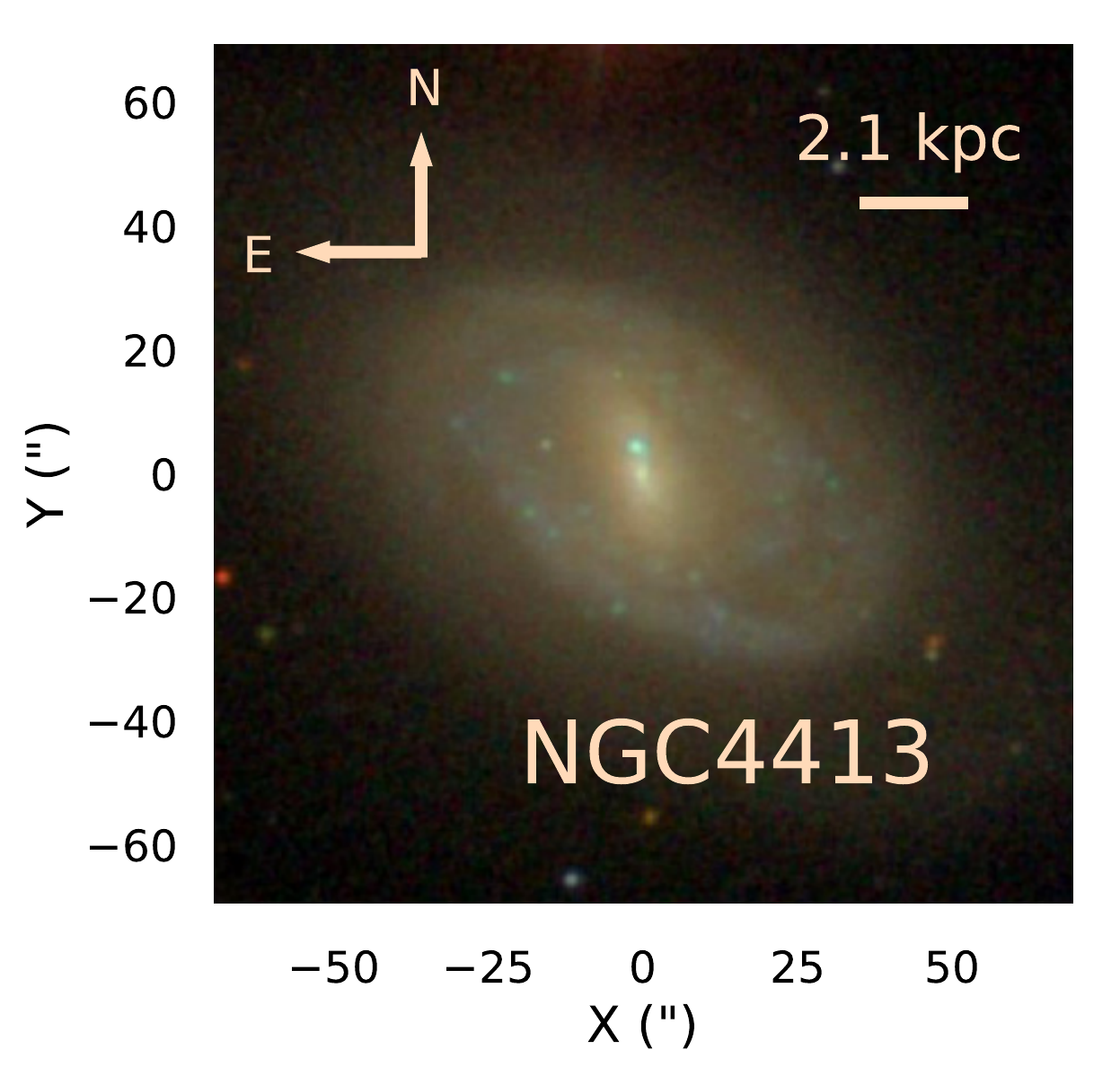}
\end{subfigure}
\hfill
\begin{subfigure}{0.19\textwidth}
\includegraphics[width=\textwidth]{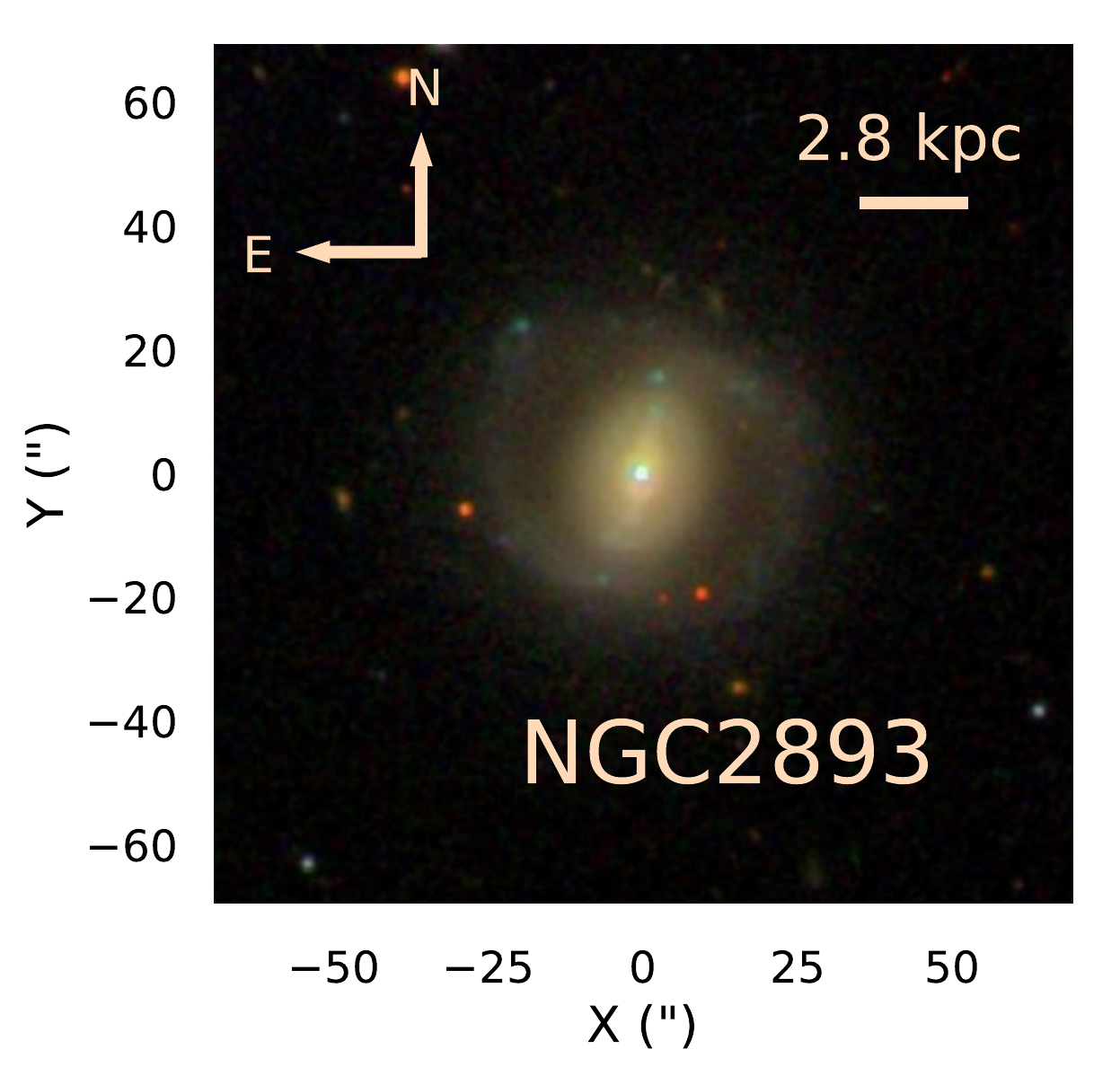}
\end{subfigure}

\vskip\baselineskip

\begin{subfigure}{0.19\textwidth}
\includegraphics[width=\textwidth]{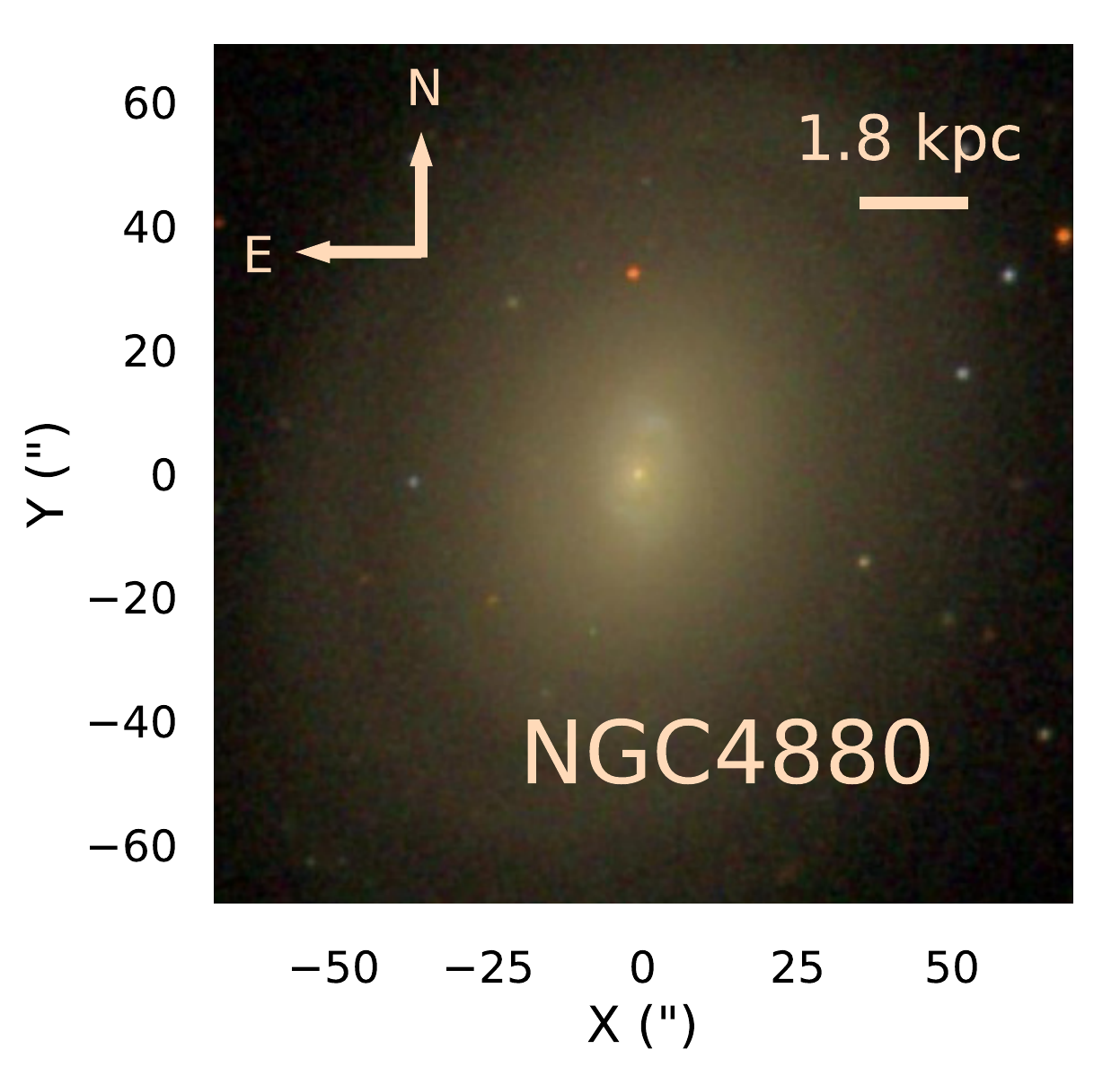}
\end{subfigure}
\hfill
\begin{subfigure}{0.19\textwidth}
\includegraphics[width=\textwidth]{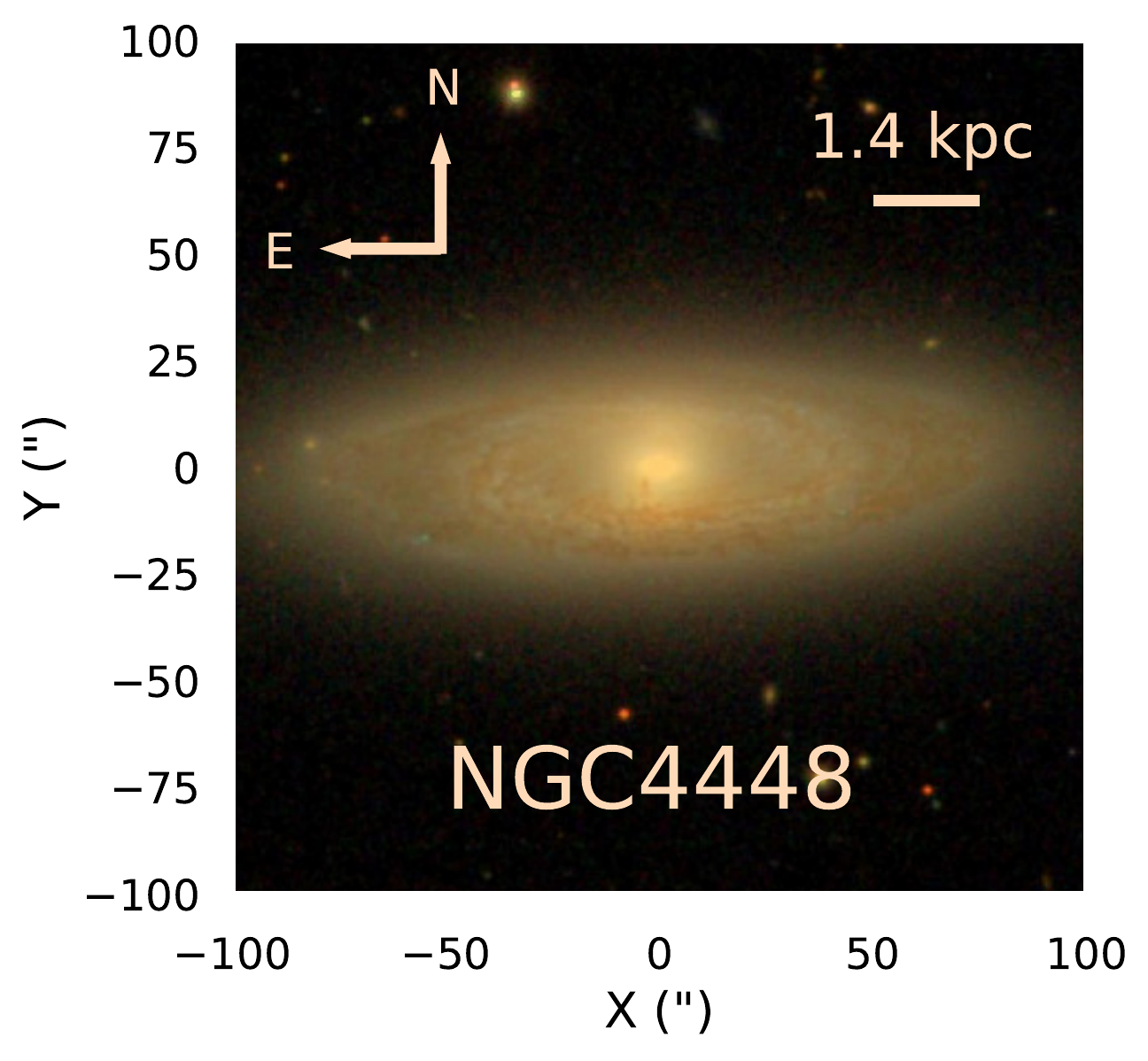}
\end{subfigure}
\hfill
\begin{subfigure}{0.19\textwidth}
\includegraphics[width=\textwidth]{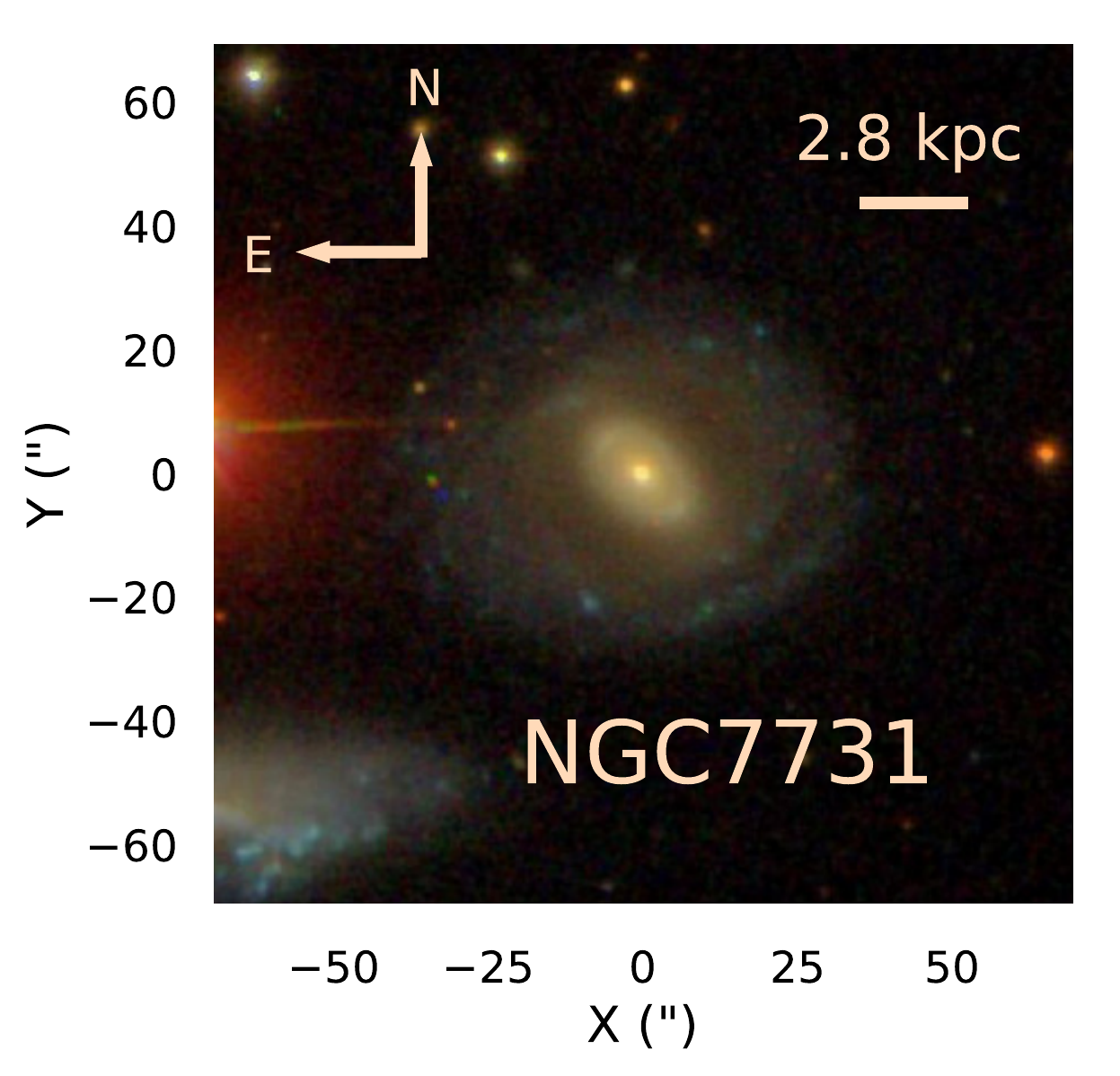}
\end{subfigure}
\hfill
\begin{subfigure}{0.19\textwidth}
\includegraphics[width=\textwidth]{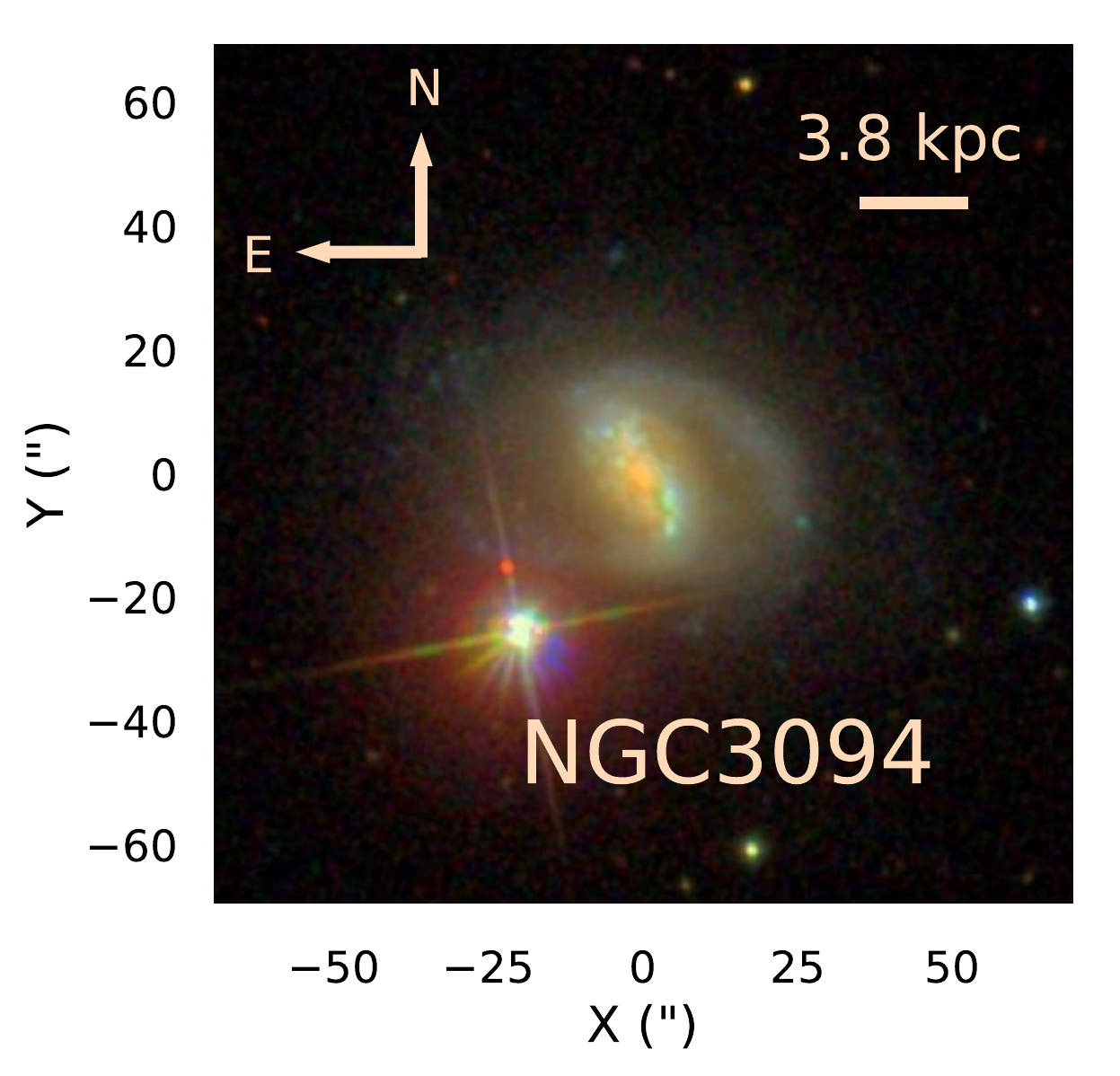}
\end{subfigure}
\hfill
\begin{subfigure}{0.19\textwidth}
\includegraphics[width=\textwidth]{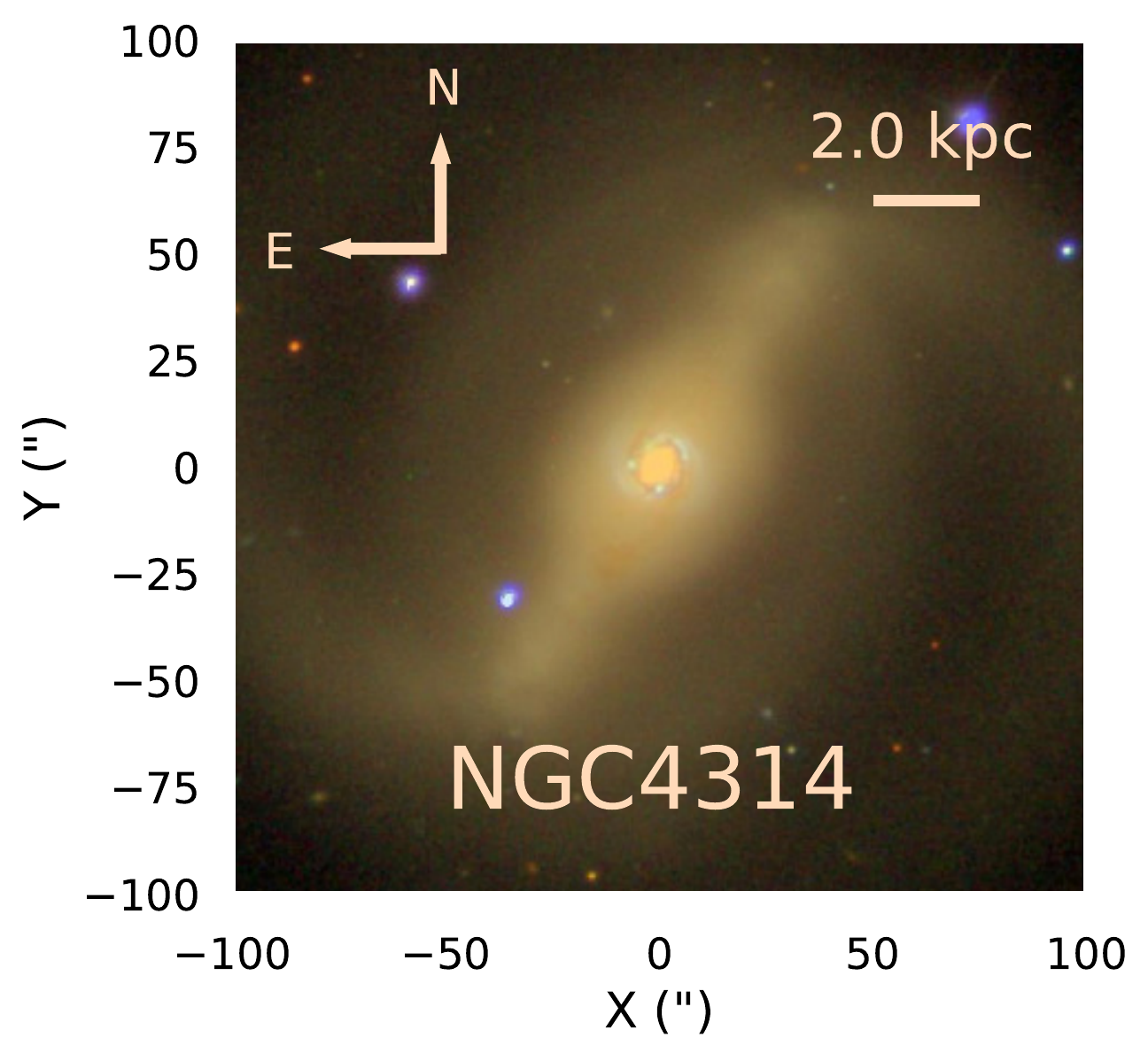}
\end{subfigure}

\vskip\baselineskip

\begin{subfigure}{0.19\textwidth}
\includegraphics[width=\textwidth]{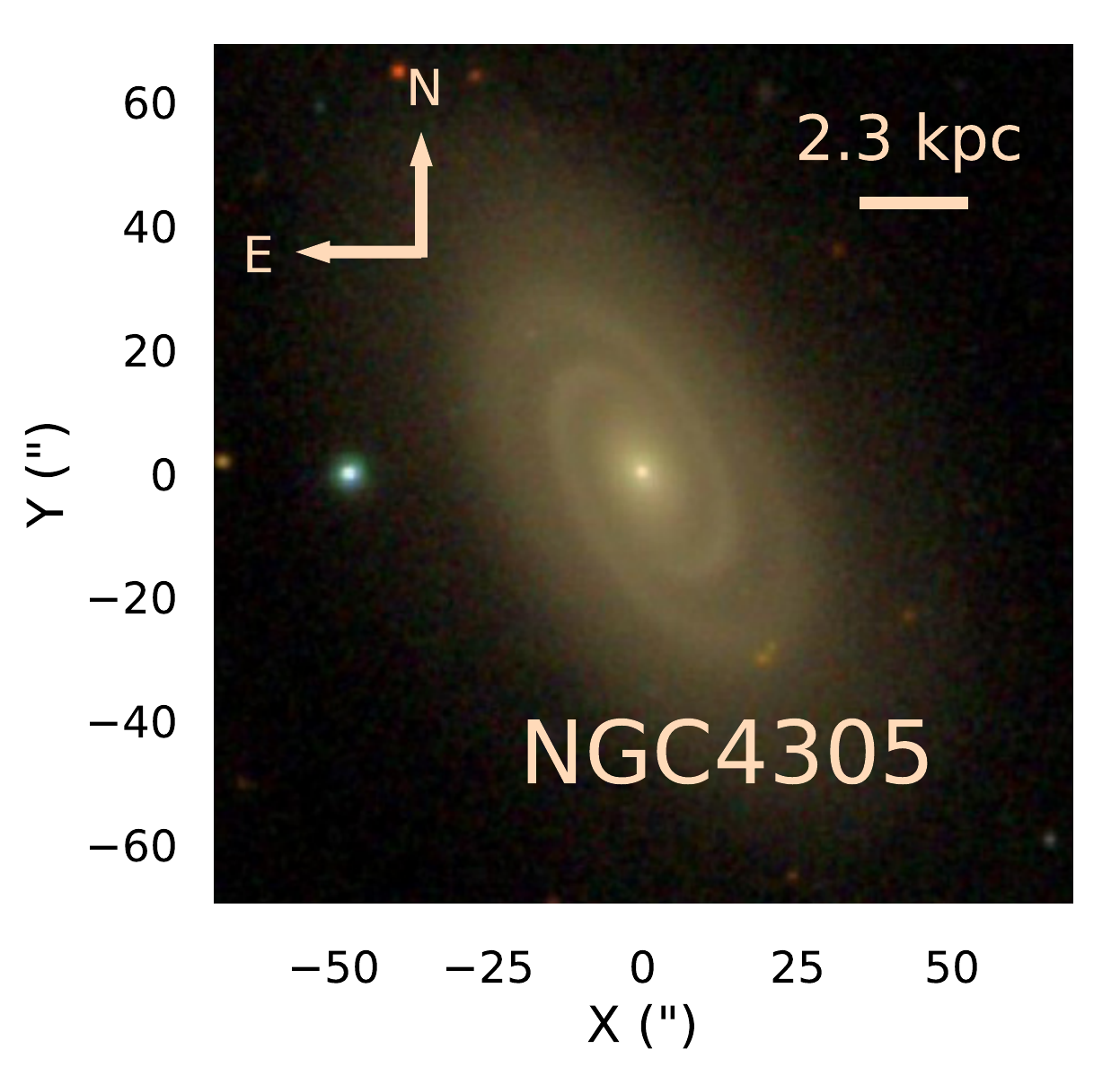}
\end{subfigure}
\hfill
\begin{subfigure}{0.19\textwidth}
\includegraphics[width=\textwidth]{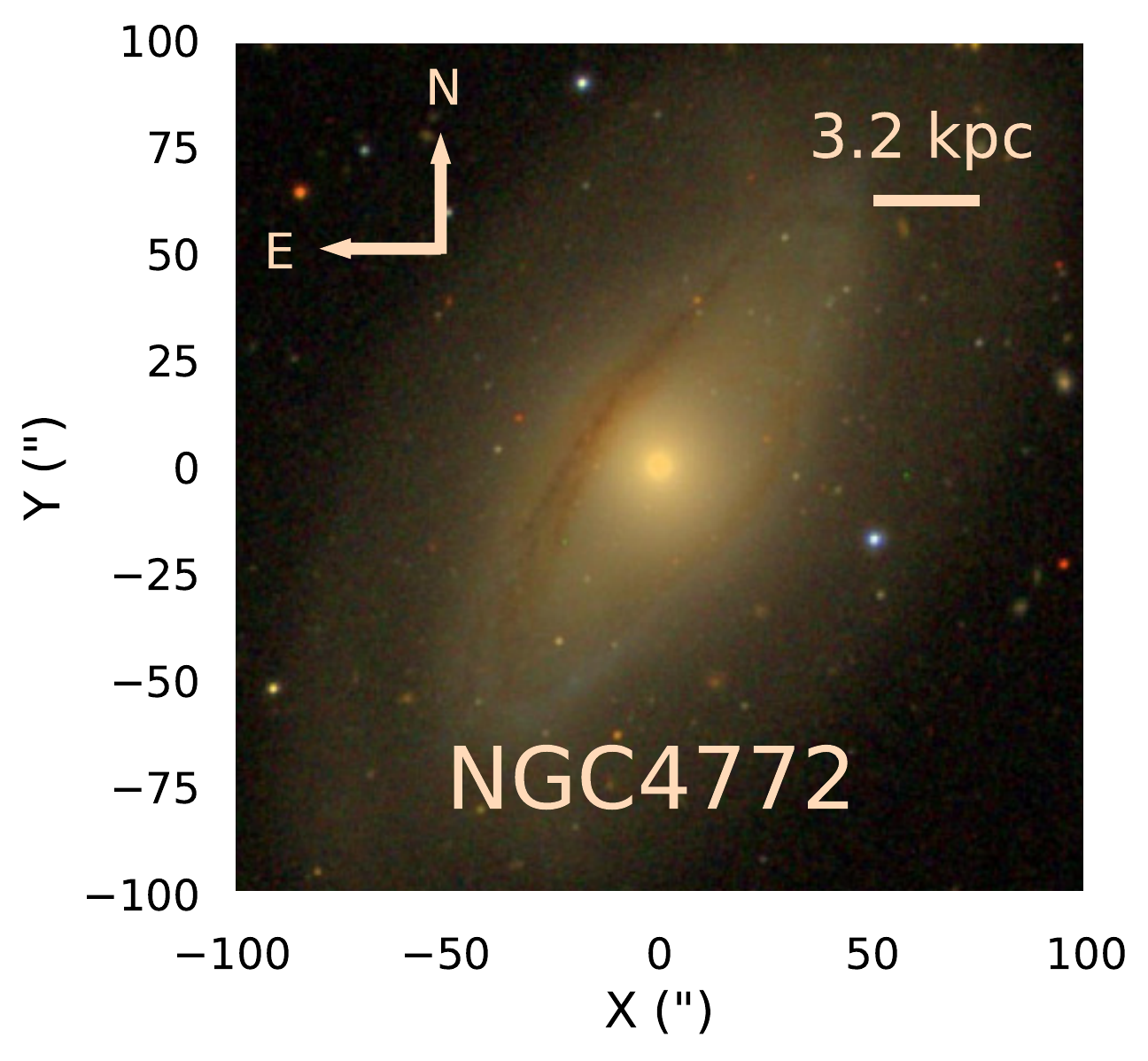}
\end{subfigure}
\hfill
\begin{subfigure}{0.19\textwidth}
\includegraphics[width=\textwidth]{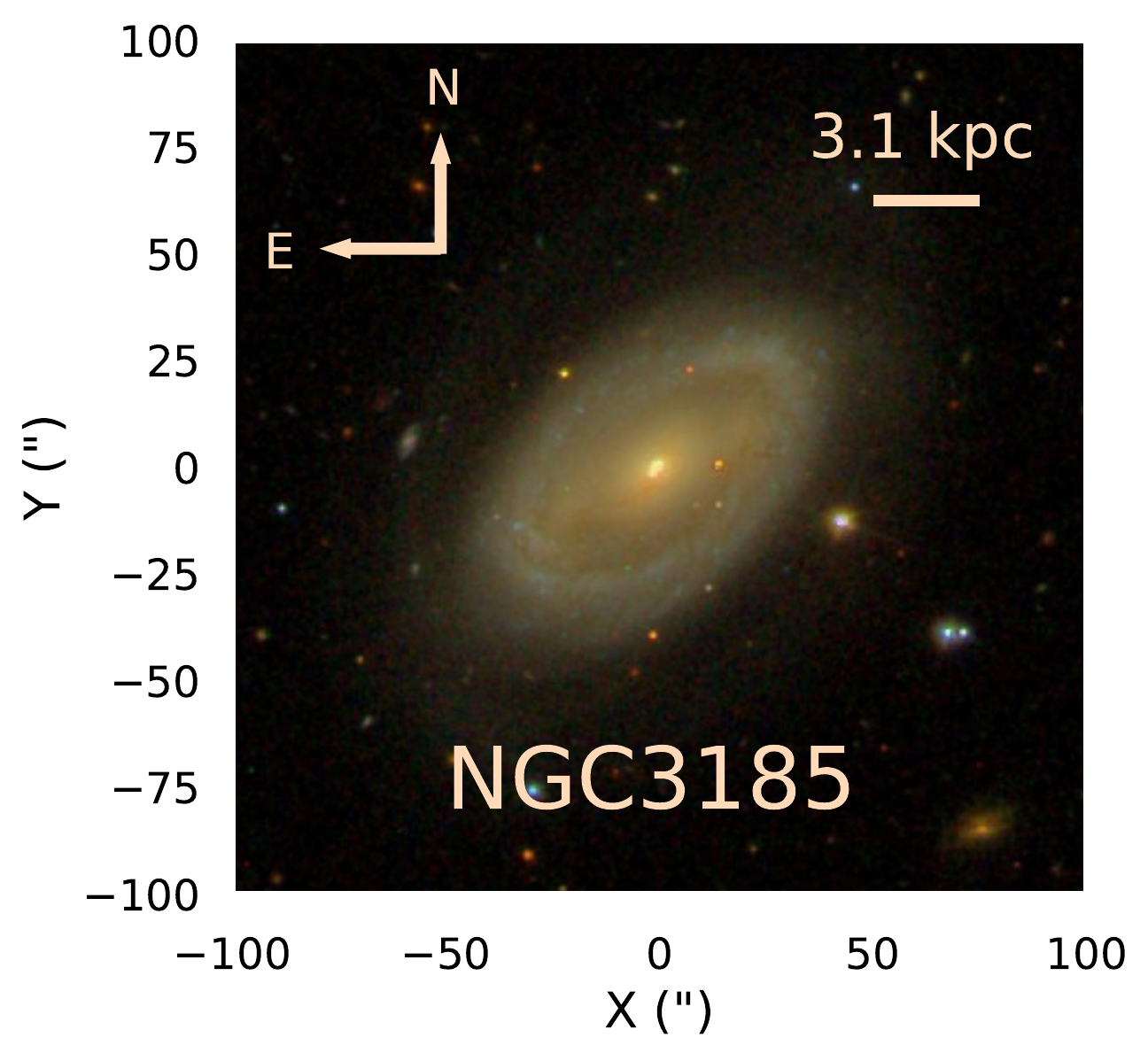}
\end{subfigure}
\hfill
\begin{subfigure}{0.19\textwidth}
\includegraphics[width=\textwidth]{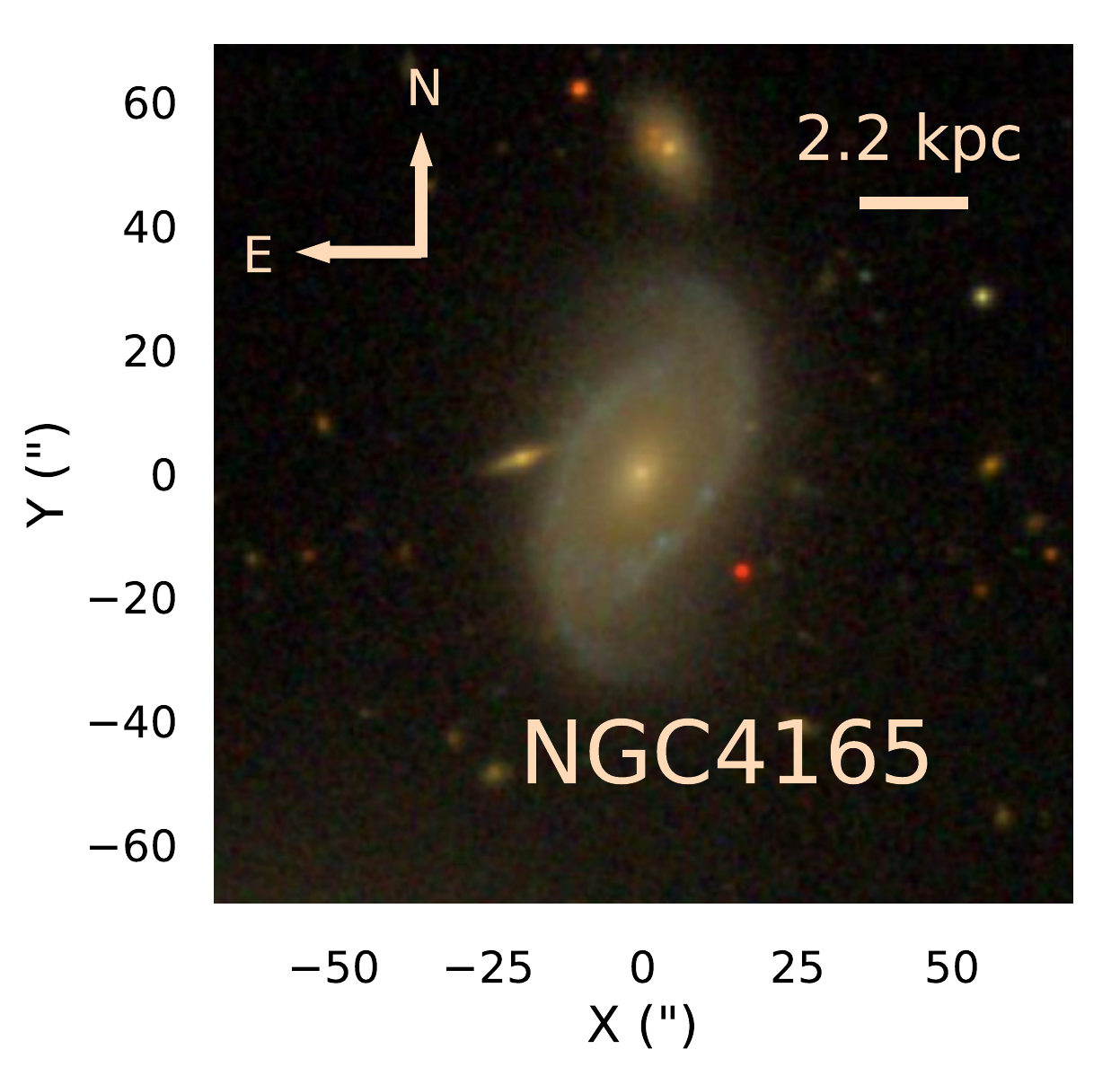}
\end{subfigure}
\hfill
\begin{subfigure}{0.19\textwidth}
\includegraphics[width=\textwidth]{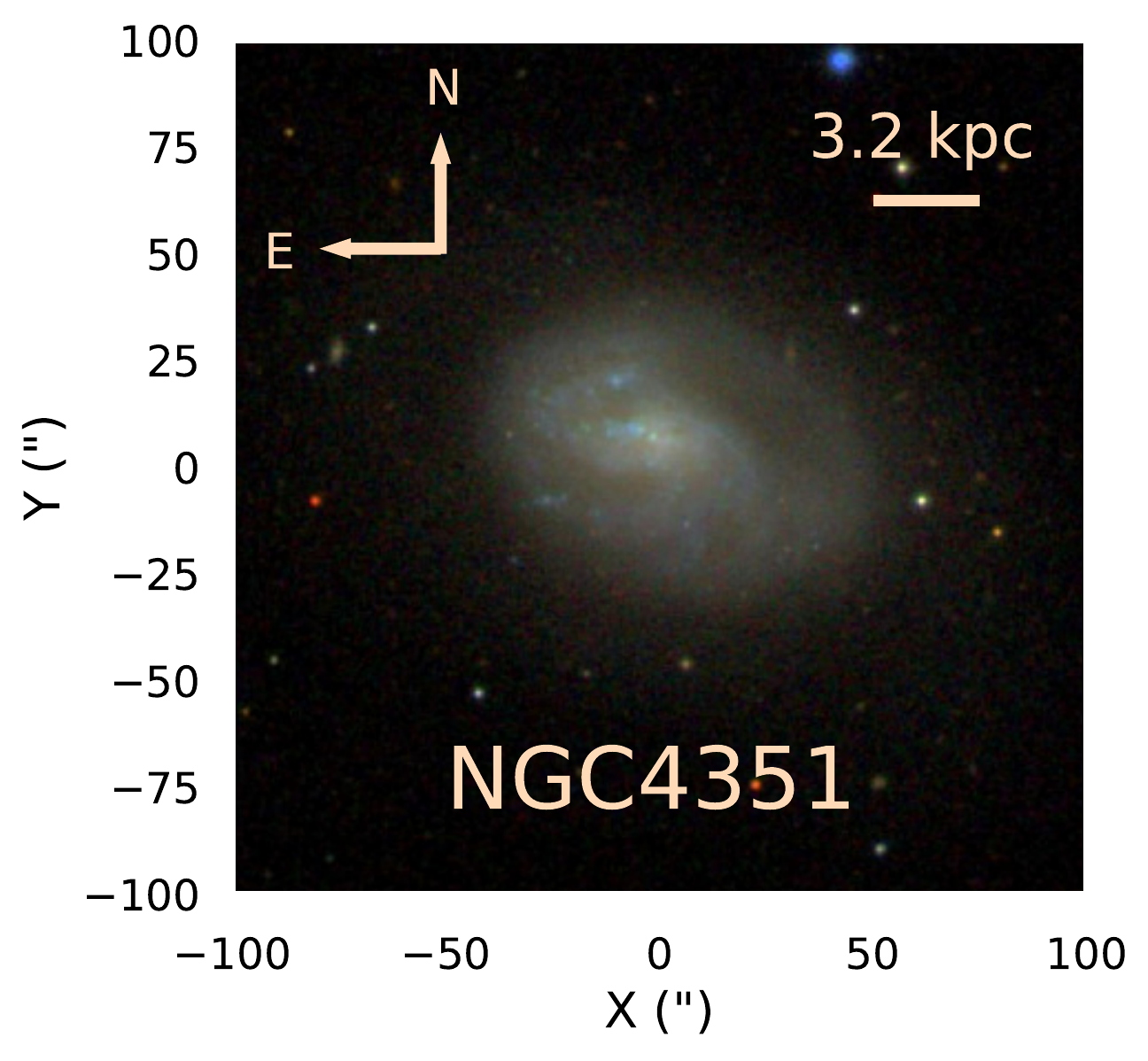}
\end{subfigure}

\vskip\baselineskip

\begin{subfigure}{0.19\textwidth}
\includegraphics[width=\textwidth]{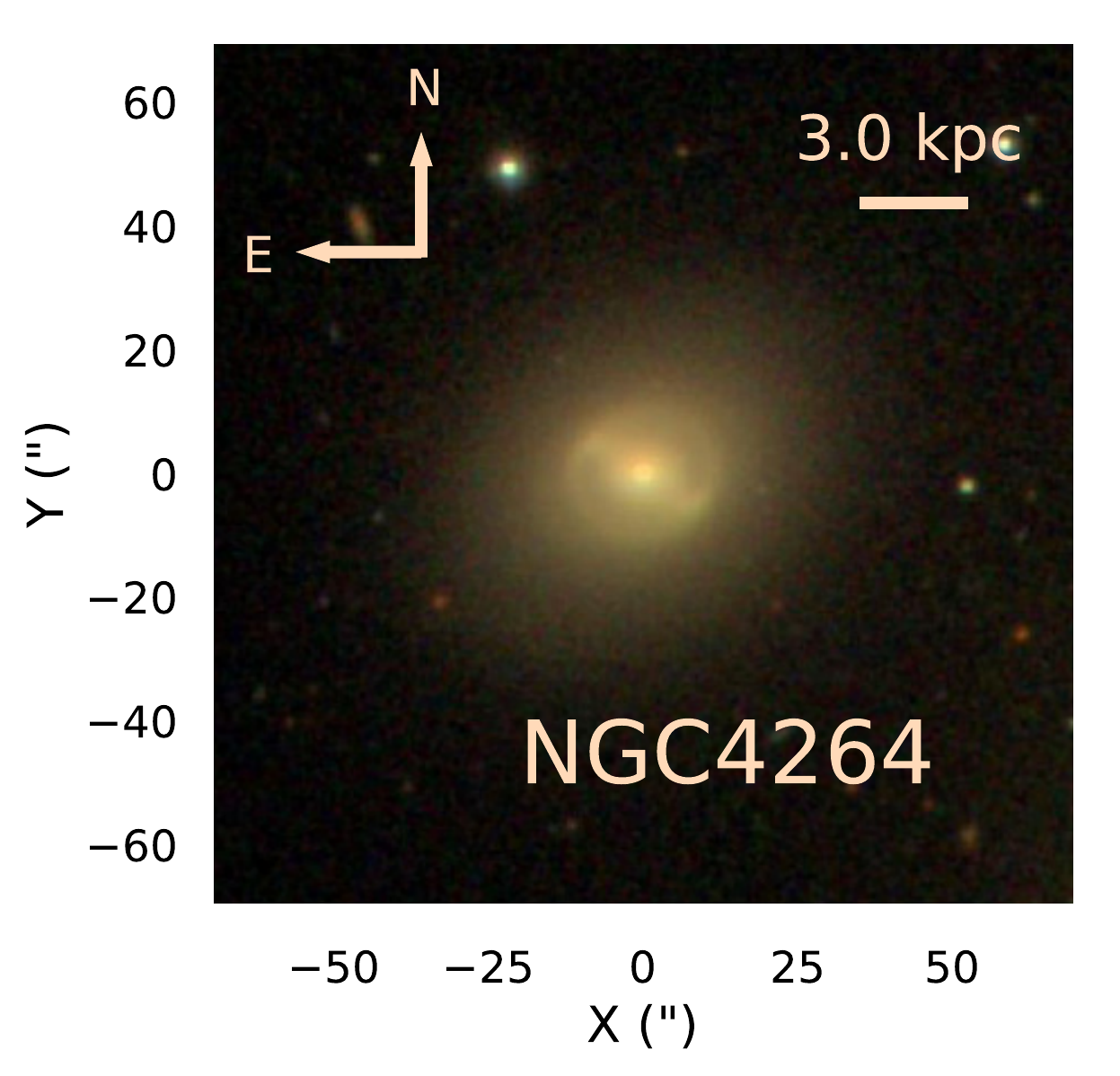}
\end{subfigure}
\hfill
\begin{subfigure}{0.19\textwidth}
\includegraphics[width=\textwidth]{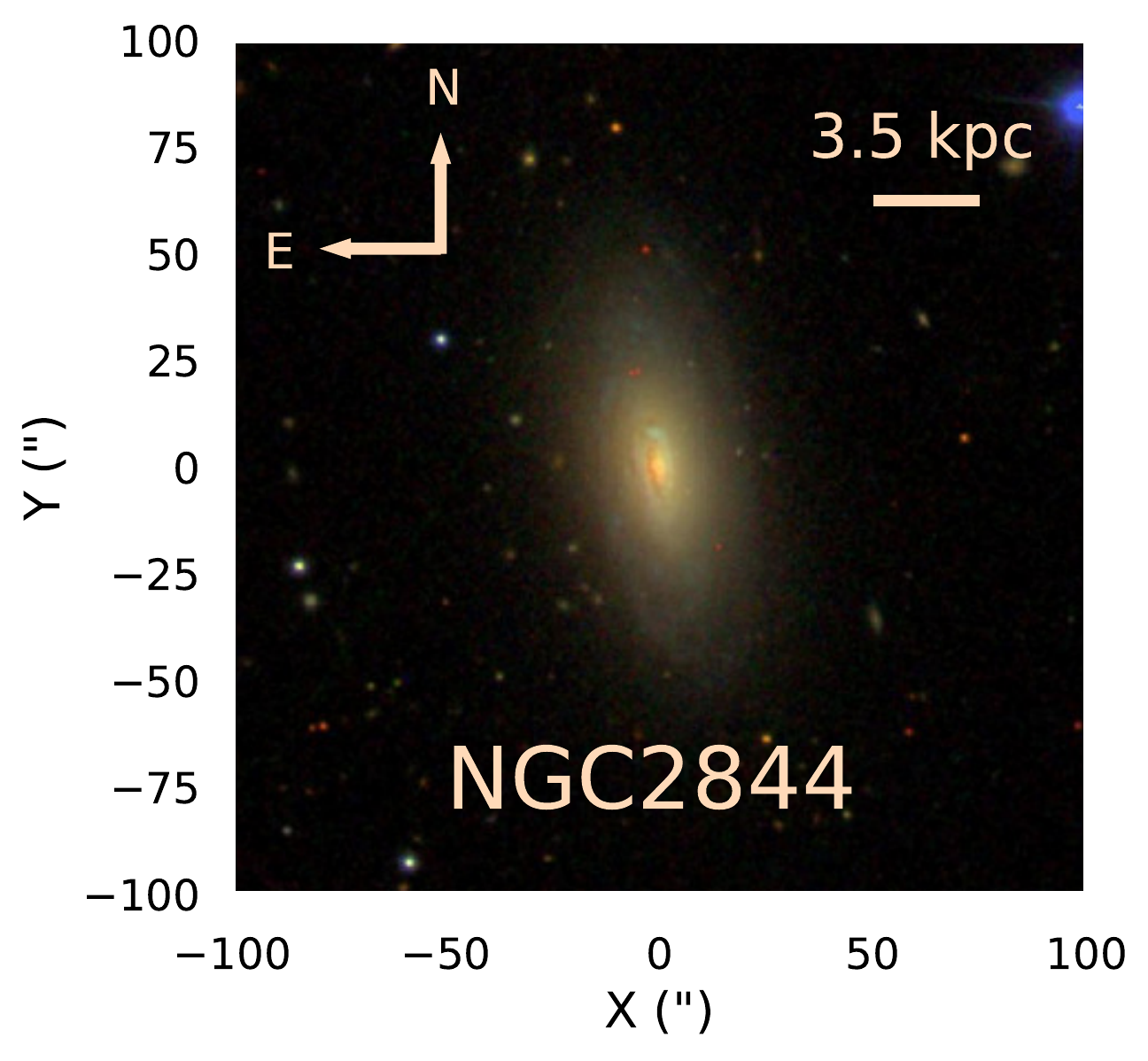}
\end{subfigure}
\hfill
\begin{subfigure}{0.19\textwidth}
\includegraphics[width=\textwidth]{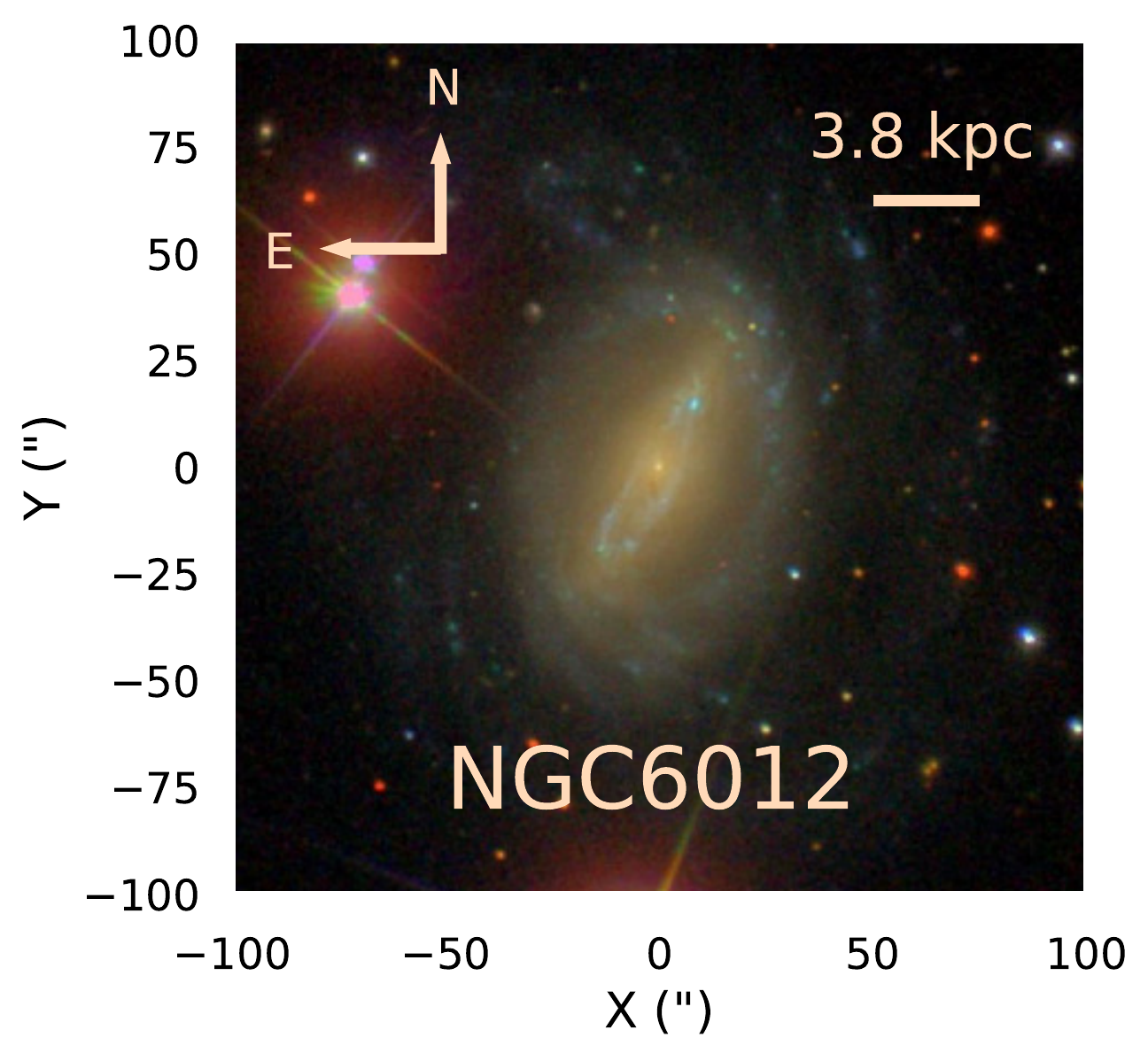}
\end{subfigure}
\hfill
\begin{subfigure}{0.19\textwidth}
\includegraphics[width=\textwidth]{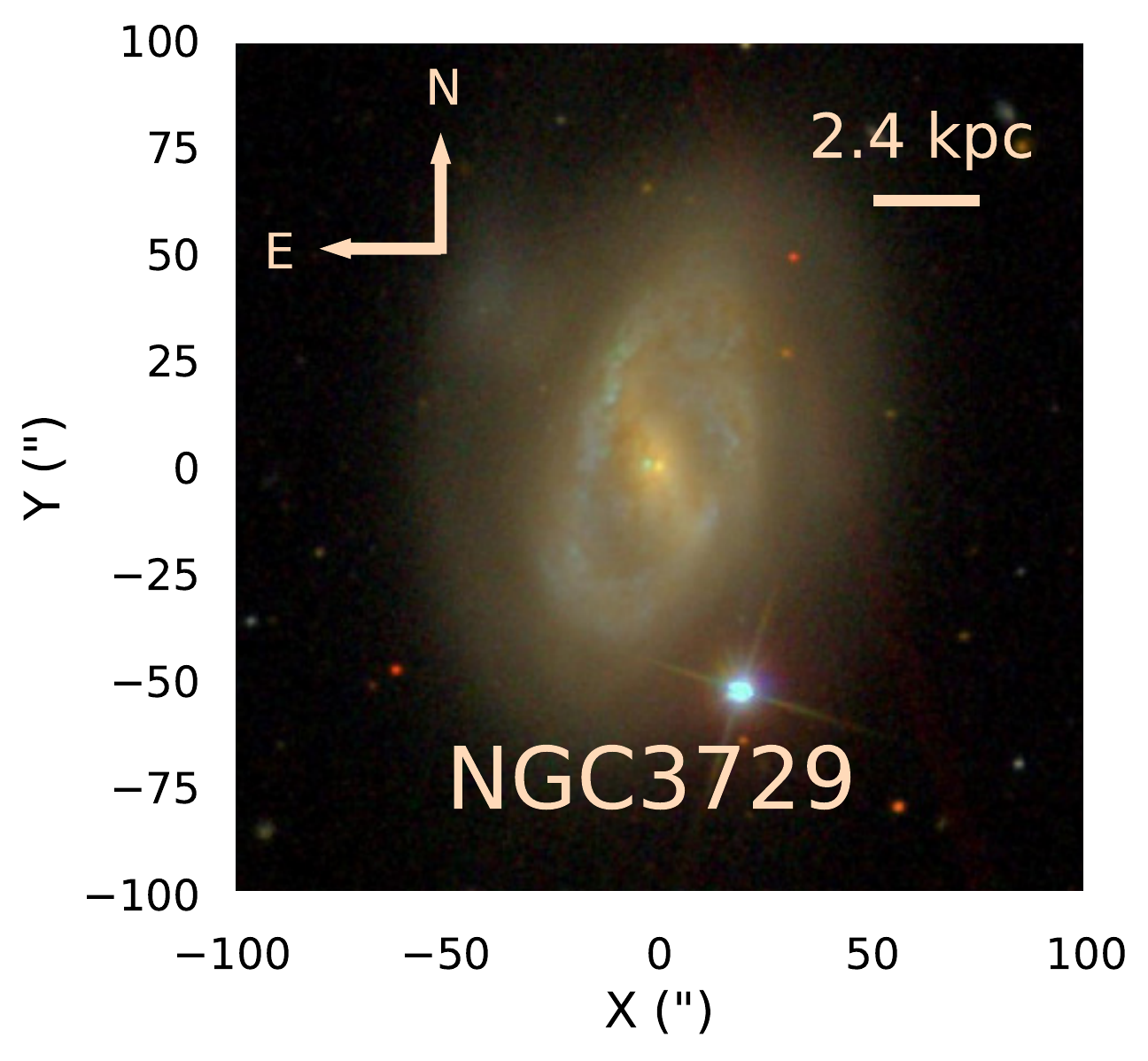}
\end{subfigure}
\hfill
\begin{subfigure}{0.19\textwidth}
\includegraphics[width=\textwidth]{newredspirals/imcut_NGC3380-eps-converted-to.pdf}
\end{subfigure}

\caption{SDSS cutout $gri$ images of the five low mass passive spiral galaxies (left) and their four galaxies from the comparison sample closest in both $z$ and stellar mass with the same T-type in the right four columns. The five passive spirals are all located in Virgo, whilst the comparison galaxies with varying amounts of star formation are located across a range of environments.}
\label{ps_lowmass}
\end{figure*}

\begin{figure*}
\centering
\includegraphics[width=\textwidth]{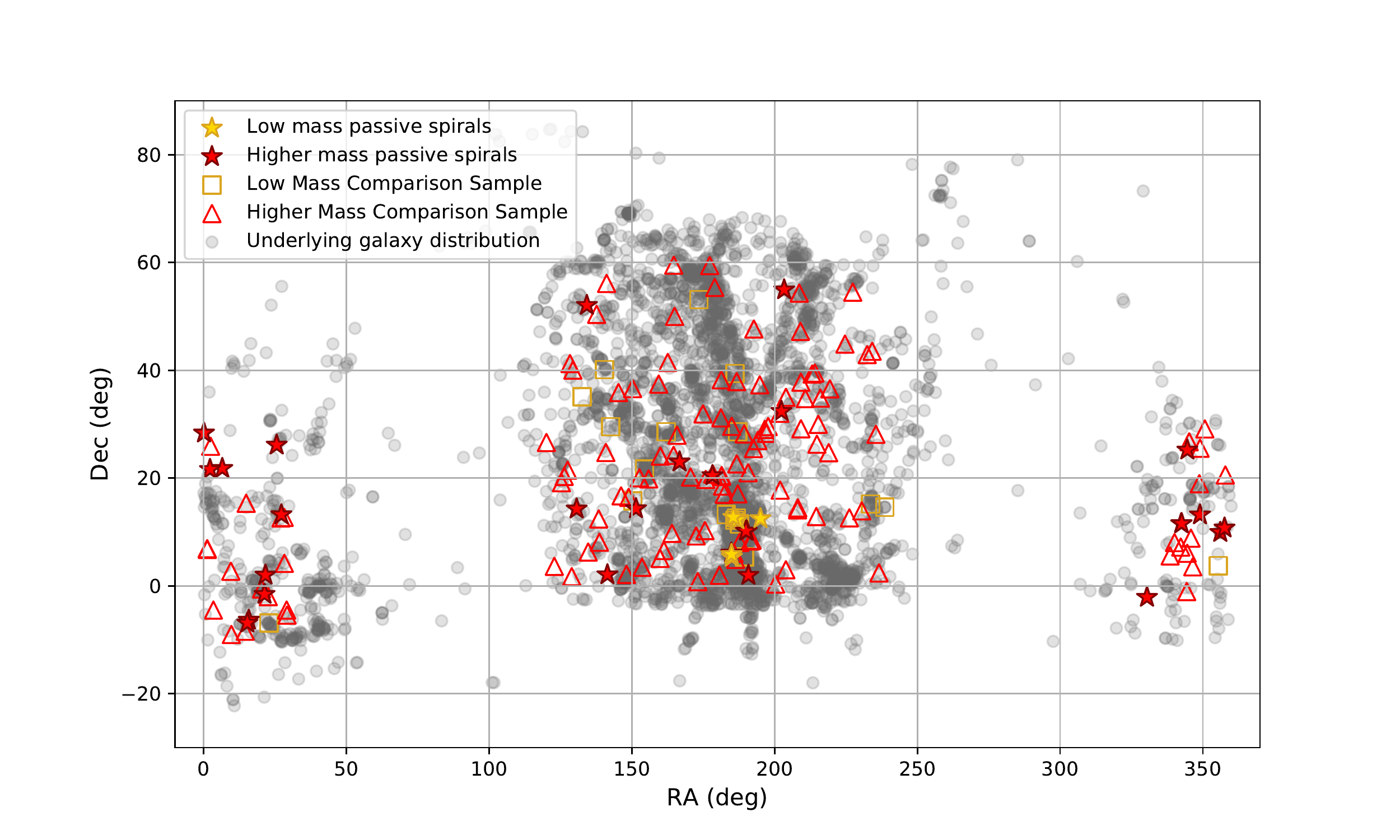} 
\caption{The position of passive spirals and comparison sample galaxies in the sky, with the underlying galaxy distribution for $z<0.01$ from the NASA Sloan Atlas to accentuate local superstructure. Low mass passive spirals are marked as gold stars, while higher mass passive spirals are red stars. The overdensity of the Virgo cluster is clearly seen in the underlying galaxy distribution, and all low mass passive spirals lie in this region.}
\label{skyplot}
\end{figure*}

 \subsubsection{Passive Spiral Bar Fractions}
Bars are an important component of disk galaxies, thought to have the ability to transfer angular momentum and gas from the disk to the central regions of a galaxy \citep[e.g.][]{Combes81, Weinberg85, Masters11}. The \citet{Masters10} red spiral sample had a bar fraction $\sim40\%$ higher than a similar sample of blue (or more obviously star forming) spirals. They suggest a correlation between bar instabilities and the quenching of star formation in optically red spirals.  
Motivated by this result, we check the bar fraction of both our passive spiral sample, and to mitigate any selection issues in bar identification, the mass, $z$, and T-type-matched control sample. 

We visually examine the SDSS colour images of each galaxy to determine bar fraction, including both large and small bars. We note that due to the quality of the SDSS images coupled with the low redshift of the sample, small bars within a galaxy are easily visible. The subtlety of these objects may have made them more difficult to see in older photographic plate images.

We find a significantly higher bar fraction in the passive spiral sample of $74\pm15\%$, compared to the comparison sample with $36\pm5\%$, shown in Table~\ref{bar_frac_table}, where errors are binomial. Local Universe bar fractions have been stated to be anywhere from $20-30\%$ \citep{deVac91,Masters11}, up to $\sim50\%$ \citep{Barazza08, Aguerri09}. Our result is in line with the red spiral bar fraction of $67 \pm 5\%$ for the \citet{Masters10} sample. 

Ansa bars are bars that terminate with two distinct enhancements of light at either end of the bar \citep[or a `handle', e.g.][]{Martinez08}, a good example of which is shown by NGC 4440 in Figure~\ref{ps_lowmass}. The origin of ansae is unknown, but it is thought to be related to the growth of galactic bars, with ansae appearing in simulations only after a few Gyr of evolution \citep[e.g.][]{Martinez06}. Ansa bars are prevalent in Sa spirals, but almost non-existent in later types \citep{Martinez07}.
Interestingly, the ansa bar fraction of the passive spiral sample is much higher than the comparison sample ($69\pm16\%$ of all barred spirals in the sample, compared to $29\pm8\%$), despite being matched in T-type. 
Bars (and ansa bars in particular) are far more common in passive spiral galaxies than comparable star forming spiral galaxies. However, whether these bars are responsible for, or a effect of, quenching is unclear.

\subsubsection{Other Quenching Mechanisms}
From the previous sections, one could paint a picture of quenching being a consequence of bars and environment in combination, perhaps with unbarred galaxies being satellites while barred galaxies are brightest group galaxies and isolated galaxies. 
This is inconsistent with the data however, as passive spiral galaxies without bars and with stellar masses of $\sim5\times10^{10}~\textrm{M}_\odot$ can be brightest group galaxies (UGC 6163), satellite galaxies (PGC 070141) and isolated galaxies (NGC 2618).

The one unbarred, truly isolated galaxy in our passive spiral sample, NGC 2618, has a stellar mass of $8.2\times10^{10}~\textrm{M}_{\odot}$. This implies a halo mass of $\sim10^{13}~\textrm{M}_{\odot}$ \citep{PHopkins14}, at which star formation is largely truncated and consistent with virial shock heating \citep[e.g.][]{Dekel06, Dolley14}.
While by definition we can invoke mass quenching to explain the most massive passive spiral galaxies, it provides few, if any, insights into the underlying astrophysics.
 Furthermore, all of our massive passive spiral galaxies are matched to star forming control galaxies with comparable stellar masses, so mass quenching is not deterministic (at least for our mass range).

Combining mass quenching with another mechanism is not particularly satisfying either, given passive spiral galaxies with masses in the $1\times10^{10}-8.5\times10^{10}~\textrm{M}_{\odot}$ range include galaxies with and without bars, isolated galaxies, and group members. Only passive spiral galaxies with masses below $1\times10^{10}~\textrm{M}_{\odot}$, which all reside in Virgo, show evidence of all being quenched by the same cluster-specific mechanism. For now, we conclude that the mechanism(s) that quench the most massive passive spiral galaxies remain a puzzle.

\begin{figure}  
\centering
\includegraphics[width=0.45\textwidth]{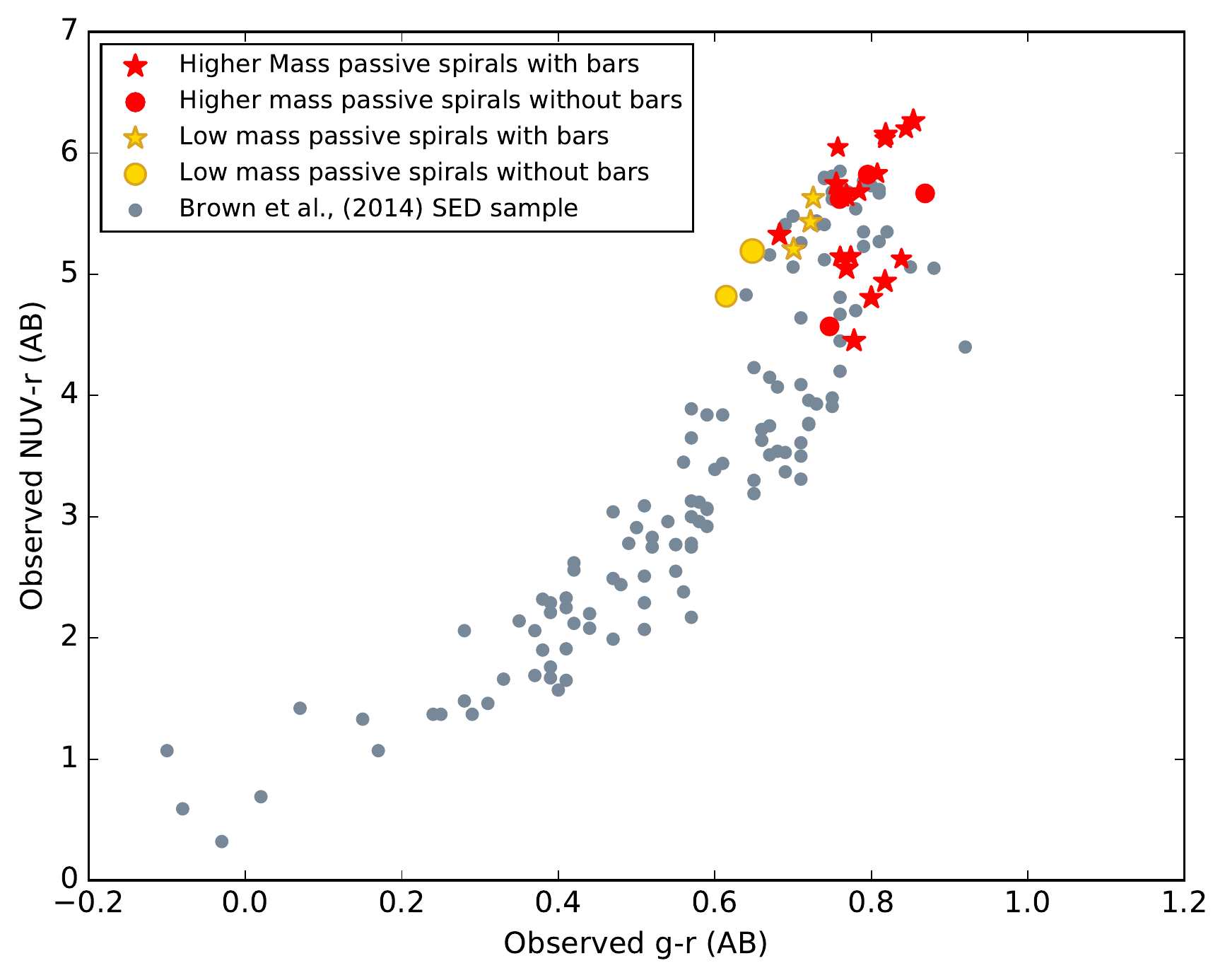}
\caption{Optical/NUV colour colour diagram of the passive spiral sample in order to determine whether we can see evidence of particular quenching mechanisms showing up in their stellar populations. Photometry for the passive spirals is taken from matched aperture photometry of NASA Sloan Atlas images, and the underlying galaxy population from \citet{Brown14}. The low mass passive spirals are mostly bluer in $g-r$ colour than the higher mass sample, indicating either younger stellar populations, or lower metallicity (or both). }
\label{color_fig}
\end{figure}

\begin{figure}
\includegraphics[width=0.49\textwidth]{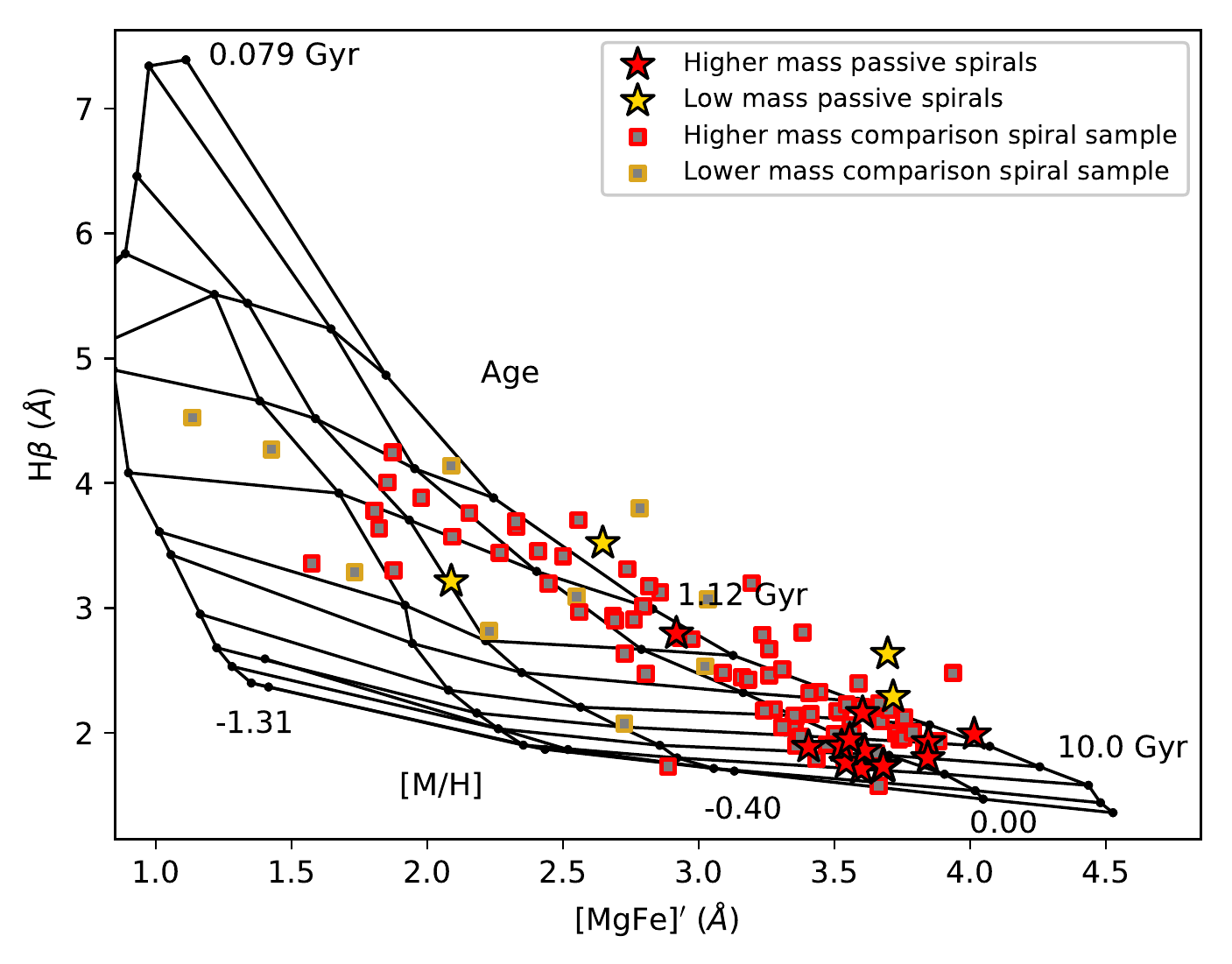} 
\caption{Spectroscopic index-index diagram for passive spiral galaxies and the mass, $z$, and T-type matched comparison sample using the Lick indices H$\beta$ and [MgFe]$^{\prime}$ measured with SDSS-II 3$^{\prime\prime}$ fibre spectroscopy of galaxy nuclei.
The single stellar population model predictions of \citet{Vazdekis10} are plotted as black lines for comparison. 
Of the passive spirals that have SDSS Lick index measurements, all but one of the higher mass galaxies have similar metallicities and ages, while lower mass galaxies have a broad spread of ages and metallicities. 
The lower mass passive spirals have on average, younger stellar populations than the high mass passive spirals. Given the lower mass passive spirals are located in a rich cluster, we may speculate that they have fallen into the cluster relatively recently. }
\label{indexindex}
\end{figure}

\section{Stellar Age and Metallicity Differences Between Low and High Mass Passive Spirals}
Given that low mass passive spiral galaxies appear to be quenched by cluster-scale environmental processes and higher mass through a combination of mechanisms, we investigate whether evidence of the quenching mechanism shows up in a galaxy's stellar population. We examine both the integrated colours and the central stellar age and metallicity derived from H$\beta$ and [MgFe]$^{\prime}$ Lick indices of the passive spiral sample. 

Figure~\ref{color_fig} is an optical and UV colour-colour diagram, with passive spiral photometry remeasured using matched-aperture photometry of archival NASA Sloan Atlas images. We ensure that any foreground features are masked in this process. The SED sample of \citet{Brown14} with reliable multi-band photometry is also shown to illustrate the general shape of the galaxy distribution in this colour space. High mass barred and unbarred passive spirals are systematically redder in $g-r$ colour than their low mass counterparts. Given the well-known age-metallicity degeneracy \citep[e.g.][]{Tremonti04}, this suggests either a younger or more metal poor stellar population in low mass passive spiral galaxies. 

In an attempt to separate out the effects of stellar age and metallicity, we also present an index-index diagram of both the passive spiral sample and the comparison sample in Figure~\ref{indexindex}. H$\beta$ and [MgFe]$^\prime$ are Lick indices sourced from the SDSS-II MPA-JHU Galspec catalogue \citep{Kauffmann03, Brinchmann04, Tremonti04} for both the passive spiral sample and the mass, $z$, and T-type-matched comparison sample, which provides emission line-subtracted line index measurements using the Lick IDS system for a large portion of the SDSS DR7 spectroscopic sample. Here we define 
\begin{equation*}
[\textrm{MgFe}]^{\prime} = \sqrt{\textrm{Mg}b~(0.72\times\textrm{Fe5270}+0.28\times\textrm{Fe5335})},
\end{equation*}
as in \citet{Gonzalez93} and \citet{Thomas03}. We use these particular indices as H$\beta$ is a good indicator of recent star formation, and [MgFe]$^{\prime}$ of metallicity. We note that these indices will be measured for the nuclear regions only, as they are derived from fibre spectra.
To convert the index measurements to estimates of stellar age and metallicity, on Figure~\ref{indexindex} we overlay the single stellar population model predictions of \citet{Vazdekis10} using Padova 2000 isochrones and a \citet{Chabrier03} initial mass function. 
We see that all but one of the high mass passive spiral galaxies with measurements in the Galspec catalogue are clustered around a similarly old stellar age and approximately Solar metallicity. This is distinct from the mass-matched comparison galaxies, whose bulges span a range of stellar ages and metallicities. These galaxies are perhaps similar to comparably massive early type galaxies.

In contrast, the four low mass passive spirals are spread across a range of metallicities and stellar ages. The low mass passive spirals have younger stellar ages than all but one of the higher mass galaxies, and while one is more metal poor than the higher mass spirals, the other three are more metal rich. From their comparatively young stellar ages, we postulate that the low mass passive spirals have fallen into Virgo and quenched relatively recently, within the past $\sim$1-2 Gyr. The low mass comparison sample are also spread throughout stellar age and metallicity space. 

The tight clustering of the higher mass passive spirals around a common age and metallicity is perhaps surprising, given their lack of coherent quenching model. We suspect the similarly old bulge stellar ages and rich metallicities mean these galaxies quenched a long time ago, and any signature of quenching is no longer visible. 

\section{Summary \& Conclusions}
We investigate what quenched star formation in passive spiral galaxies, using a sample of 35 $z<0.033$ passive spiral galaxies and a comparison sample matched in mass, $z$, and T-Type.

All five low mass ($\textrm{M}_{\star}<1\times10^{10}~\textrm{M}_{\odot}$) passive spiral galaxies in our sample are members of the Virgo cluster, and thus environment driven quenching is the most likely explanation for these galaxies. 
A large spread in both [MgFe]$^{\prime}$ and H$\beta$ Lick indices implies a range of central metallicities and stellar ages, though the oldest of the low mass passive spirals is younger than all but one of the high mass passive spiral sample. Given the implied ages and metallicities from the \citet{Vazdekis10} models and the similar environments of the low mass passive spiral sample, we suspect these galaxies have fallen into the Virgo cluster and quenched more recently than their higher mass passive spiral counterparts. Ram-pressure stripping and/or strangulation may be the relevant quenching mechanisms.

The bar fraction of passive spiral galaxies is high: $74\pm15\%$, compared to $36\pm5\%$ for a mass, $z$, and T-type-matched comparison sample of spirals.
The bars of passive spiral galaxies feature ansae $69\pm16\%$ of the time, much more frequently than the comparison sample at $29\pm8\%$.
From this we conclude that bars or the mechanism(s) responsible for creating them may also be quenching star formation in passive spirals. This is consistent with a bar quenching scenario, where gas is funnelled via a bar to the central regions of the galaxy, promoting pseudobulge growth, and inducing a starburst, followed by eventual quenching \citep[e.g.][]{Friedli95, Knapen02, Jogee05}.

Higher mass passive spiral galaxies are amongst the oldest and most metal rich spiral galaxies.
While many high mass passive galaxies have bars and all low mass passive spiral galaxies are Virgo satellite galaxies, a simple combination of bars and environment driven quenching does not explain passive spiral galaxies. Passive spiral galaxies without bars can be brightest group galaxies (e.g. UGC 6163), satellite galaxies (e.g. NGC 345), be interacting (e.g. NGC 2648) or isolated galaxies (e.g. NGC 2618). We thus conclude no one mechanism is responsible for quenching all passive spiral galaxies in our sample. Bars (and ansa bars) seem heavily involved  for many galaxies, and environment driven quenching (perhaps ram-pressure stripping or harassment) best explains the lowest mass passive spiral galaxies.

While future studies with larger sample sizes will be able to address this question in a more statistical way, large-scale galaxy integral field spectroscopic surveys may also be employed to determine stellar populations and metallicities across an entire galaxy.
Surveys such as Mapping Nearby Galaxies at APO \citep[MaNGA;][]{Bundy15} and the Sydney-AAO Multi-object Integral field Spectrograph galaxy survey \citep[SAMI;][]{Croom12} will provide insight into the star formation and quenching history of low and high mass passive spiral galaxies, in turn, confirming the relevant quenching mechanisms and timescales.

\section*{Acknowledgements}
We wish to thank the anonymous referee for their insightful comments that improved the quality of this manuscript.
AFM acknowledges the support of an Australian Postgraduate Award and a Monash Graduate Education Postgraduate Publications Award.
Funding for the SDSS III has been provided by
the Alfred P. Sloan Foundation, the U.S. Department of Energy Office of
Science, and the Participating Institutions.  This publication makes use of data products from the Wide-field Infrared Survey Explorer, which is a joint project of the University of California, Los Angeles, and the Jet Propulsion Laboratory/California Institute of Technology, funded by the National Aeronautics and Space Administration. 
This publication makes use of data products from the Wide-field Infrared Survey Explorer, which is a joint project of the University of California, Los Angeles, and the Jet Propulsion Laboratory/California Institute of Technology, funded by the National Aeronautics and Space Administration.
This publication makes use of data products from the Two Micron All Sky Survey, which is a joint project of the University of Massachusetts and the Infrared Processing and Analysis Center/California Institute of Technology, funded by the National Aeronautics and Space Administration and the National Science Foundation.

\appendix 

\section{Passive Spiral Comparison Sample Properties and Higher Mass Sample Images}
\begin{figure*}  
\centering
\begin{subfigure}{0.19\textwidth}
\includegraphics[width=\textwidth]{newredspirals/imcut_NGC4260-eps-converted-to.pdf}
\end{subfigure}
\hfill
\begin{subfigure}{0.19\textwidth}
\includegraphics[width=\textwidth]{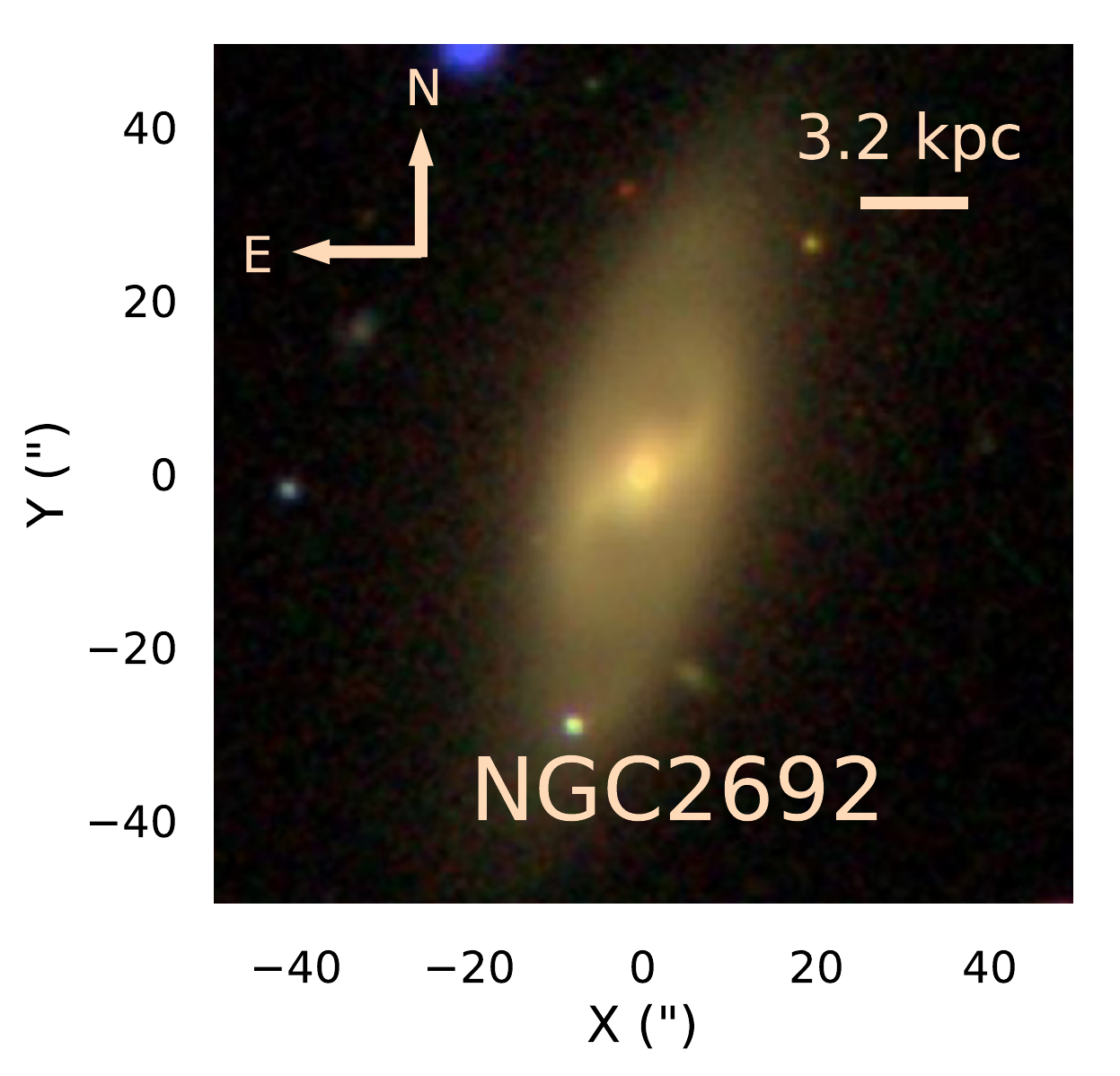}
\end{subfigure}
\hfill
\begin{subfigure}{0.19\textwidth}
\includegraphics[width=\textwidth]{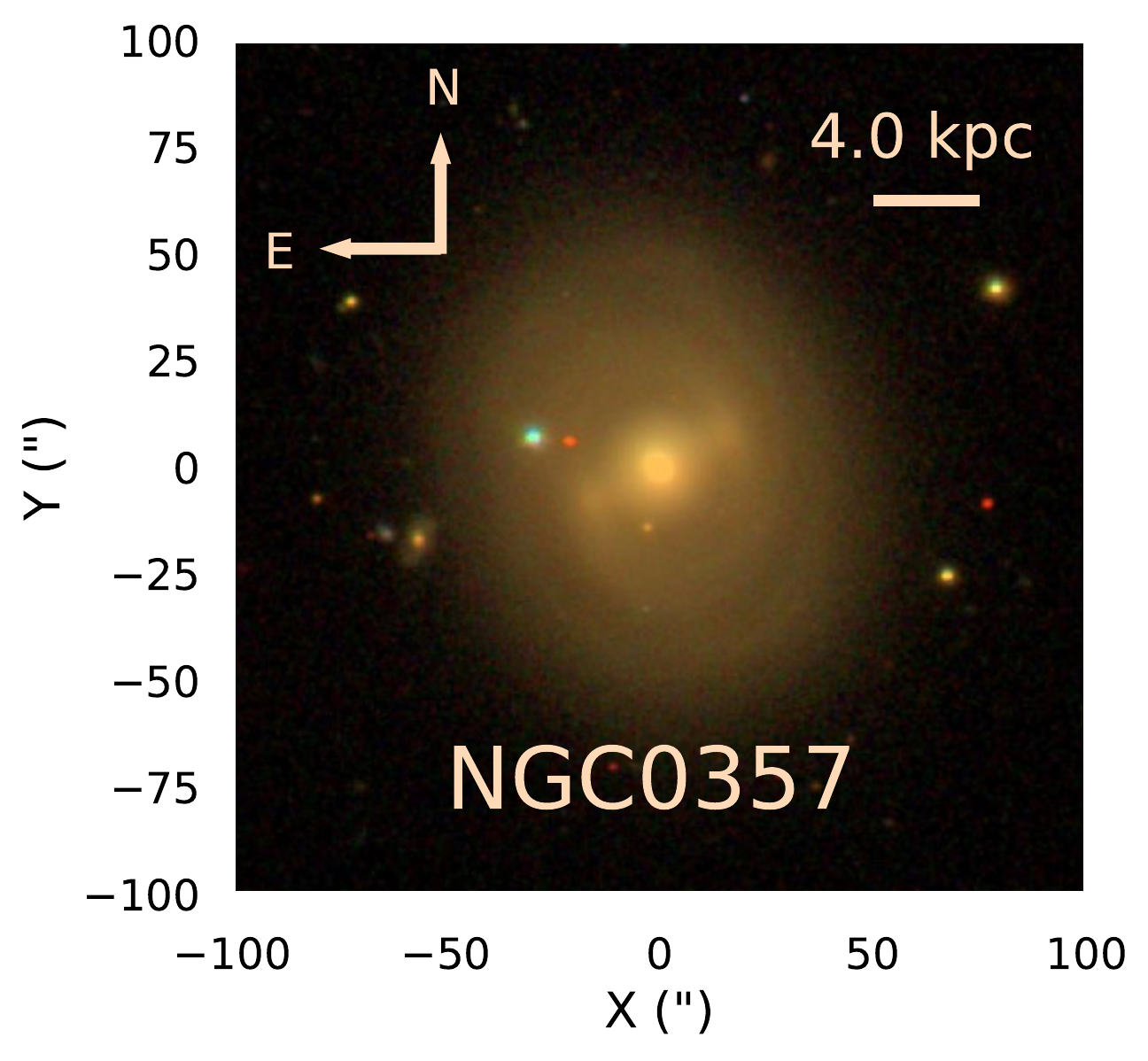}
\end{subfigure}
\hfill
\begin{subfigure}{0.19\textwidth}
\includegraphics[width=\textwidth]{newredspirals/imcut_NGC7743-eps-converted-to.pdf}
\end{subfigure}
\hfill
\begin{subfigure}{0.19\textwidth}
\includegraphics[width=\textwidth]{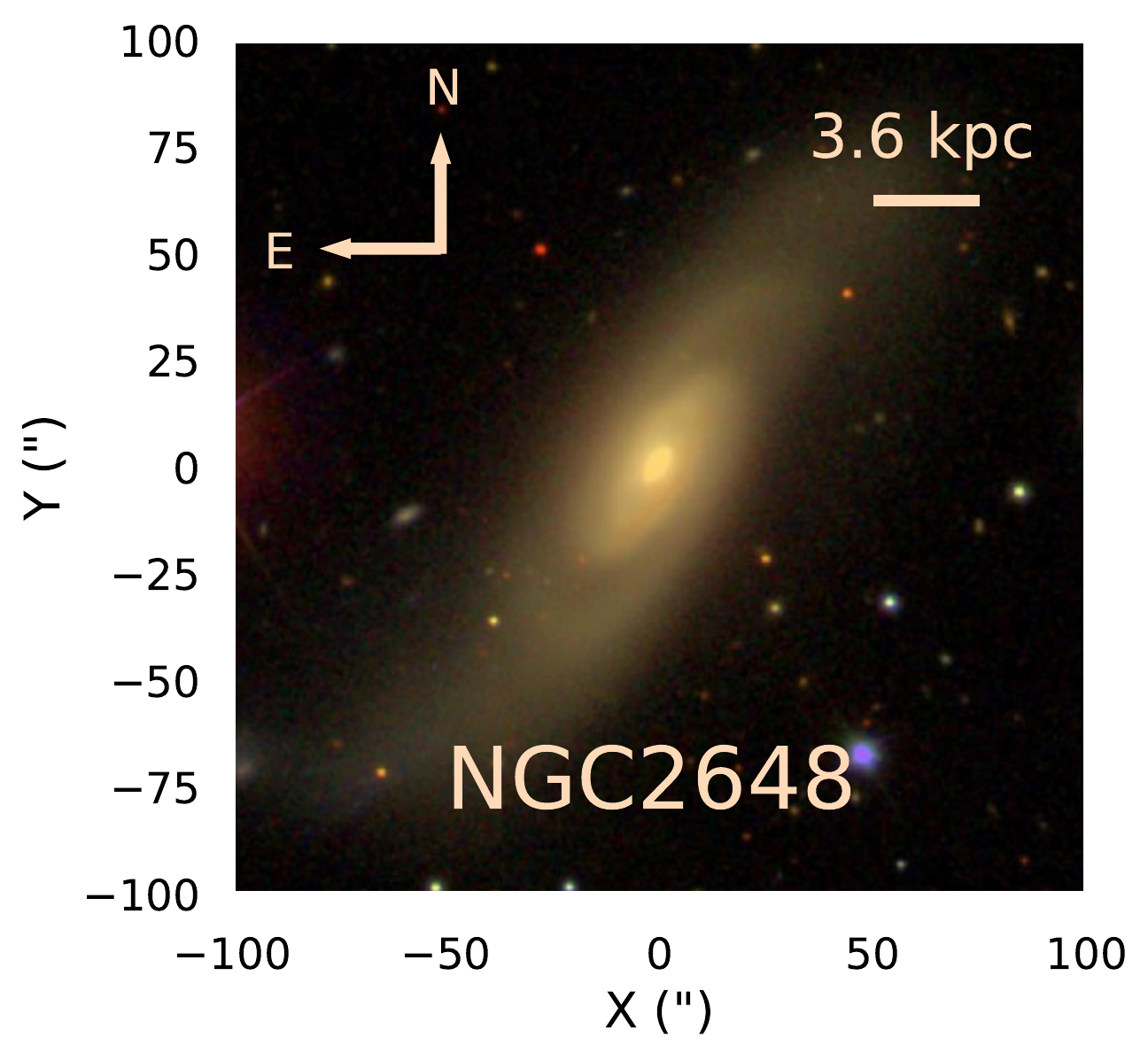}
\end{subfigure}

\vskip\baselineskip

\begin{subfigure}{0.19\textwidth}
\includegraphics[width=\textwidth]{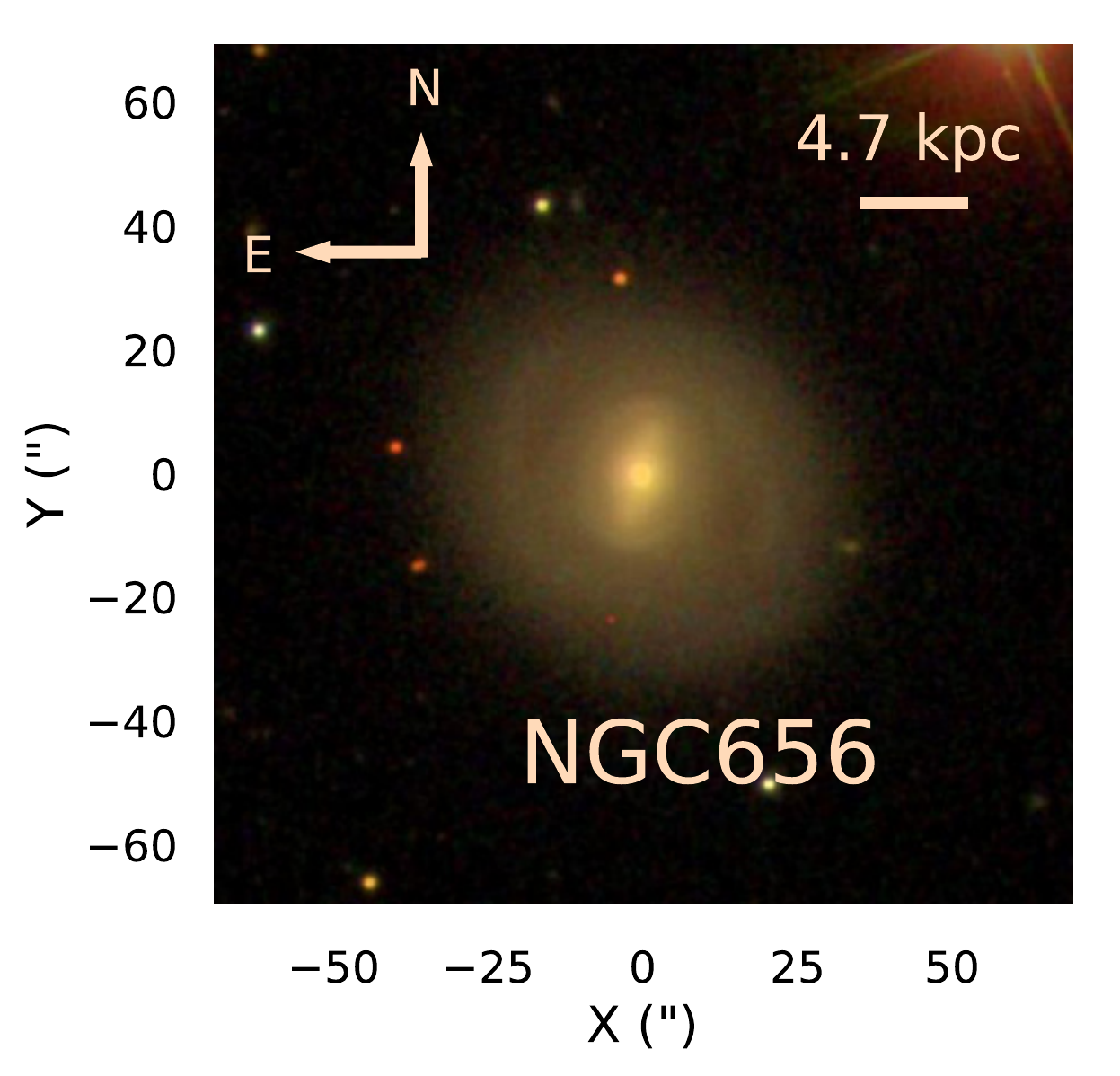}
\end{subfigure}
\hfill
\begin{subfigure}{0.19\textwidth}

\includegraphics[width=\textwidth]{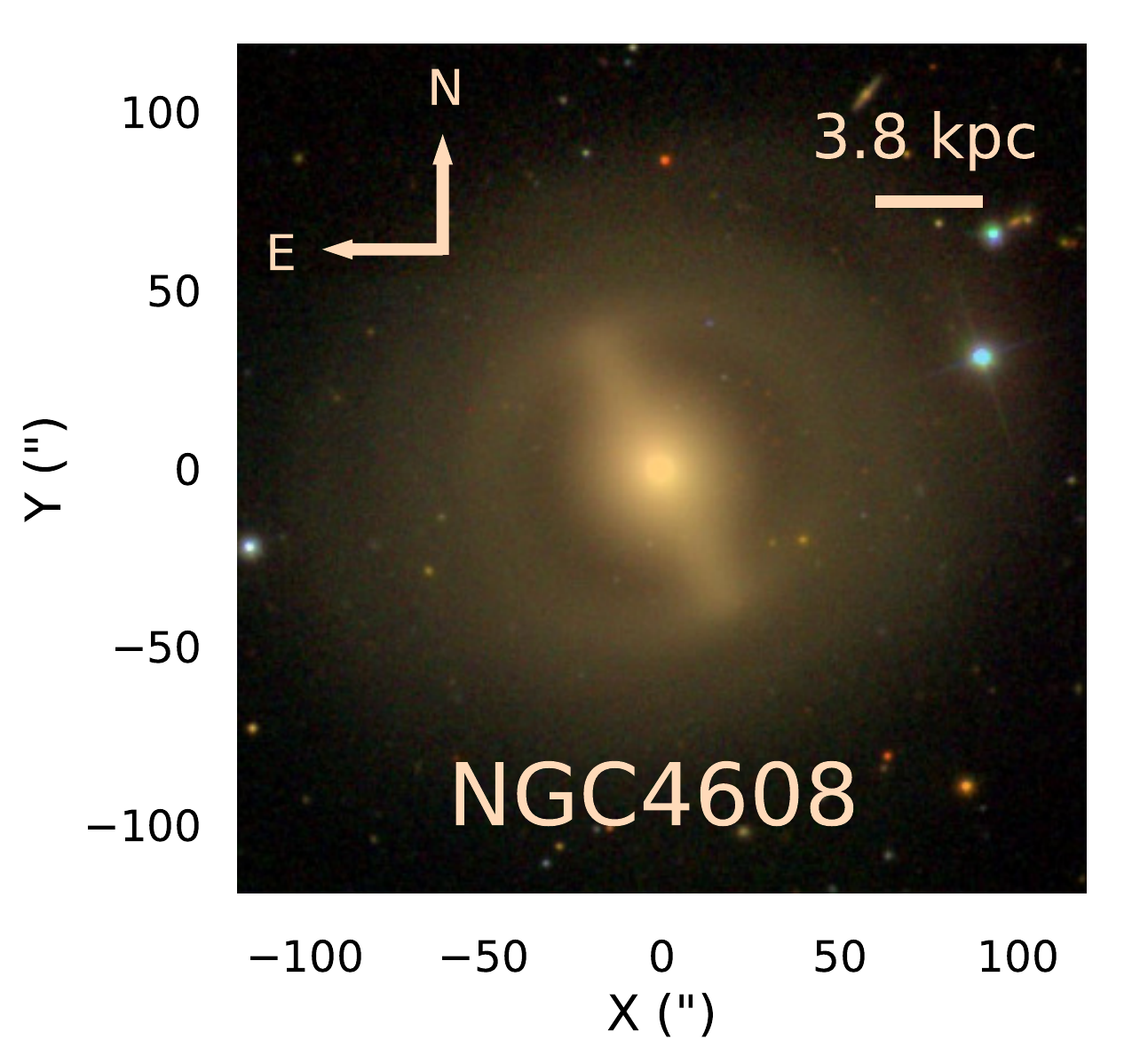}
\end{subfigure}
\hfill
\begin{subfigure}{0.19\textwidth}
\includegraphics[width=\textwidth]{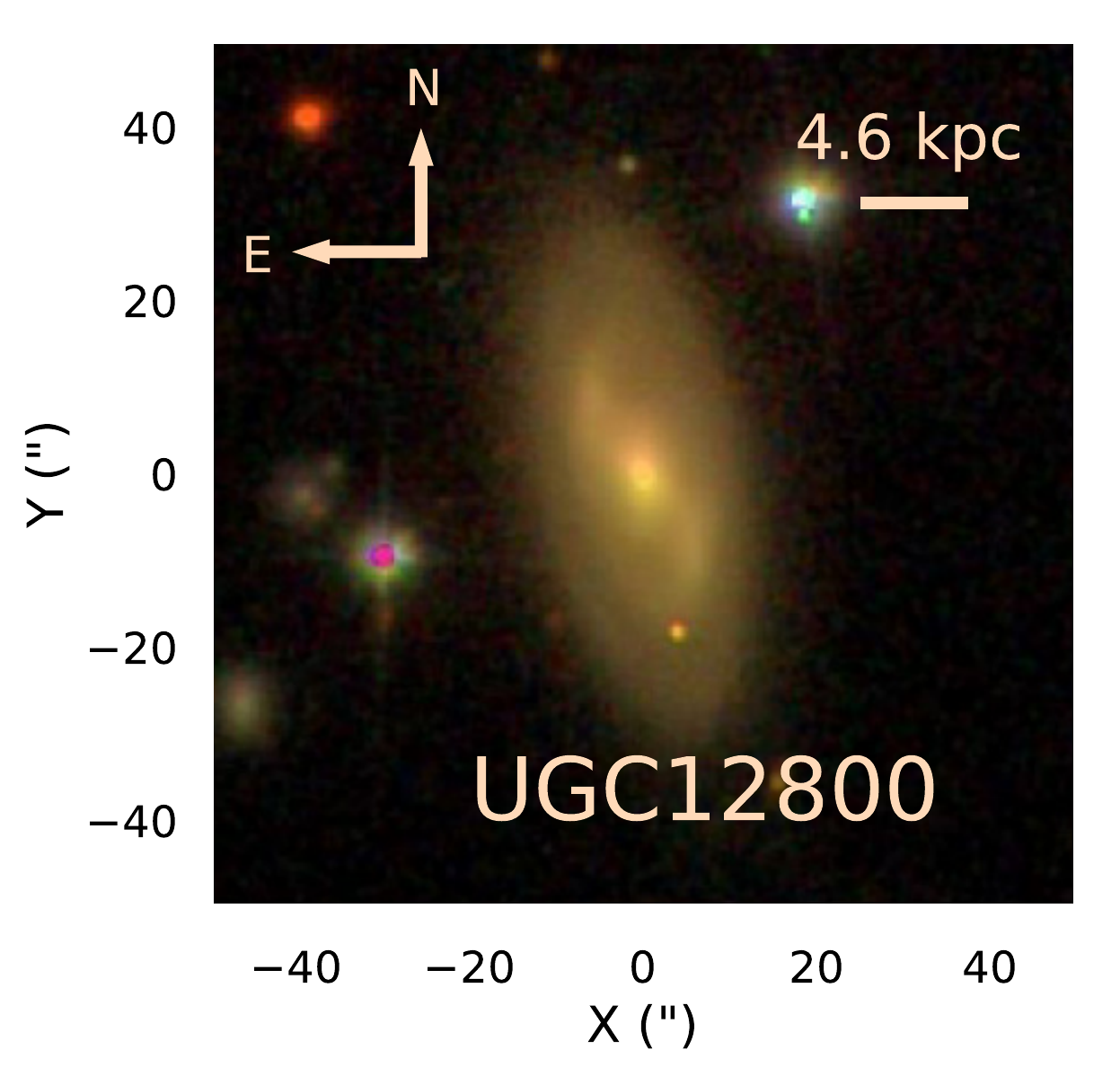}
\end{subfigure}
\hfill
\begin{subfigure}{0.19\textwidth}
\includegraphics[width=\textwidth]{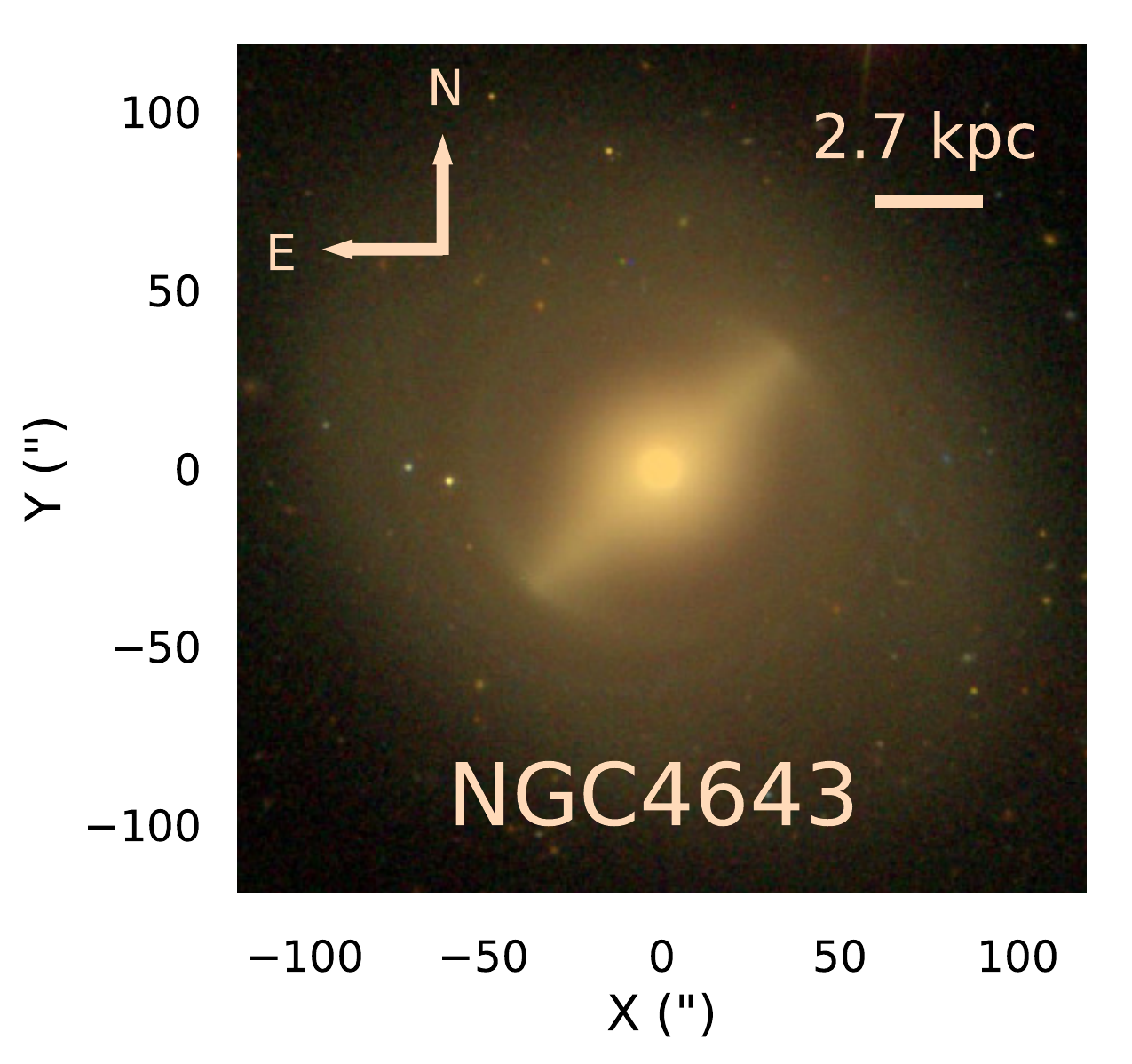}
\end{subfigure}
\hfill
\begin{subfigure}{0.19\textwidth}
\includegraphics[width=\textwidth]{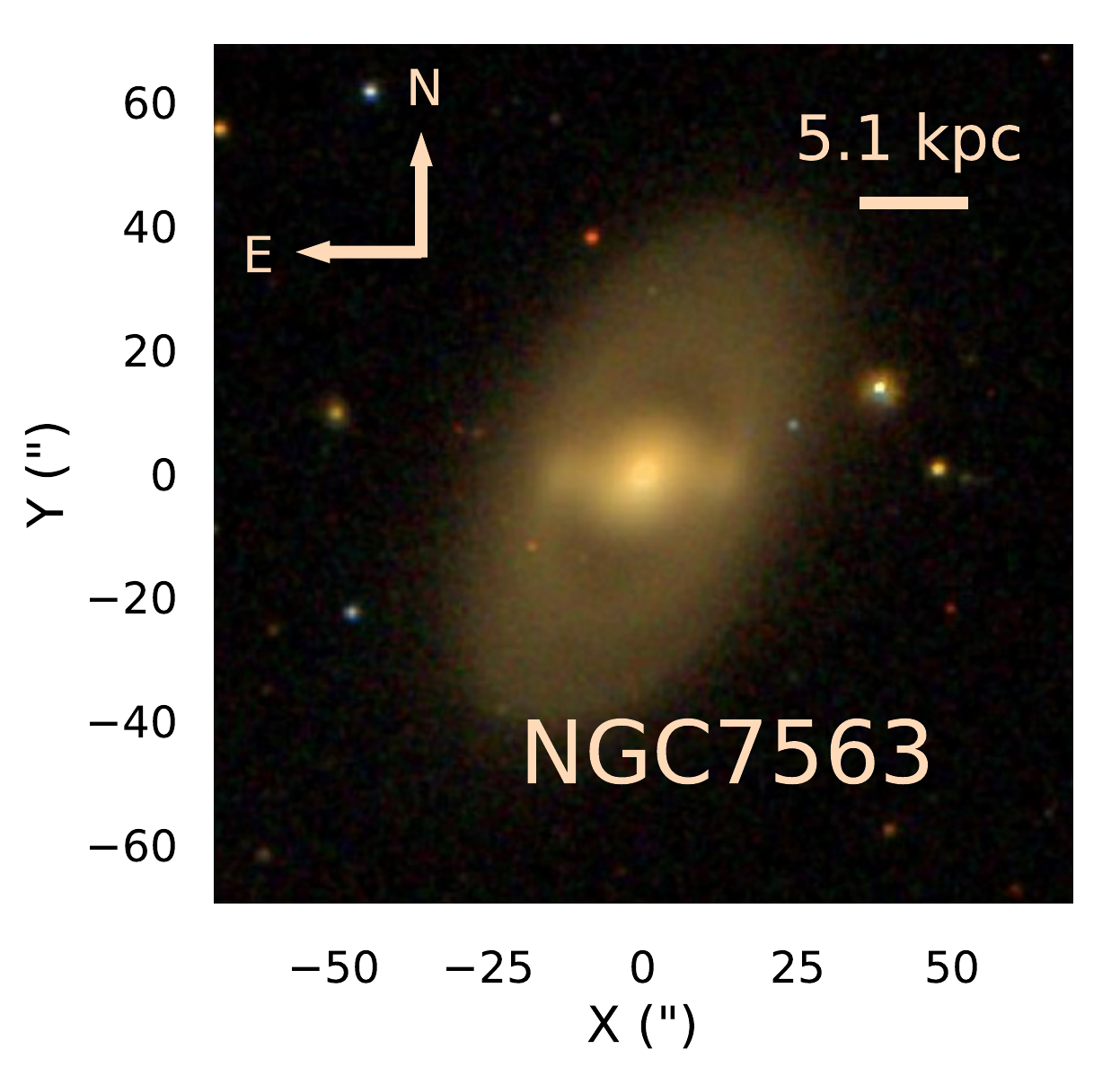}
\end{subfigure}

\vskip\baselineskip

\begin{subfigure}{0.19\textwidth}
\includegraphics[width=\textwidth]{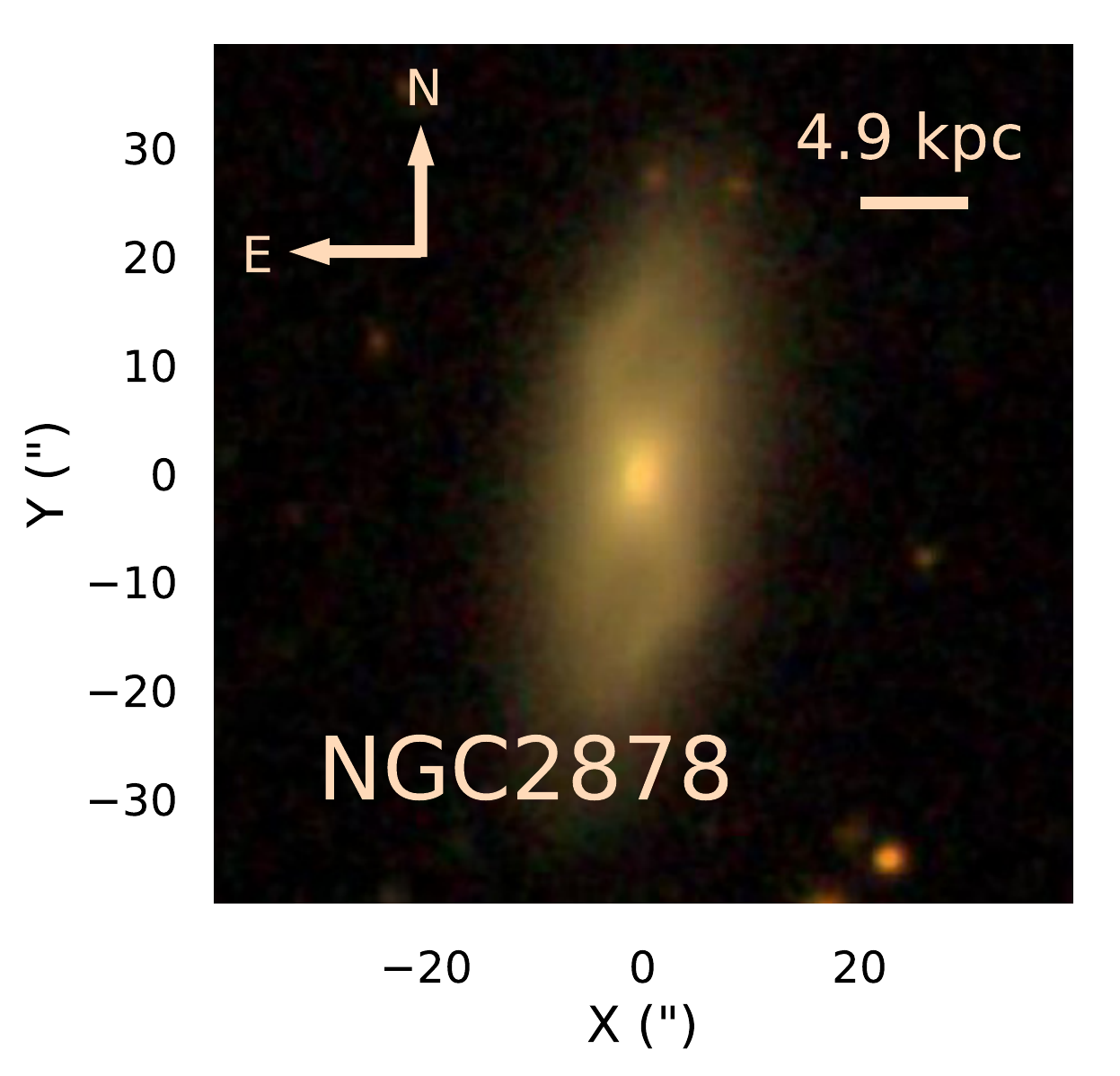}
\end{subfigure}
\hfill
\begin{subfigure}{0.19\textwidth}
\includegraphics[width=\textwidth]{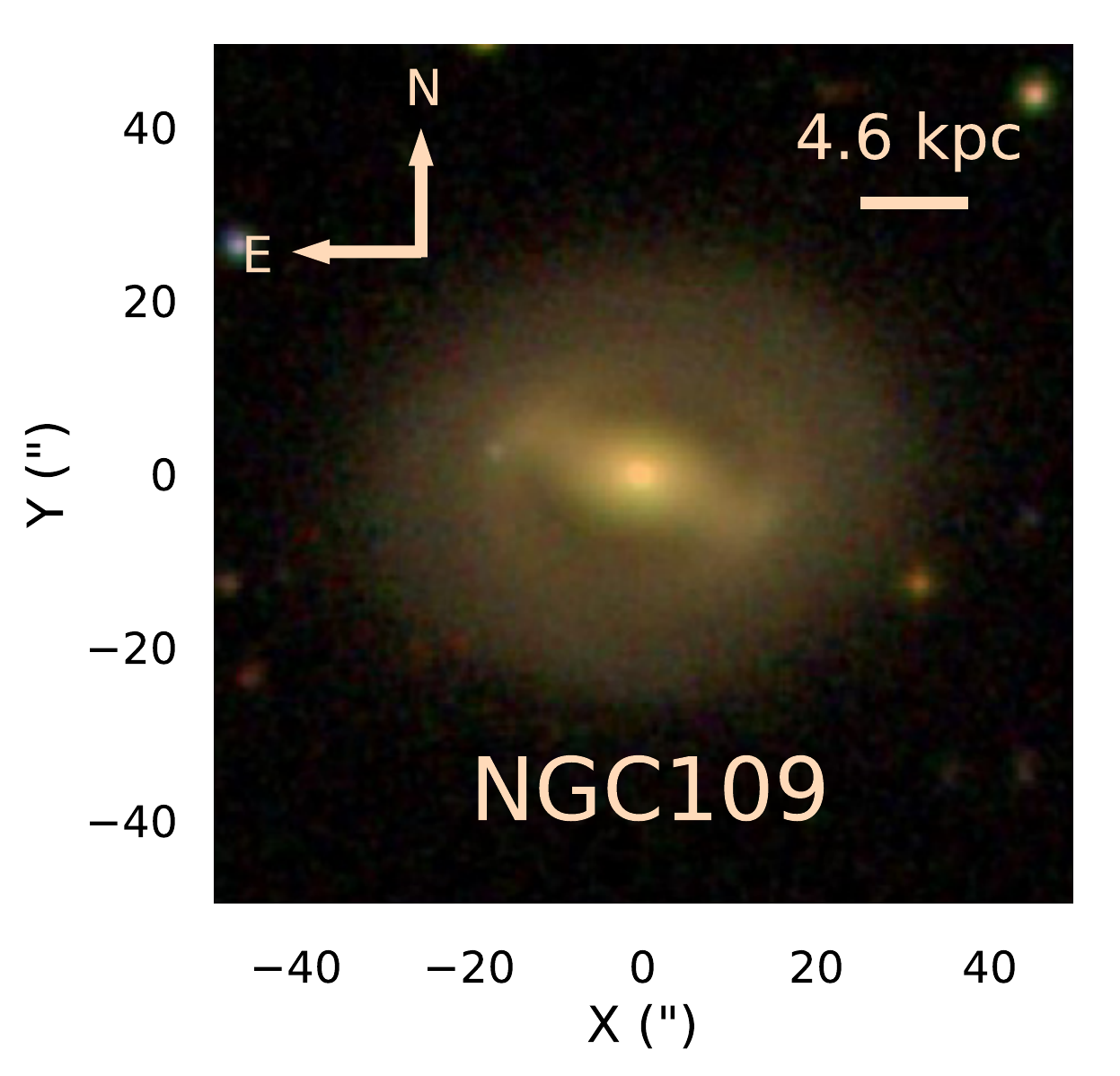}
\end{subfigure}
\hfill
\begin{subfigure}{0.19\textwidth}
\includegraphics[width=\textwidth]{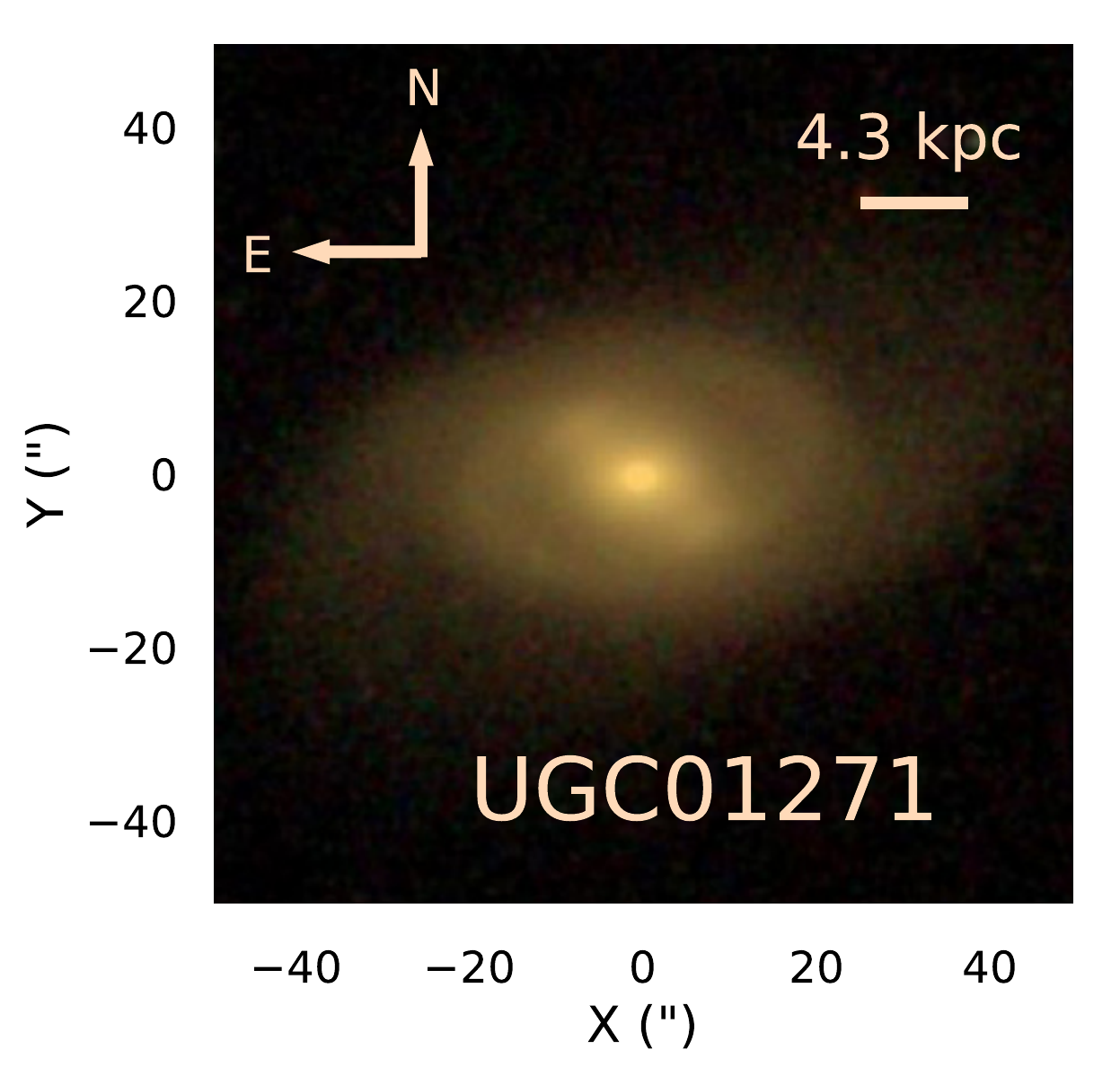}
\end{subfigure}
\hfill
\begin{subfigure}{0.19\textwidth}
\includegraphics[width=\textwidth]{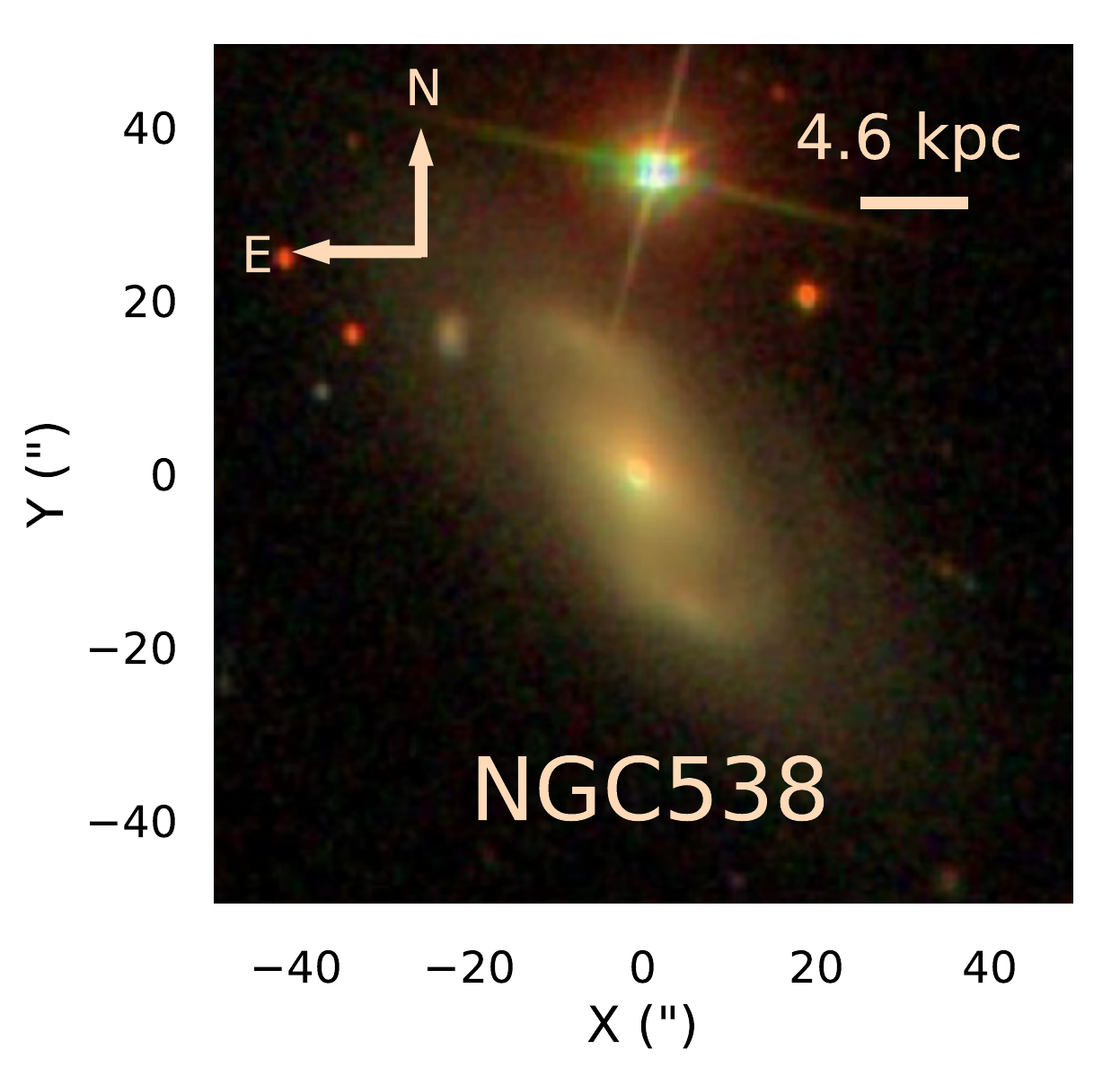}
\end{subfigure}
\hfill
\begin{subfigure}{0.19\textwidth}
\includegraphics[width=\textwidth]{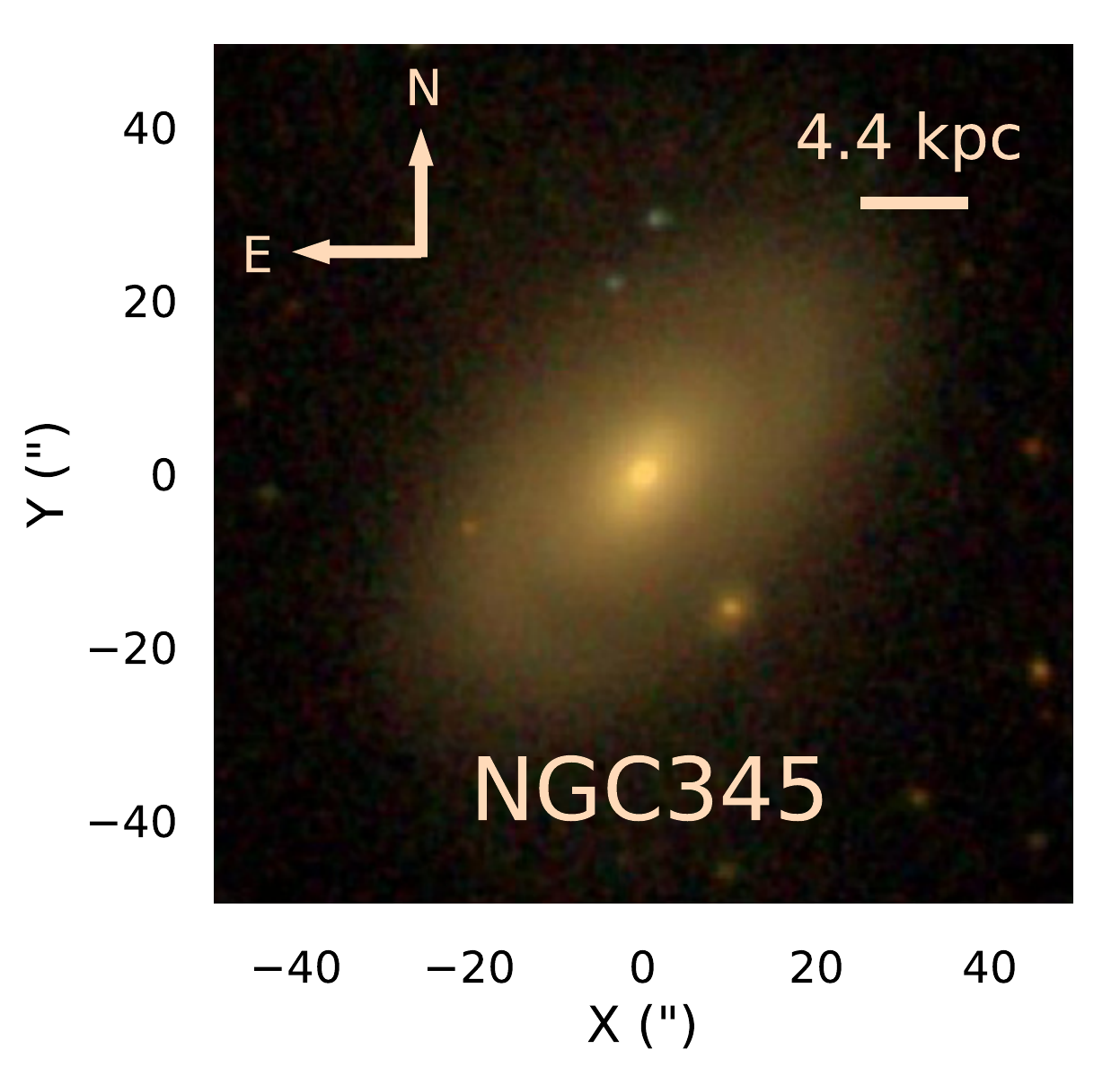}
\end{subfigure}

\vskip\baselineskip

\begin{subfigure}{0.19\textwidth}
\includegraphics[width=\textwidth]{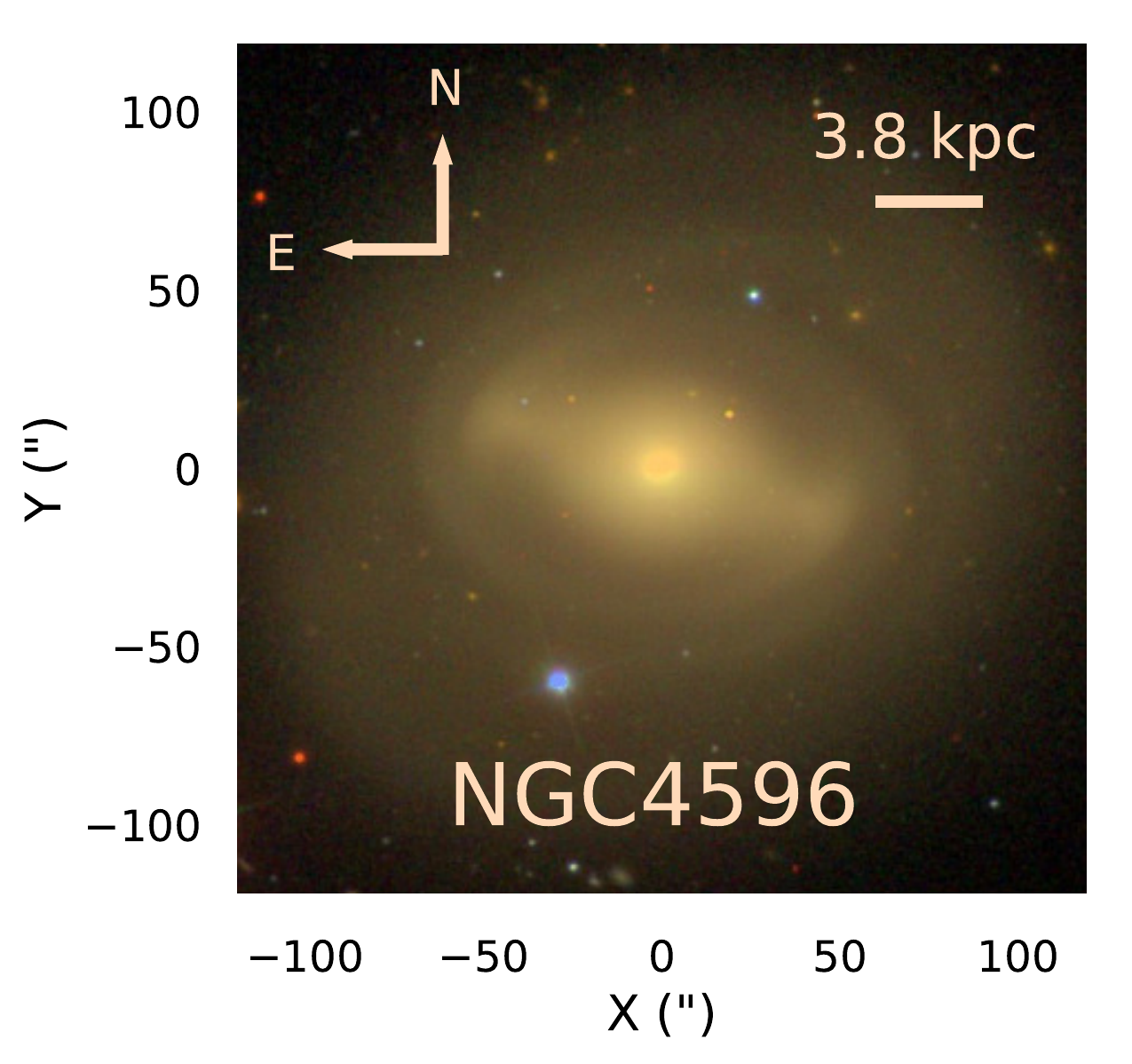}
\end{subfigure}
\hfill
\begin{subfigure}{0.19\textwidth}
\includegraphics[width=\textwidth]{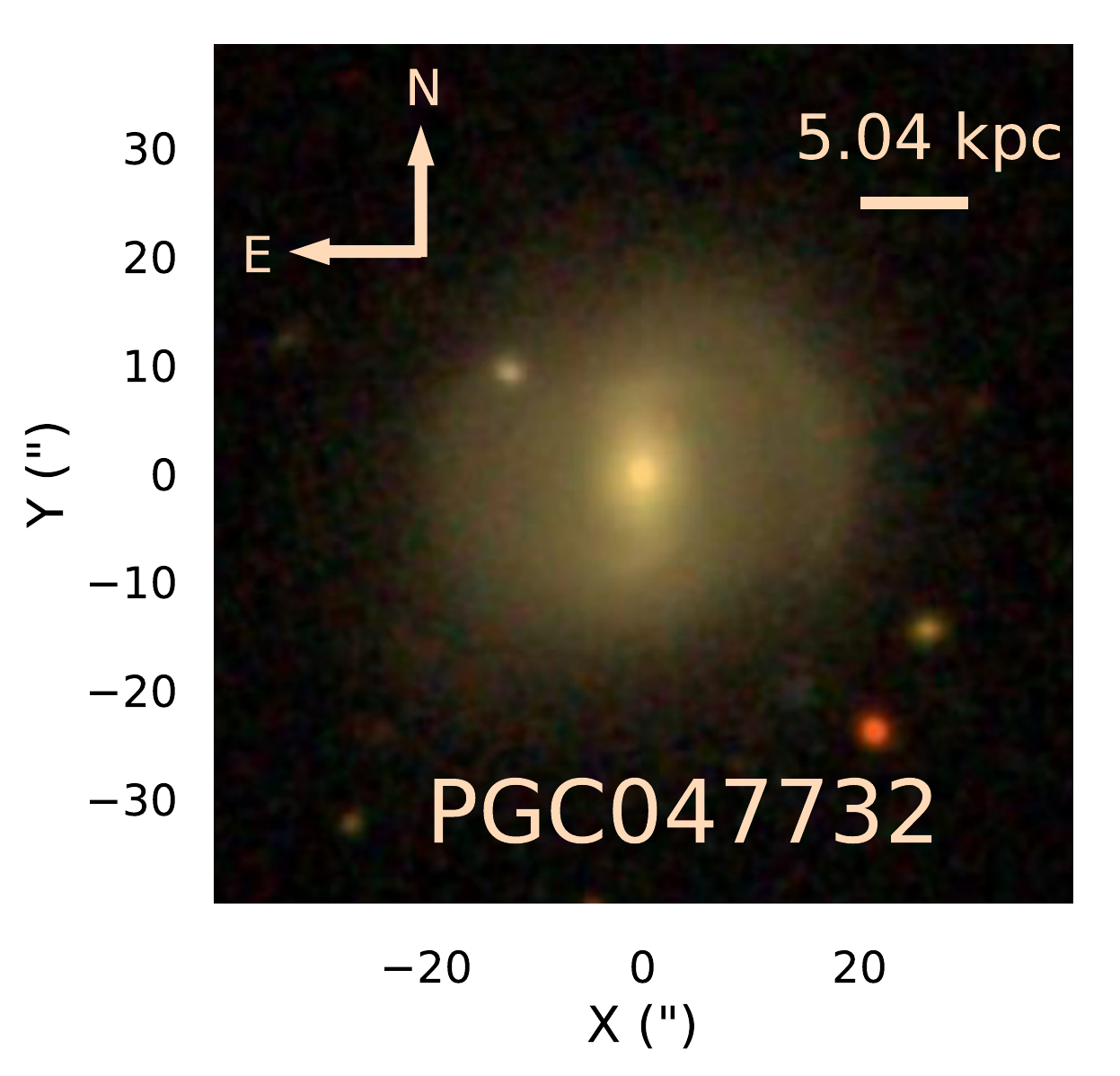}
\end{subfigure}
\hfill
\begin{subfigure}{0.19\textwidth}
\includegraphics[width=\textwidth]{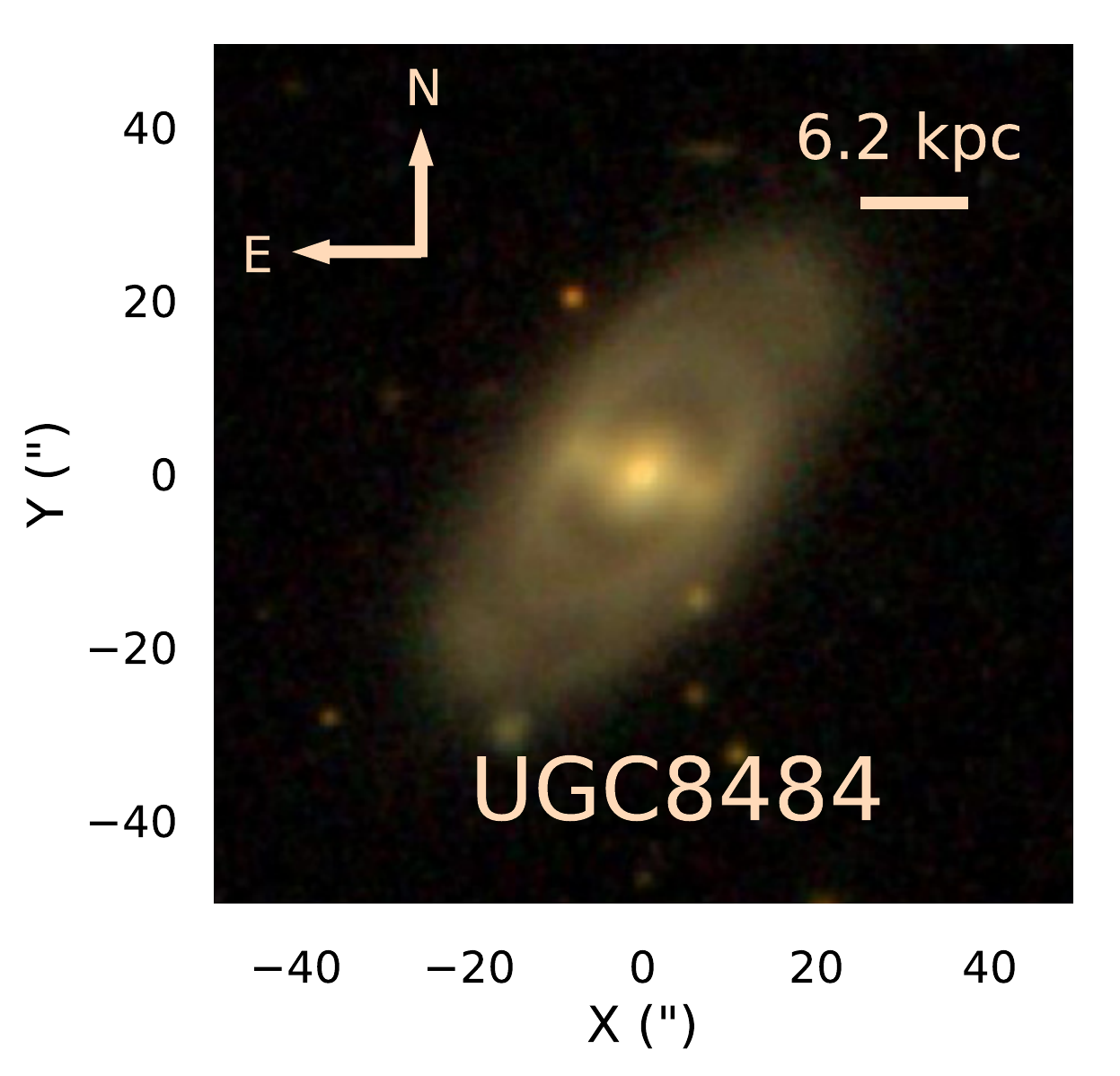}
\end{subfigure}
\hfill
\begin{subfigure}{0.19\textwidth}
\includegraphics[width=\textwidth]{newredspirals/imcut_NGC0015-eps-converted-to.pdf}
\end{subfigure}
\hfill
\begin{subfigure}{0.19\textwidth}
\includegraphics[width=\textwidth]{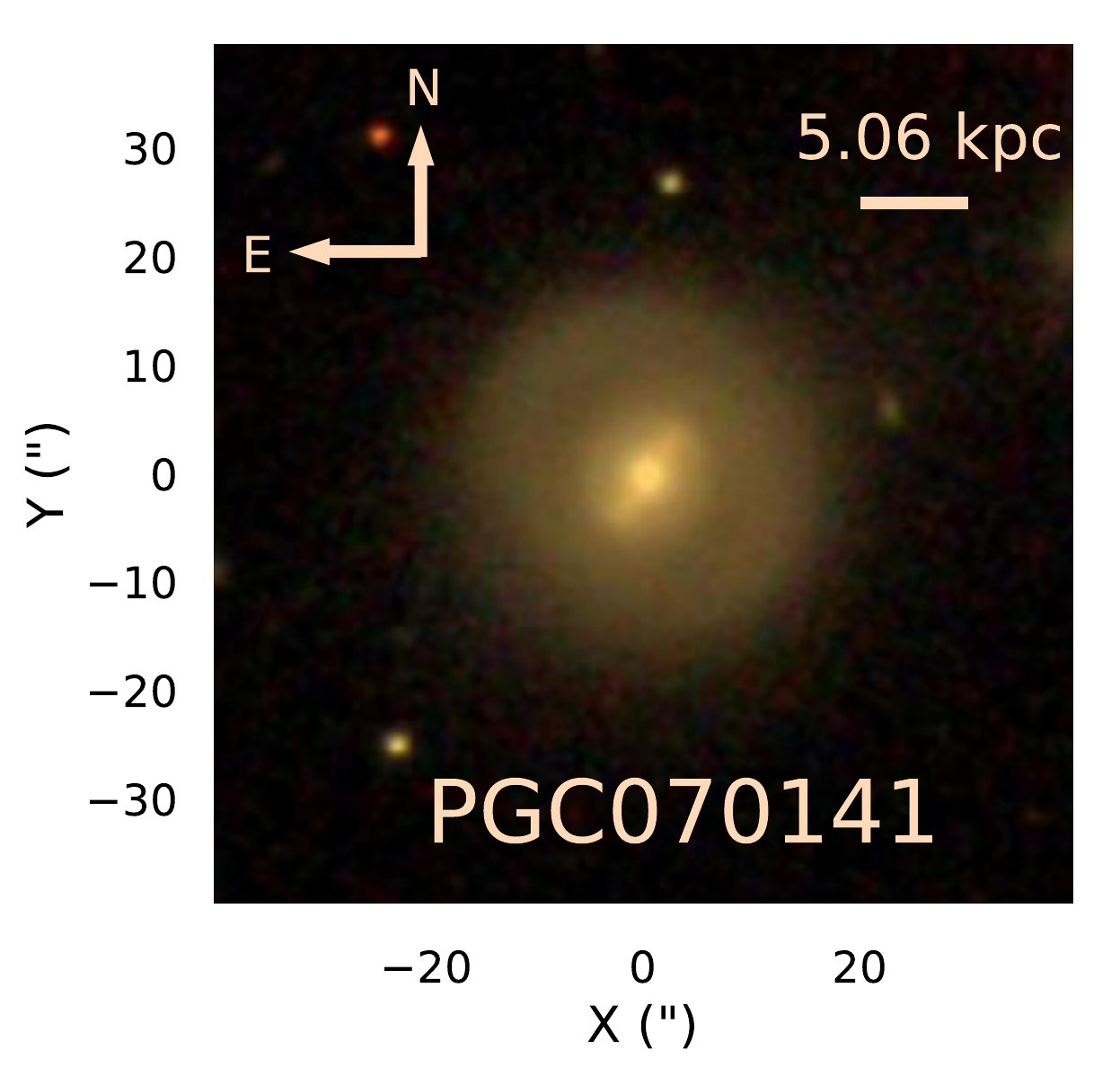}
\end{subfigure}

\vskip\baselineskip

\begin{subfigure}{0.19\textwidth}
\includegraphics[width=\textwidth]{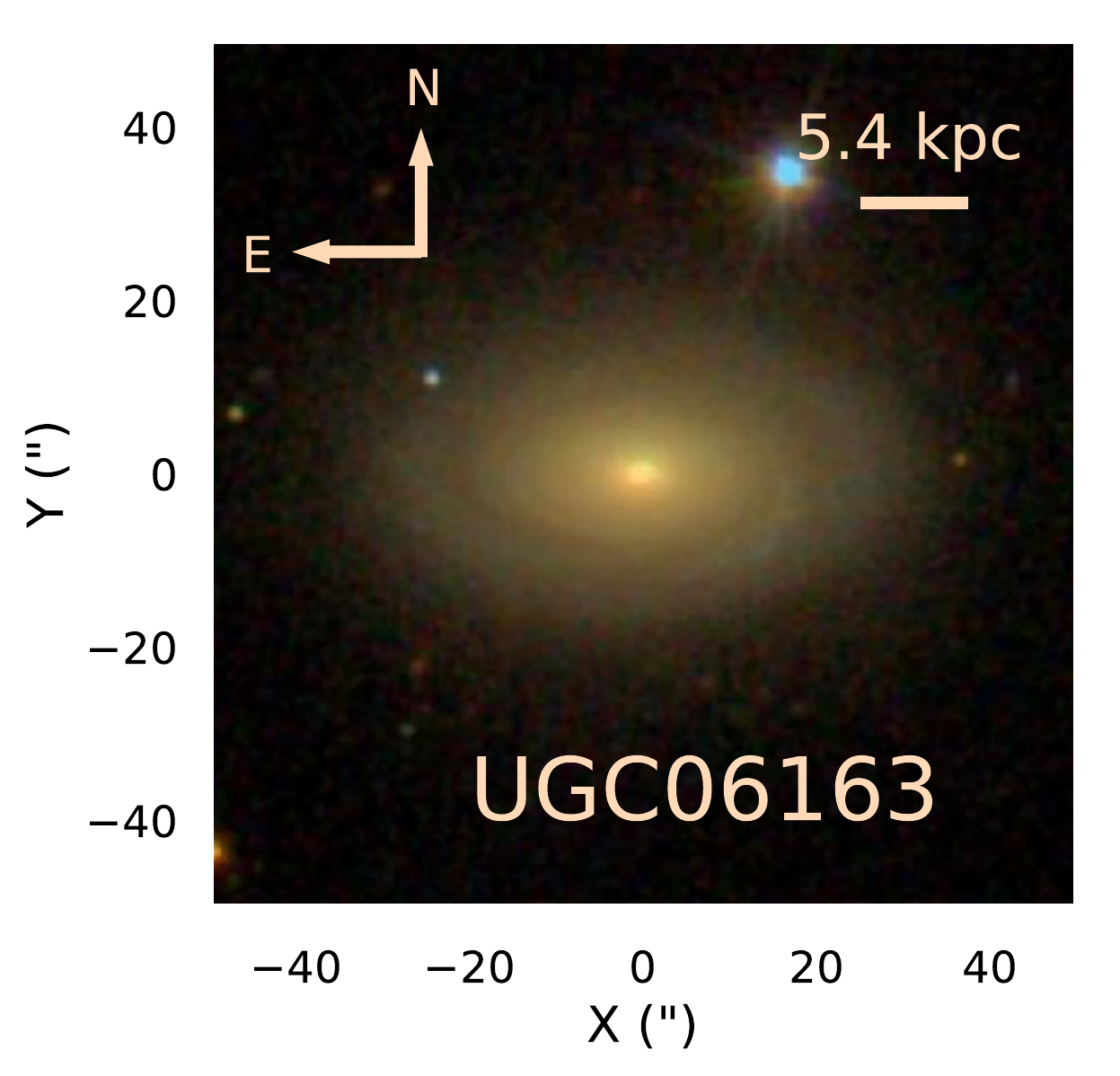}
\end{subfigure}
\hfill
\begin{subfigure}{0.19\textwidth}
\includegraphics[width=\textwidth]{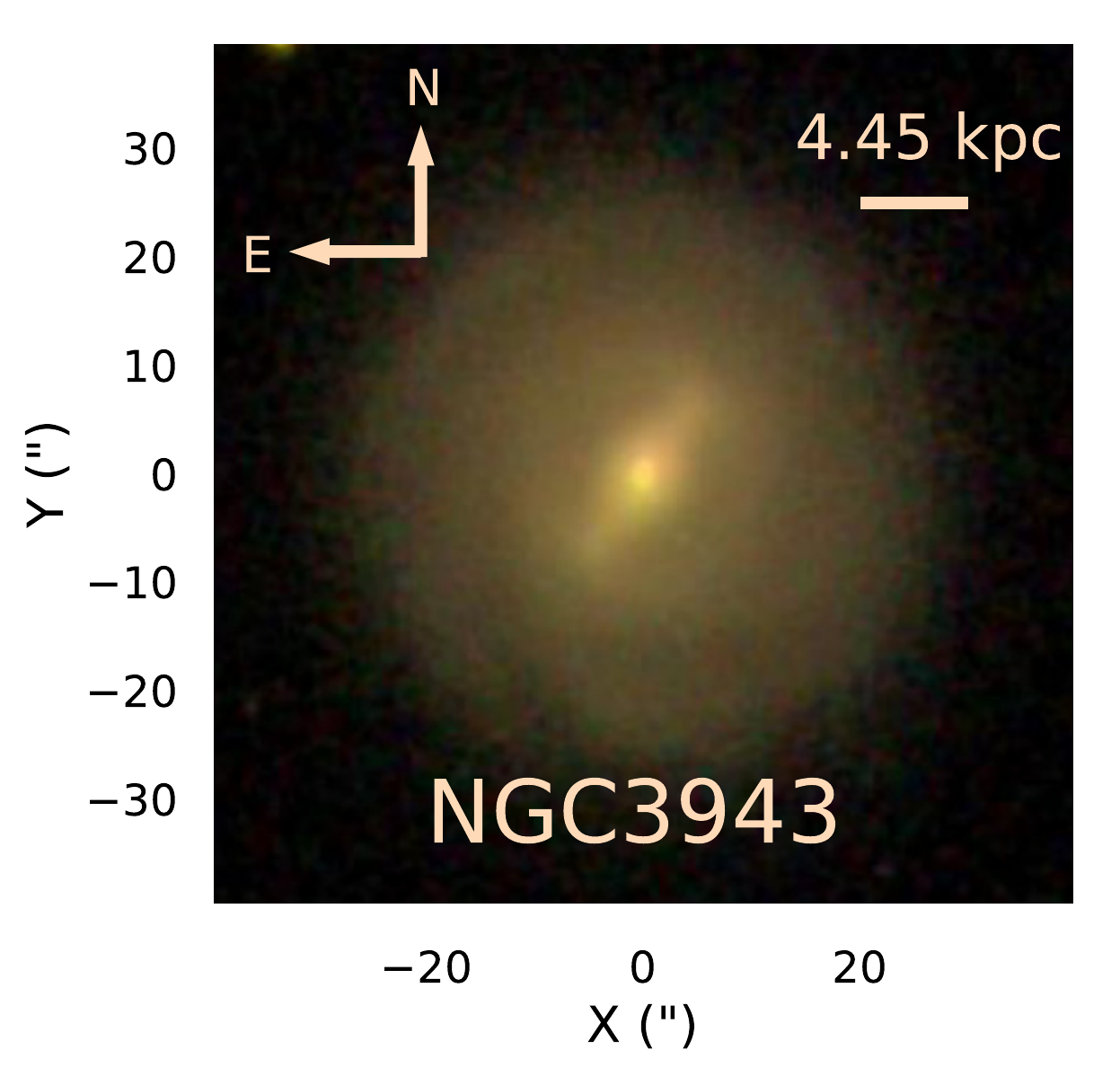}
\end{subfigure}
\hfill
\begin{subfigure}{0.19\textwidth}
\includegraphics[width=\textwidth]{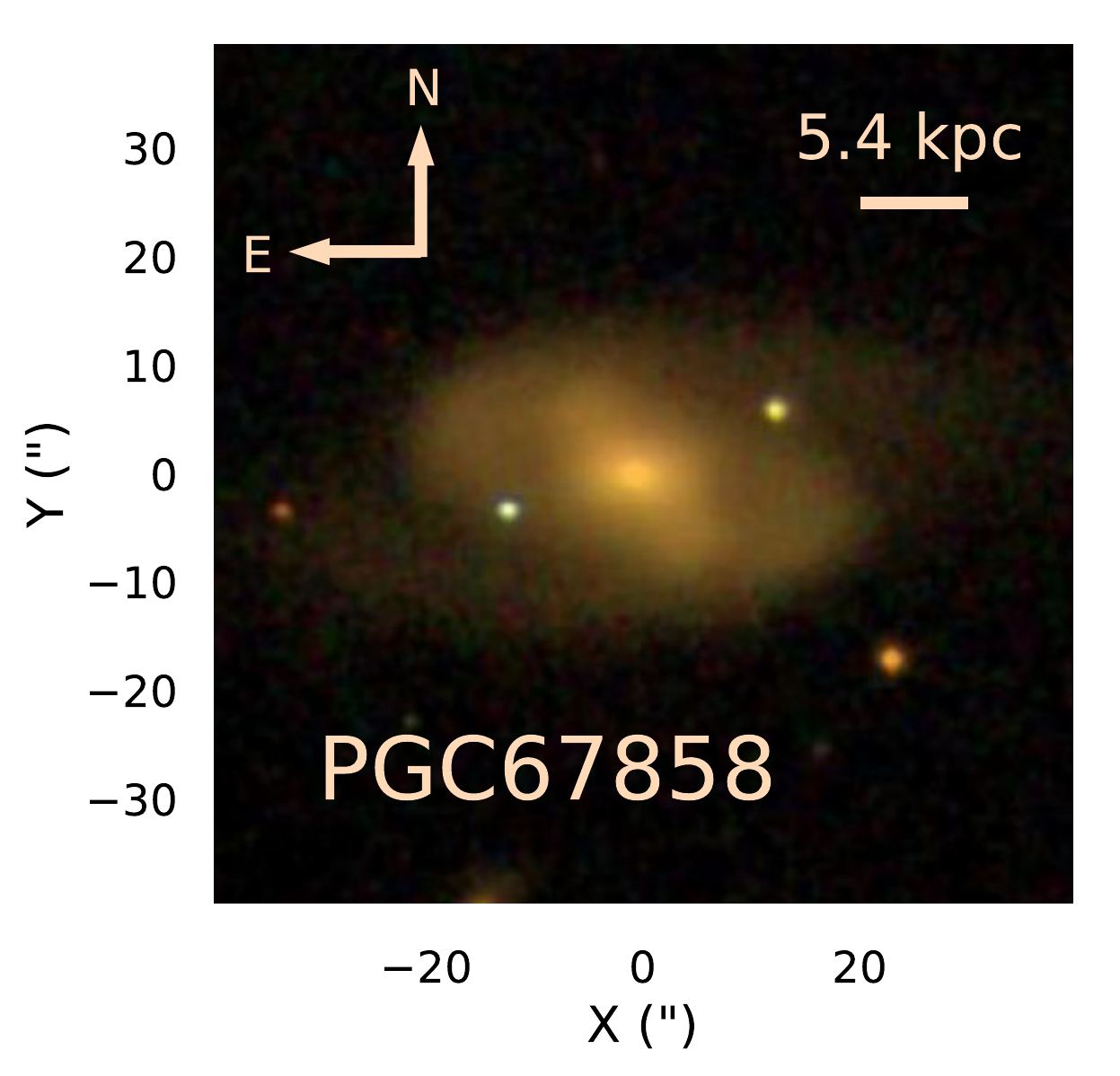}
\end{subfigure}
\hfill
\begin{subfigure}{0.19\textwidth}
\includegraphics[width=\textwidth]{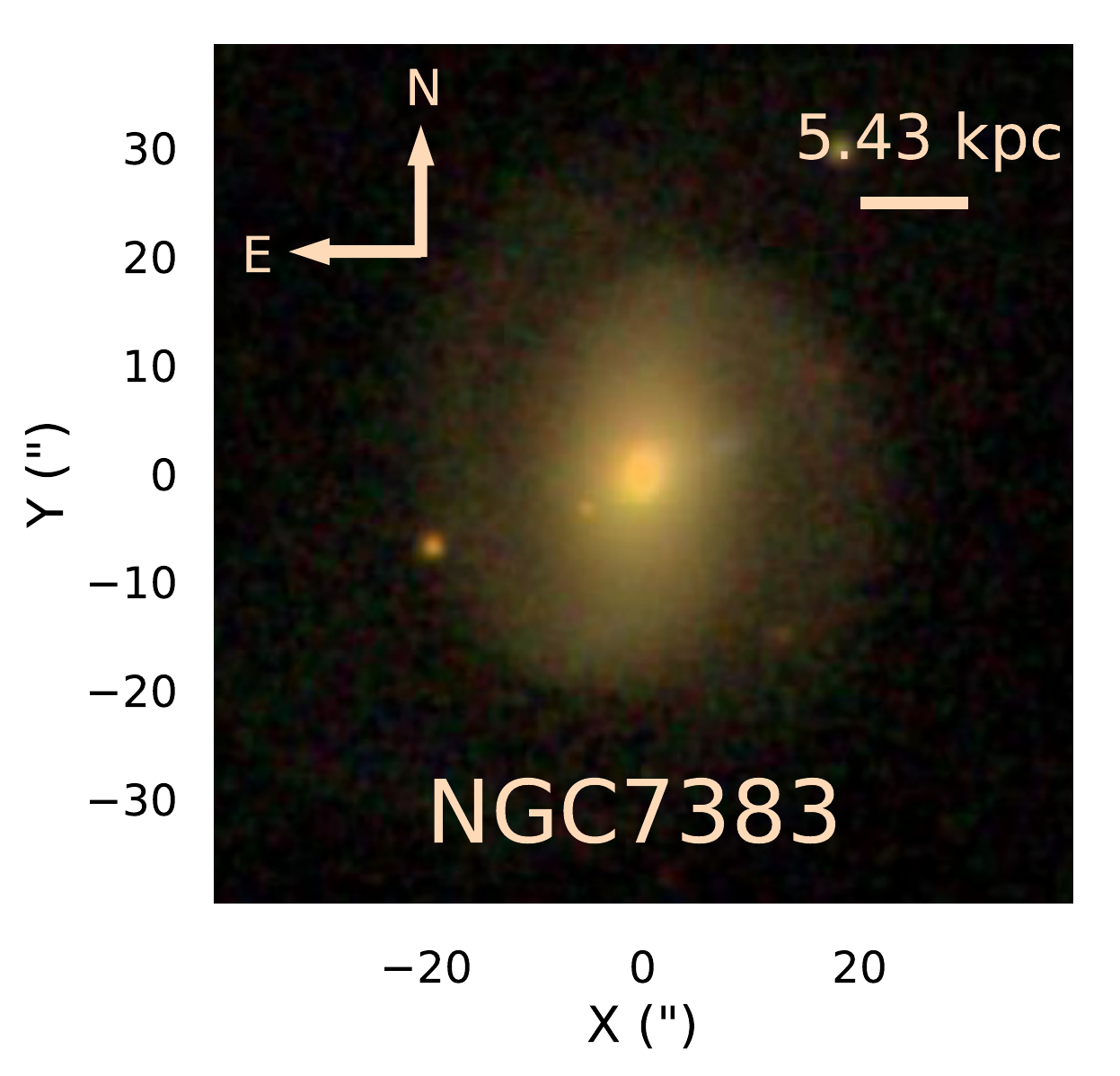}
\end{subfigure}
\hfill
\begin{subfigure}{0.19\textwidth}
\includegraphics[width=\textwidth]{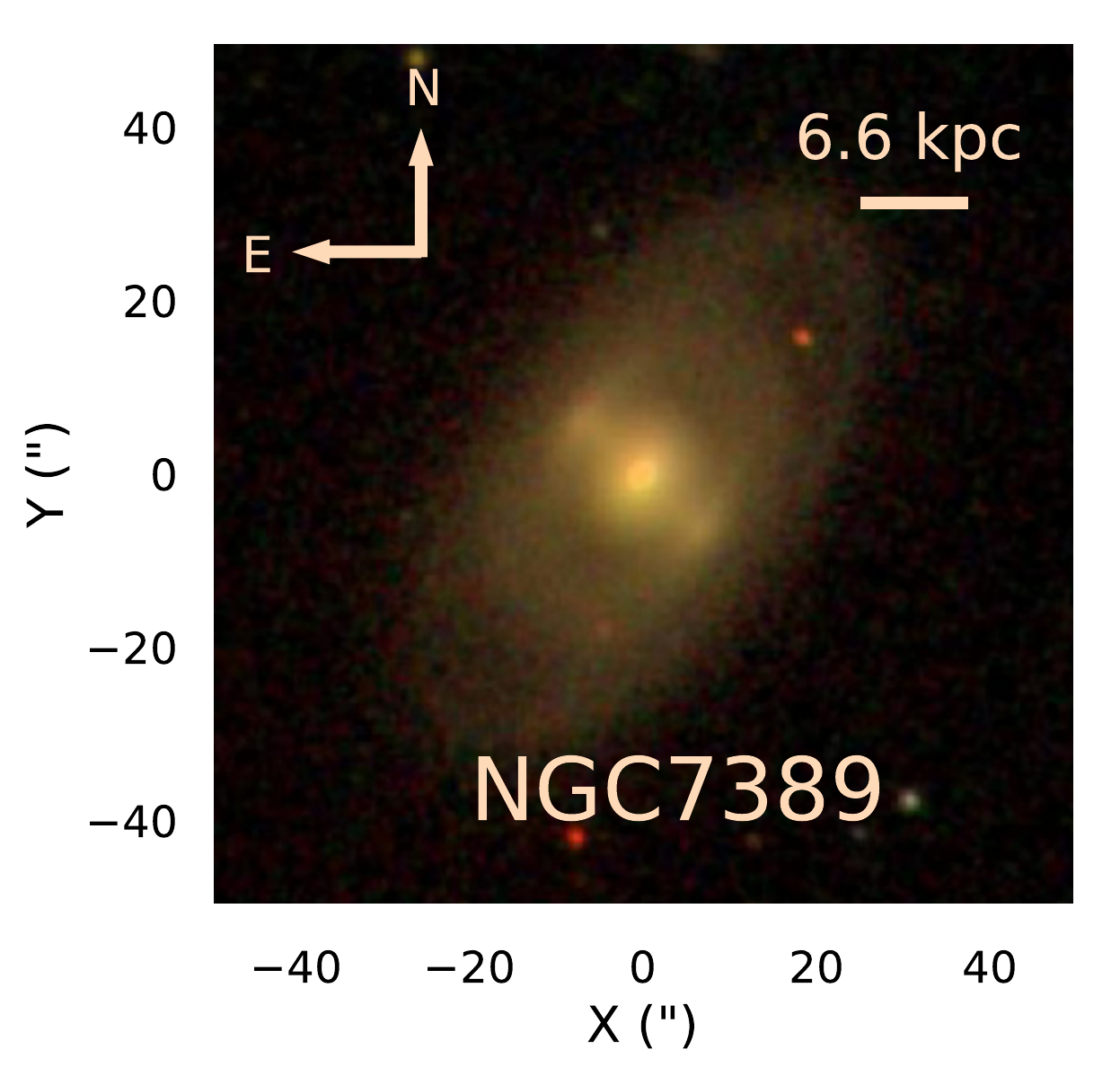}
\end{subfigure}

\vskip\baselineskip

\begin{subfigure}{0.19\textwidth}
\includegraphics[width=\textwidth]{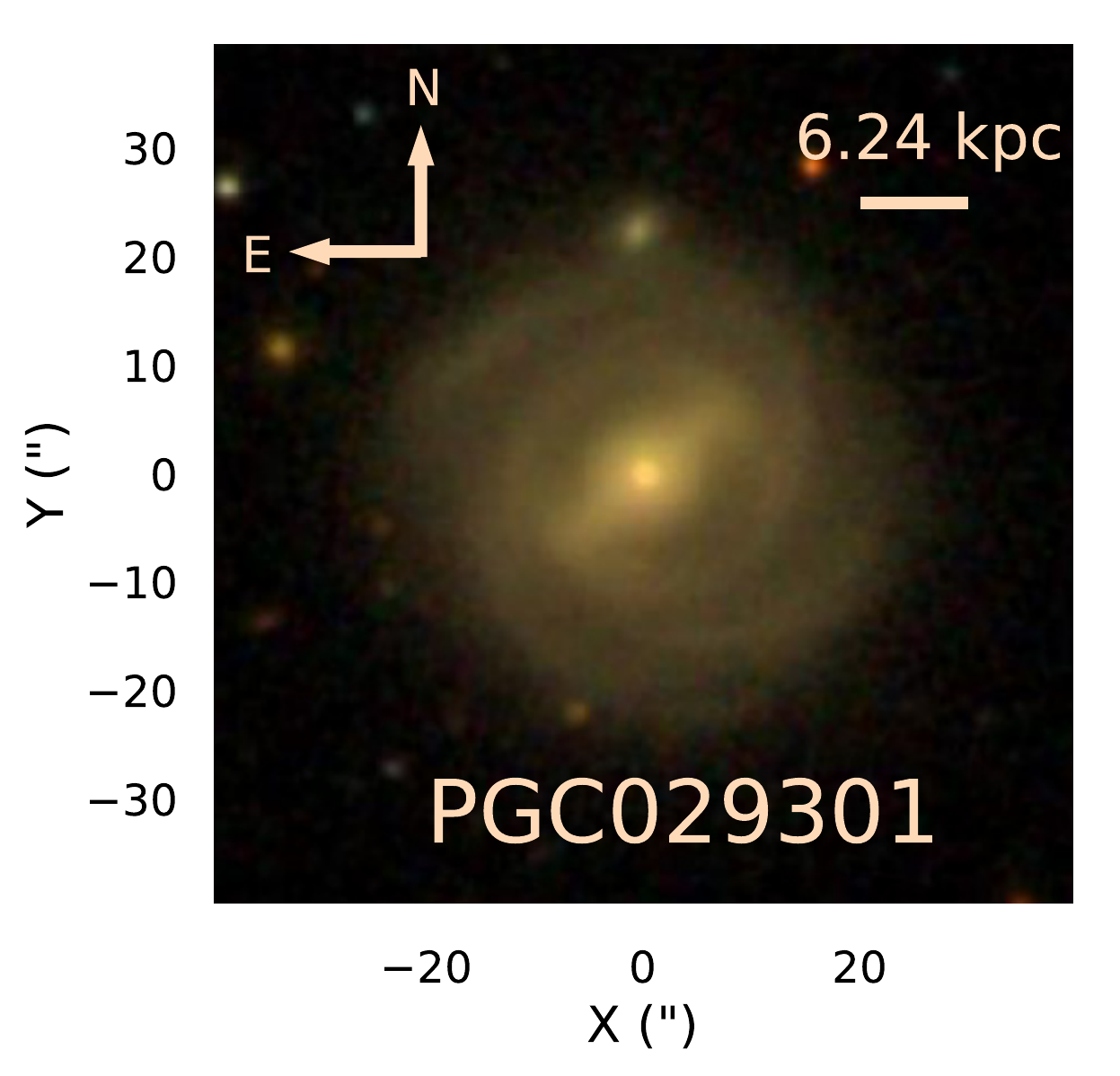}
\end{subfigure}
\hfill
\begin{subfigure}{0.19\textwidth}
\includegraphics[width=\textwidth]{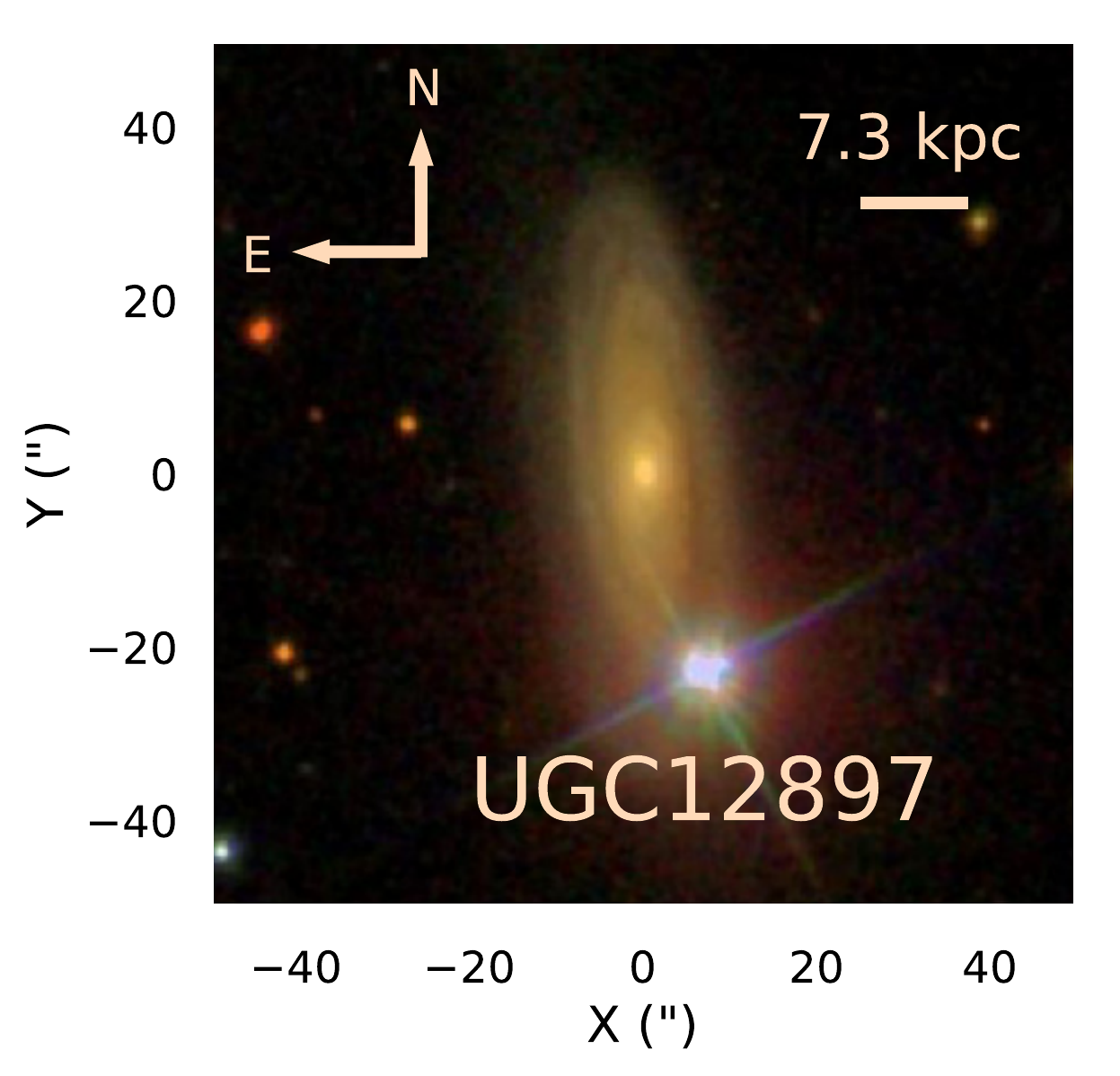}
\end{subfigure}
\hfill
\begin{subfigure}{0.19\textwidth}
\includegraphics[width=\textwidth]{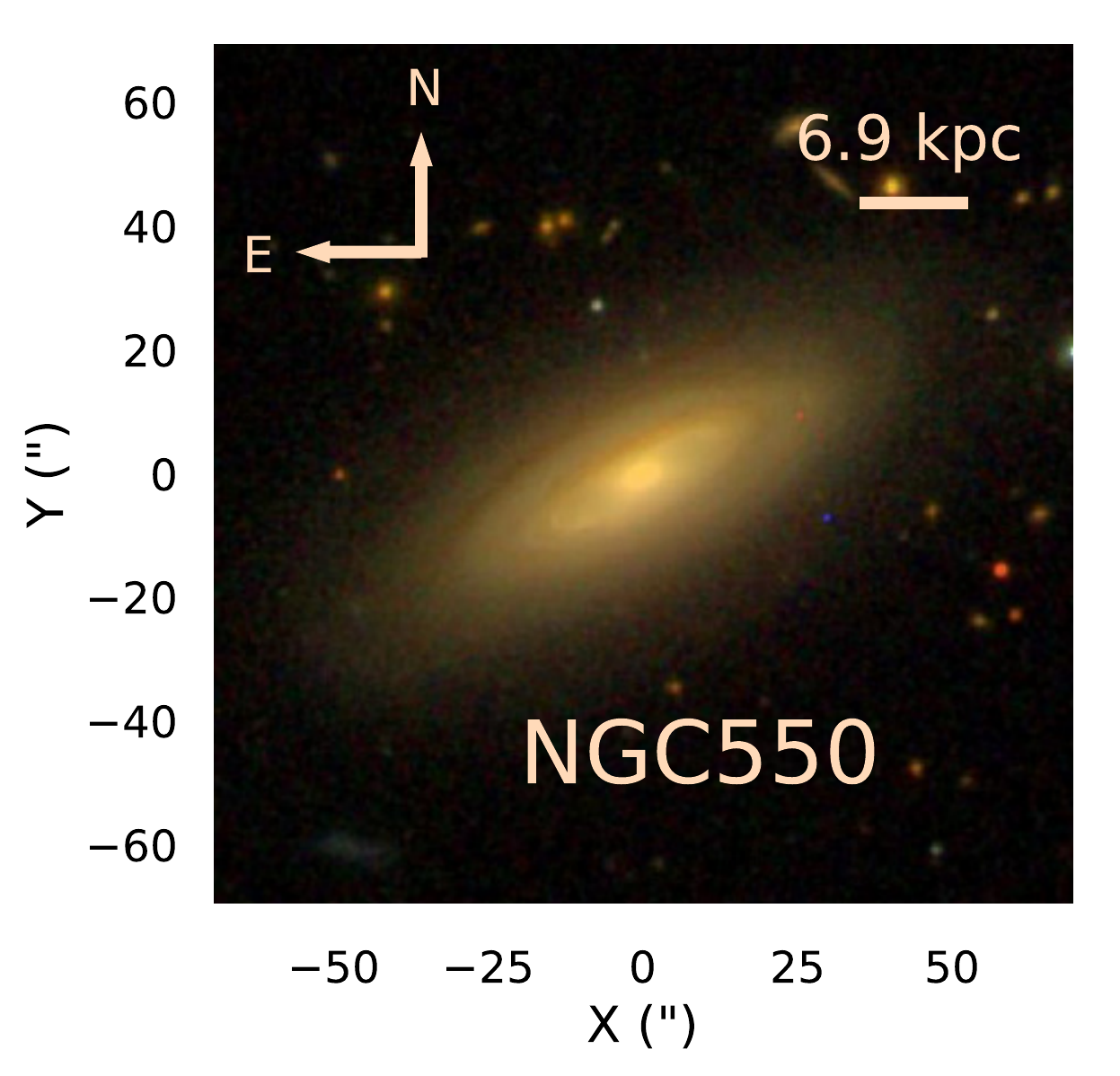}
\end{subfigure}
\hfill
\begin{subfigure}{0.19\textwidth}
\includegraphics[width=\textwidth]{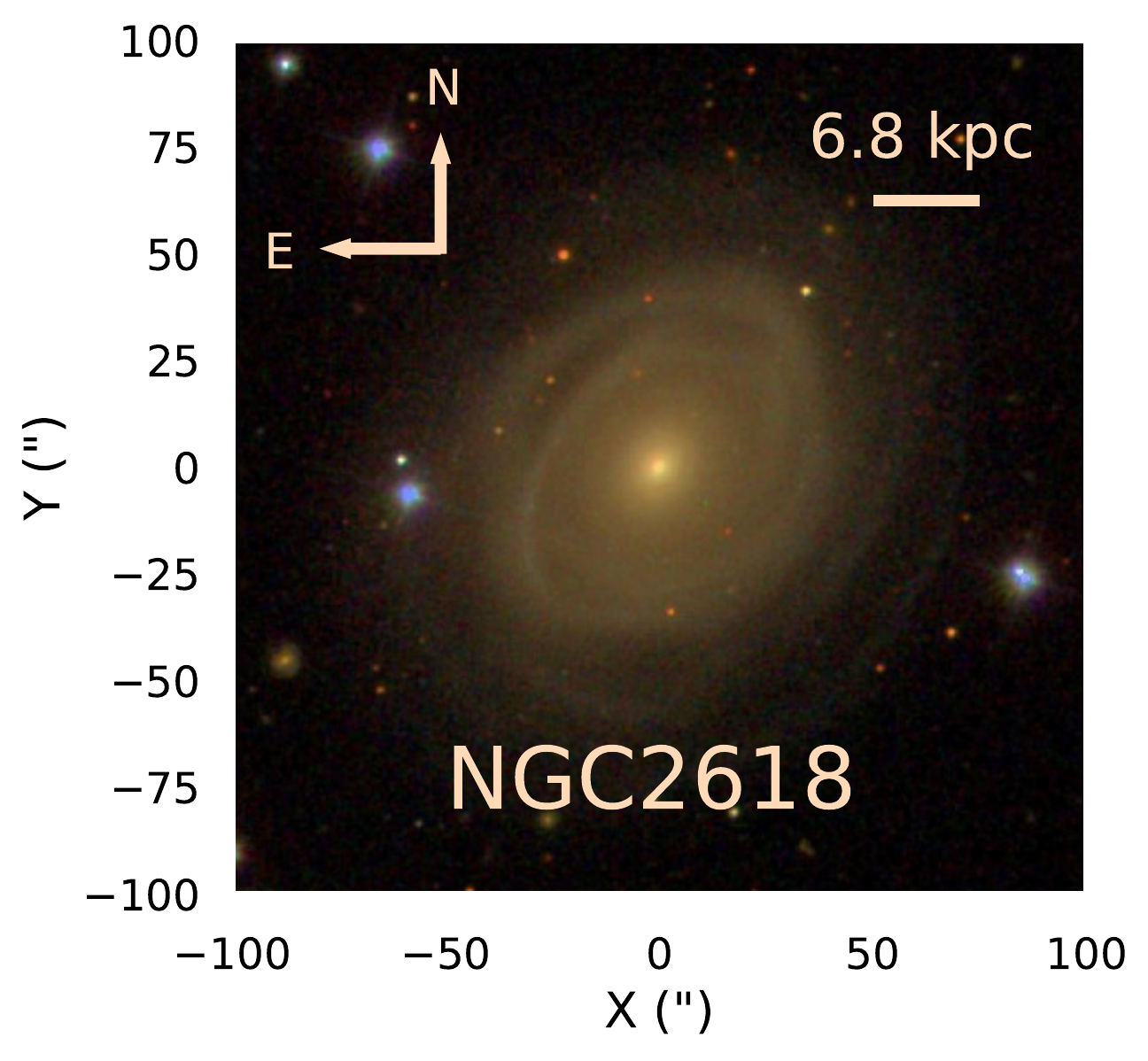}
\end{subfigure}
\hfill
\begin{subfigure}{0.19\textwidth}
\includegraphics[width=\textwidth]{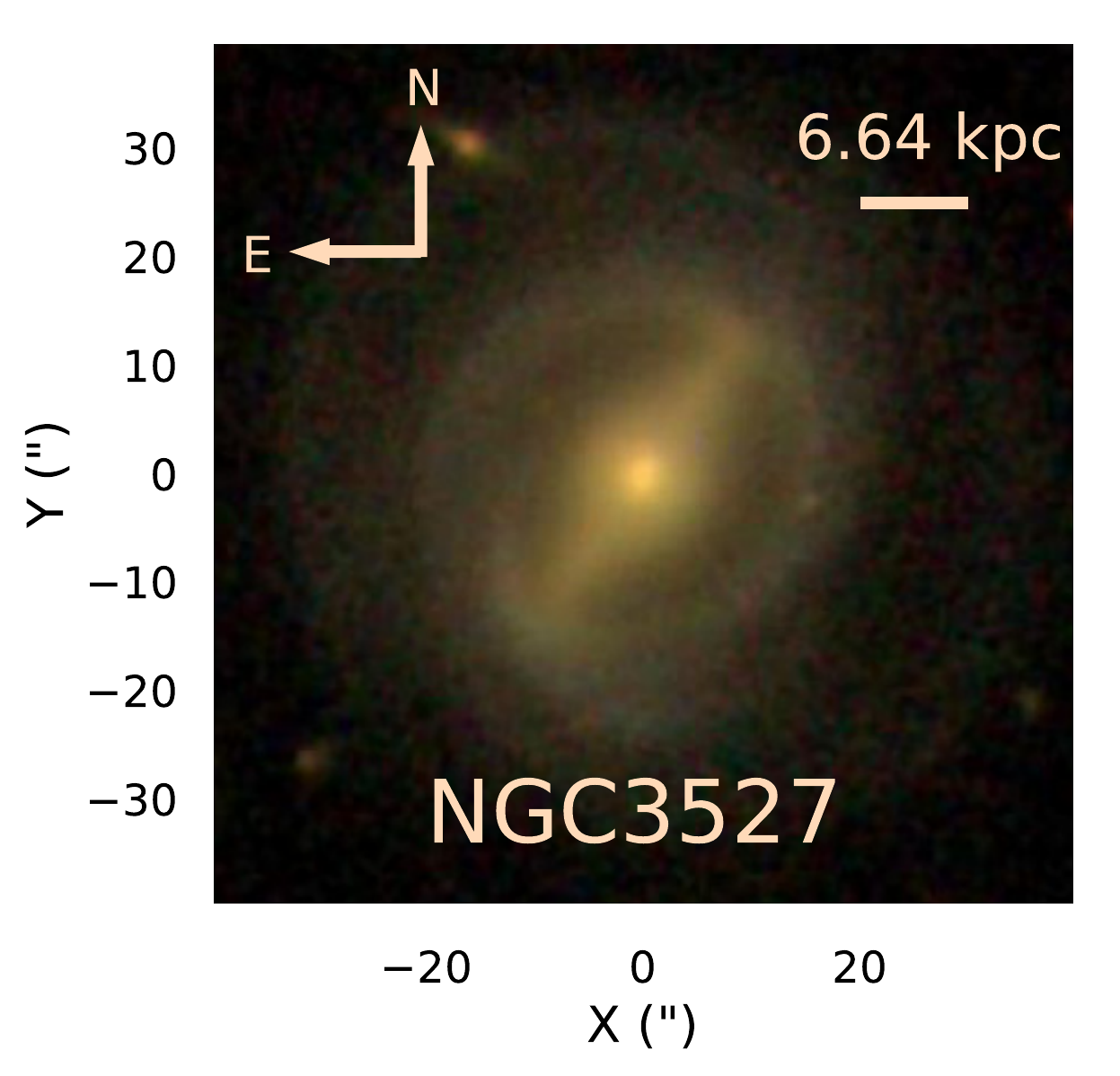}
\end{subfigure}

\caption{SDSS cutout $gri$ images of the thirty galaxies in the higher mass passive spiral sample.}
\label{ps_highmass}
\end{figure*}

\onecolumn
\begin{landscape}
\begin{footnotesize}
\begin{longtable}{l l c c c c c c c c c c}
\caption{The mass, $z$, and T-type-matched comparison sample of all spiral galaxies and their properties. The left most column denotes the passive spiral galaxy, followed by its four comparisons matched most closely in mass, $z$, and T-type from the sample of \citet{Bonne15}. The horizontal line separates the five low mass passive spiral galaxies and their comparison galaxies from their higher-mass counterparts.}\\
\hline
\textbf{Passive} & \textbf{Comparison} & \textbf{RA}         & \textbf{Dec}       & \textbf{z}$^{1}$    & \textbf{$\textrm{D}$} $^{2}$     & \textbf{Mass}$^{3}$         & \textbf{T-type}$^{4}$  &\textbf{$\textrm{N}_{\textrm{group}}^{5}$}  &\textbf{Galaxy}$^{5}$  &  \textbf{Bar?}$^{6}$ & \textbf{Ansa}$^{6}$ \\
\textbf{Spiral}&\textbf{Galaxy} & \textbf{(J2000)}  & \textbf{(J2000)}  &  &      \textbf{(Mpc)}             &   \textbf{($\textrm{M}_{\odot}$)}       &         &                  & \textbf{Environment}    &  & \textbf{Bar?} \\
\hline
NGC 4440        & PGC  41376 & 187.7381 & 11.4836  & 0.0060 & 25.63        & 7.12$\times 10^{8}$  & 1      & 197       & satellite & yes   & no        \\
  		        & PGC  40396 & 186.1508 & 39.3830 & 0.0045 & 19.14        & 3.86$\times 10^{9}$  & 1      & 1         & isolated   & no   & --        \\
  		        & PGC  5679  & 22.9035  & -6.8937 & 0.0050 & 21.23        & 3.80$\times 10^{9}$  & 1      & 8        & satellite   & no   & --        \\
  		        & PGC  42174 & 189.4517  & 5.3684  & 0.0061 & 17.50*           & 4.93$\times 10^{9}$  & 1      & 1         & isolated   & no   & --        \\
NGC 4277        & PGC  32287 & 162.0507  & 28.6018 & 0.0079 & 33.90        & 5.86$\times 10^{9}$  & 1      & 5      & satellite   & no   & --        \\
 		        & PGC  24829 & 132.5492 & 35.0764 & 0.0095 & 43.35*  & 4.92$\times 10^{9}$  & 1      & 1         & isolated   & no   & --        \\
		     & PGC  40705 & 186.6344 & 12.6108 & 0.0059 & 22.47*  &  7.10$\times 10^{8}$   & 1      & 197      & satellite & yes   & no        \\
		       & PGC  26979 & 142.5707 & 29.5400 & 0.0077 & 33.20        & 2.93$\times 10^{9}$  & 1      & 1        & isolated   & yes   & no        \\
NGC 4880        & PGC  40988 & 187.0643 & 28.6203   & 0.0027 & 11.73        & 6.33$\times 10^{9}$  & 1      & 15       & satellite  & no   & --        \\
		       & PGC  72128 & 355.3712 & 3.7400  & 0.0078 & 33.59         & 7.86$\times 10^{9}$  & 1      & 2         & BGG   & yes   & no        \\
 		       & PGC  30445 & 155.8775 & 19.8650  & 0.0052 & 23.57*  & 1.46$\times 10^{10}$ & 1      & 11       & satellite  & no   & --        \\
		        & PGC  29009 & 150.3580 & 15.7700 & 0.0106  & 35.50*  & 8.36$\times 10^{9}$  & 1      & 1         & isolated   & yes   & no        \\			
NGC 4305        & PGC  43798 & 193.3716 & 2.1683  & 0.0061 & 43.57*  & 9.67$\times 10^{9}$  & 1      & 7      & satellite   & no   & --        \\
  		        & PGC  30059 & 154.4107  & 21.6882 & 0.0060 & 27.54*  & 8.23$\times 10^{9}$  & 1      & 11       & satellite  & yes   & no        \\
  		        & PGC  38885 & 183.0492  & 13.2464 & 0.0062 & 57.27* & 3.28$\times 10^{9}$  & 1      & 9        & satellite   & no   & --        \\
  		        & PGC  40306 & 186.0065 & 12.2050 & 0.0063  & 29.33*  & 5.02$\times 10^{9}$  & 1      & 197      & satellite & no   & --        \\
NGC 4264        & PGC  26501 & 140.4501 & 40.1512 & 0.0068  & 43.75*          & 5.50$\times 10^{9}$   & 1      & 1         & isolated   & no   & --        \\
 		      & PGC  56334 & 238.5581 & 14.6012 & 0.0073 & 19.99*  & 7.56$\times 10^{9}$  & 1      & 1         & isolated   & yes   & no        \\
  		        & PGC  55480 & 233.6350 & 15.1938 & 0.0082 & 22.77*  & 8.30$\times 10^{9}$  & 1      & 3         & BGG   & no   & --        \\
  		        & PGC  35711 & 173.4556 & 53.1255 & 0.0047  & 22.69*  & 8.27$\times 10^{9}$  & 1      & 65     & satellite  & no   & --        \\
\hline
NGC 4260        & PGC  43254 & 192.0955 & 8.4872  & 0.0060  & 28.24*  & 1.95$\times 10^{10}$ & 1      & 1         & isolated   & no   & --        \\
  		       & PGC  41383 & 187.7489 & 8.0779  & 0.0062 & 26.62        & 1.56$\times 10^{10}$ & 1      & 197       & satellite & no   & --        \\
  		        & PGC  38031 & 180.6761 & 1.9768   & 0.0010 & 28.43*        & 1.98$\times 10^{10}$ & 1      & 1         & isolated   & no   & --        \\
  		        & PGC  36158 & 174.9271 & 31.9094  & 0.0114 & 46.73*  & 1.96$\times 10^{10}$ & 1      & 2         & BGG   & no   & --        \\
NGC 2692        & PGC  49431 & 208.6214 & 54.3307 & 0.0167 & 72.25        & 2.12$\times 10^{10}$ & 1      & 1        & isolated  & yes   & no        \\
  		        & PGC  70348 & 345.8153 & 8.8737   & 0.0142  & 49.72*  & 2.01$\times 10^{10}$ & 1      & 3         & satellite   & no   & --        \\
  		        & PGC  34883 & 170.5614 & 20.2085 & 0.0173  & 74.64        & 2.11$\times 10^{10}$  & 1      & 5        & satellite   & no   & --        \\
  		        & PGC  72639 & 357.8615  & 20.5862 & 0.0158  & 46.93*  & 2.14$\times 10^{10}$ & 1      & 1         & isolated   & no   & --        \\
NGC 357         & PGC  29855 & 153.5629 & 3.4661   & 0.0041  & 14.62*  & 2.21$\times 10^{10}$ & 1      & 3        & satellite   & no   & --        \\
 		        & PGC  36907 & 177.2668 & 59.4327 & 0.0121 & 52.13        & 2.17$\times 10^{10}$ & 1      & 7        & satellite   & yes   & no        \\
  		        & PGC  33371 & 165.7967 & 27.9725   & 0.0075  & 17.87*  & 1.84$\times 10^{10}$ & 1      & 2         & BGG   & yes   & no        \\
 		        & PGC  54110 & 227.3815 & 54.5064 & 0.0122 & 65.03*  & 2.04$\times 10^{10}$ & 1      & 2         & BGG   & yes   & yes        \\	
NGC 7743        & PGC  35440 & 172.5311 & 9.2766   & 0.0038 & 19.79*  & 1.59$\times 10^{10}$ & 1      & 2         & BGG   & no   & --        \\
  		       & PGC  39724 & 184.9611 & 29.6147   & 0.0036  & 20.67*  & 1.56$\times 10^{10}$ & 1      & 15      & satellite  & no   & --        \\
		       & PGC  6993  & 28.3054  & 4.1958  & 0.0045 & 19.13        & 1.49$\times 10^{10}$ & 1      & 1         & isolated   & yes   & no        \\
		       & PGC  3563  & 14.9171  & 15.3310 & 0.0164 & 71.59*  & 2.22$\times 10^{10}$ & 1      & 1         & isolated   & yes   & no        \\			
 NGC 2648        & PGC  70098 & 344.3314  & -1.0490 & 0.0086  & 36.77        & 2.63$\times 10^{10}$  & 1      & 1         & isolated   & yes   & no        \\
  		        & PGC  49604 & 209.2333 & 29.1643 & 0.0101  & 42.86*  & 2.59$\times 10^{10}$ & 1      & 1         & isolated   & yes   & yes        \\
  		        & PGC  28631 & 148.8900 & 16.4328 & 0.0152 & 56.68*  & 2.42$\times 10^{10}$ & 1      & 2       & Satellite   & no   & --        \\
  		        & PGC  70118 & 344.5018 & 6.0698   & 0.0141  & 60.98         & 2.49$\times 10^{10}$  & 1      & 1         & isolated   & yes   & yes        \\
NGC 656         & PGC  6982  & 28.2487  & 12.7085 & 0.0135  & 57.97*       & 2.33$\times 10^{10}$ & 1      & 10       & satellite  & no   & --        \\
  		         & PGC  37497 & 179.0299 & 55.3907 & 0.0208  & 90.32        & 2.60$\times 10^{10}$  & 1      & 1         & isolated   & no   & --        \\
  		        & PGC  23855 & 127.5105 & 21.4885 & 0.0172 & 58.57*  & 2.36$\times 10^{10}$ & 1      & 3       & satellite   & no   & --       \\
  		        & PGC  55993 & 236.5682 & 2.4155  & 0.0140 & 39.97*  & 2.84$\times 10^{10}$ & 1      & 2         & BGG   & no   & --        \\
NGC 4608        & PGC  43451 & 192.6109 & 25.5008 & 0.0047 & 20.99*  & 3.00$\times 10^{10}$  & 1      & 3         & BGG   & no   & --       \\
  		       & PGC  26008 & 138.3983  & 12.4408 & 0.0185   & 82.43*  & 2.86$\times 10^{10}$  & 1      & 1         & isolated   & no   & --        \\
		        & PGC  31701 & 159.8827 & 5.1075  & 0.0301  & 131.54         & 2.84$\times 10^{10}$ & 1      & 1         & isolated   & no   & --        \\
		       & PGC  32472 & 162.5900 & 41.4640  & 0.0258 & 112.22        & 3.14$\times 10^{10}$ & 1      & 1         & isolated   & no   & --        \\			
UGC 12800  & PGC  29631 & 152.6664 & 20.0702 & 0.0154  & 60.16*       & 3.04$\times 10^{10}$ & 1      & 1         & isolated   & no   & --       \\
  			   & PGC  22962 & 122.8462 & 3.6331  & 0.0153 & 66.11        & 2.95$\times 10^{10}$ & 1      & 1         & isolated   & yes   & no        \\
  			  & PGC  35594 & 173.0900 & 0.8040  & 0.0230 & 99.85        & 3.04$\times 10^{10}$ & 1      & 3         & BGG   & no   & --        \\
  			  & PGC  27926 & 146.2379  & 16.7074 & 0.0226 & 97.96       & 3.07$\times 10^{10}$ & 1      & 1         & isolated   & yes   & yes        \\
NGC 4643        & PGC  43074 & 191.5646 & 8.3484  & 0.0243 & 100.43*  & 3.73$\times 10^{10}$ & 1      & 1         & isolated   & yes   & yes        \\
  		        & PGC  45757 & 197.7571  & 29.5783 & 0.0265  & 113.35*  & 3.68$\times 10^{10}$ & 1      & 2         & satellite   & yes   & no        \\
 		        & PGC  45542 & 197.1137 & 28.3206& 0.0242  & 104.97        & 2.68$\times 10^{10}$ & 1      & 136       & satellite & yes   & yes        \\
 		       & PGC  69449 & 340.0711 & 8.0537  & 0.0229 & 99.59        & 2.62$\times 10^{10}$ & 1      & 4         & BGG   & yes   & no        \\
NGC 7563        & PGC  49563 & 209.0694 & 47.2356 & 0.0075 & 26.05*  & 3.28$\times 10^{10}$ & 1      & 1         & isolated   & no   & --        \\
  		        & PGC  70455 & 346.4512 & 3.5451  & 0.0145  & 39.77*  & 3.47$\times 10^{10}$ & 1      & 3         & BGG   & no   & --       \\
  		        & PGC  49598 & 209.2126 & 37.797  & 0.0120 & 51.59        & 2.30$\times 10^{10}$ & 1      & 9      & satellite   & yes   & yes        \\
  		        & PGC  7322  & 29.2408  & -5.4029 & 0.0149 & 64.19        & 3.45$\times 10^{10}$  & 1      & 1         & isolated   & yes   & yes        \\			 
NGC 2878        & PGC  38288 & 181.3663 & 20.3088 & 0.0271 & 117.99        & 3.29$\times 10^{10}$ & 2      & 30       & satellite  & no   & --        \\
  		        & PGC  38338 & 181.4588 & 20.4770 & 0.0252 & 109.54        & 3.29$\times 10^{10}$ & 2      & 30       & satellite  & no   & --       \\
  		        & PGC  26606 & 140.8979 & 24.7616 & 0.0275 & 109.38*       & 4.25$\times 10^{10}$ & 2      & 4         & BGG   & no   & --        \\
  		        & PGC  26665 & 141.1621 & 56.1296 & 0.0264 & 114.88        & 2.45$\times 10^{10}$ & 2      & 1         & isolated   & yes   & yes        \\		       		        		       
NGC 109         & PGC  52261 & 219.3423 & 36.5678 & 0.0159 & 50.71*  & 3.38$\times 10^{10}$ & 1      & 7       & satellite   & yes   & no        \\
  		         & PGC  25225 & 134.7103 & 6.2931  & 0.0153 & 65.80       & 3.41$\times 10^{10}$ & 1      & 1         & isolated   & yes   & no        \\
  		        & PGC  6633  & 27.1378  & 12.6138 & 0.0165   & 71.28       & 3.52$\times 10^{10}$ & 1      & 10       & satellite  & yes   & no        \\
  		        & PGC  70819 & 348.7644 & 18.9734   & 0.0151  & 64.94      & 3.26$\times 10^{10}$ & 1      & 4       & satellite   & no   & --        \\
UGC 1271     & PGC  23630 & 126.3003 & 20.3348 & 0.0168 & 70.20*  & 2.80$\times 10^{10}$ & 1      & 21       & satellite  & yes   & no        \\
  			    & PGC  50986 & 214.0995 & 39.5023 & 0.0212 & 92.02         & 3.67$\times 10^{10}$ & 1      & 1         & isolated   & yes   & yes        \\
  			 & PGC  24230 & 129.3608 & 40.0355 & 0.0253  & 109.84        & 3.63$\times 10^{10}$ & 1      & 4       & satellite   & no   & --        \\
  			  & PGC  46633 & 200.3463 & 0.3426  & 0.0212   & 63.84*  & 3.17$\times 10^{10}$ & 1      & 1         & isolated   & no   & --        \\
NGC 538         & PGC  44557 & 194.7871 & 37.3103& 0.0177  & 76.48        & 3.60$\times 10^{10}$ & 2      & 2      & satellite   & no   & --        \\
  		        & PGC  38634 & 182.3871   & 17.0142 & 0.0251  & 109.08        & 3.57$\times 10^{10}$ & 2      & 4       & satellite   & yes   & no        \\
 		        & PGC  33040 & 164.6870 & 59.5107 & 0.0230 & 101.42*  & 3.56$\times 10^{10}$  & 2      & 1         & isolated   & no   & --        \\
  		        & PGC  24152 & 128.9521 & 1.7217  & 0.0161 & 53.20*  & 2.91$\times 10^{10}$ & 2      & 3         & BGG   & no   & --      \\
NGC 345         & PGC  4906  & 20.3193  & -0.5445 & 0.0157 & 67.85       & 5.07$\times 10^{10}$ & 1      & 1         & isolated   & yes   & yes        \\
  		        & PGC  47180 & 201.9456 & 17.7789 & 0.0245 & 106.49       & 4.07$\times 10^{10}$ & 1      & 1         & isolated   & yes   & no        \\
  		        & PGC  23441 & 125.3526 & 19.1477 & 0.0284 & 126.28*  & 4.07$\times 10^{10}$ & 1      & 1         & isolated   & no   & --        \\
  		        & PGC  698   & 2.4725   & 25.9238 & 0.0132 & 55.18*  & 3.93$\times 10^{10}$ & 1      & 2         & BGG   & no   & --   \\
NGC 4596        & PGC  23993 & 128.3242 & 41.2595 & 0.0243 & 105.57        & 4.18$\times 10^{10}$ & 1      & 8        & satellite   & no   & --        \\
		       & PGC  40490 & 186.3254 & 4.9251  & 0.0064 & 21.73*  & 3.75$\times 10^{10}$ & 1      & 39       & satellite  & no   & --       \\
  		       & PGC  42743 & 190.7249  & 20.9897 & 0.0244  & 115.34*  & 4.22$\times 10^{10}$ & 1      & 1         & isolated   & yes   & yes        \\
		      & PGC  3486  & 14.5992  & -8.4078 & 0.0128  & 55.19        & 4.46$\times 10^{10}$ & 1      & 2         & BGG   & no   & --       \\
PGC 047732  & PGC  42137 & 189.3377 & 28.2081 & 0.0281 & 124.29*  & 4.21$\times 10^{10}$ & 2      & 2         & BGG   & no   & --       \\
  			  & PGC  71258 & 350.6911 & 29.1379 & 0.0172 & 89.79*  & 4.26$\times 10^{10}$ & 2      & 5        & satellite   & no   & --        \\
  			  & PGC  51439 & 216.0316 & 34.8589 & 0.0148  & 63.81       & 4.36$\times 10^{10}$ & 2      & 1         & isolated   & no   & --        \\
  			  & PGC  28984 & 150.2631 & 36.6186  & 0.0240 & 104.24       & 3.93$\times 10^{10}$ & 2      & 1         & isolated   & no   & --        \\ 
UGC 8484    & PGC  38271 & 181.3114 & 38.2355   & 0.0257 & 111.76        & 4.47$\times 10^{10}$ & 3      & 1         & isolated   & no   & --        \\
		   & PGC  38441 & 181.7644 & 18.5317  & 0.0268 & 107.39*  & 4.00$\times 10^{10}$  & 3      & 1         & isolated   & no   & --       \\
 		    & PGC  53817 & 226.1691 & 12.6335 & 0.0307 & 114.35*  & 4.51$\times 10^{10}$ & 3      & 1         & isolated   & no   & --        \\
		    & PGC  51283 & 215.3047 & 29.9936 & 0.0325 & 123.30*  & 4.43$\times 10^{10}$ & 3      & 1         & isolated   & no   & --        \\ 
NGC 15        & PGC  366   & 1.3079   & 6.7720  & 0.0189  & 94.16*   & 4.21$\times 10^{10}$  & 1      & 3        & satellite   & no   & --      \\
                       & PGC  55243 & 232.4375   & 42.9187 & 0.0194 & 59.56*   & 4.59$\times 10^{10}$ & 1      & 5         & BGG   & yes   & no        \\
  		        & PGC  49244 & 208.0347 & 14.1163  & 0.0253 & 91.88*         & 4.53$\times 10^{10}$ & 1      & 5        & satellite   & no   & --        \\
  		        & PGC  28452 & 148.1235 & 2.1544  & 0.0197 & 63.54*  & 4.30$\times 10^{10}$ & 1      & 1         & isolated   & no   & --      \\
PGC 070141   & PGC  36436 & 175.8523  & 19.7498 & 0.0231 & 100.09        & 5.07$\times 10^{10}$ & 1      & 61       & satellite  & no   & --        \\
  		  	& PGC  69780 & 342.2421 & 7.2190  & 0.0226 & 98.01        & 4.87$\times 10^{10}$ & 1      & 1         & isolated   & yes   & no        \\
  			   & PGC  33012 & 164.6050  & 24.2264 & 0.0241 & 107.39*  & 4.77$\times 10^{10}$ & 1      & 2         & BGG   & no   & --       \\
  			   & PGC  40768 & 186.7311  & 37.9089  & 0.0247 & 107.19        & 4.70$\times 10^{10}$ & 1      & 1         & isolated   & yes   & yes        \\		         
UGC 6163     & PGC  44144 & 194.1160 & 26.9875  & 0.0240 & 104.27         & 5.16$\times 10^{10}$ & 1      & 136      & satellite & yes   & no        \\
 			   & PGC  50042 & 210.7543 & 34.7579 & 0.0145   & 63.04*  & 4.95$\times 10^{10}$ & 1      & 6       & satellite   & no   & --        \\
 			     & PGC  40783 & 186.7478 & 22.6395 & 0.0254 & 110.75        & 4.97$\times 10^{10}$ & 1      & 1         & isolated   & yes   & yes        \\
			    & PGC  47953 & 203.9511  & 2.9989  & 0.0243 & 105.44        & 5.10$\times 10^{10}$ & 1      & 4       & satellite   & no   & --        \\
NGC 3943        & PGC  33126 & 164.9711   & 50.0153 & 0.0276   & 120.22        & 4.97$\times 10^{10}$ & 2      & 4       & satellite   & yes   & no        \\
  		        & PGC  45358 & 196.5720 & 29.0631  & 0.0260 & 110.37*  & 5.18$\times 10^{10}$ & 2      & 136       & satellite & no   & --        \\
  		        & PGC  55601 & 234.1757  & 43.5394 & 0.0198   & 85.74        & 4.88$\times 10^{10}$ & 2      & 7        & satellite   & yes   & no        \\
  		        & PGC  50750 & 213.1589 & 39.3102 & 0.0271  & 117.96        & 5.25$\times 10^{10}$ & 2      & 1         & isolated   & no   & --        \\
PGC 067858   & PGC  27666 & 145.3192 & 35.8822 & 0.0251 & 103.41*  & 5.16$\times 10^{10}$ & 3      & 5         & BGG   & no   & --        \\
  			& PGC  70877 & 349.0029 & 25.5567  & 0.0251 & 109.19        & 4.72$\times 10^{10}$ & 3      & 9       & satellite   & no   & --        \\
  			 & PGC  7259  & 29.0908  & -4.4676 & 0.0161 & 64.43*  & 5.05$\times 10^{10}$ & 3      & 1         & isolated   & no   & --      \\
  			 & PGC  51108 & 214.6165 & 12.8830  & 0.0271 & 119.32*  & 4.98$\times 10^{10}$ & 3      & 1         & isolated   & yes   & no        \\			
NGC 7383        & PGC  22445 & 120.0874  & 26.6135 & 0.0284  & 120.31*       & 5.15$\times 10^{10}$ & 1      & 4         & BGG   & yes   & no        \\
		      & PGC  40192 & 185.7985 & 6.0722  & 0.0267  & 116.05         & 5.20$\times 10^{10}$ & 1      & 2         & BGG   & yes   & no        \\
		       & PGC  47961 & 203.9583 & 34.9988 & 0.0270 & 117.57        & 5.08$\times 10^{10}$ & 1      & 16       & satellite  & no   & --        \\
		       & PGC  31572 & 159.4078 & 37.4557 & 0.0251 & 109.19        & 5.96$\times 10^{10}$ & 1      & 3      & satellite   & no   & --        \\ 
NGC 7389        & PGC  38227 & 181.1808 & 31.1772 & 0.0274  & 119.51       & 5.70$\times 10^{10}$ & 3      & 2         & BGG   & no   & --       \\
 			& PGC  31729 & 159.9914 & 24.0913 & 0.0236 & 90.48*  & 5.66$\times 10^{10}$ & 3      & 1         & isolated   & no   & --        \\
  			 & PGC  70250 & 345.2045 & 26.7409 & 0.0244 & 92.37*  & 5.99$\times 10^{10}$ & 3      & 9       & satellite   & yes   & no        \\
  		       & PGC  37264 & 178.3347 & 20.7516 & 0.0234  & 102.41*  & 5.77$\times 10^{10}$ & 3      & 18        & satellite  & yes   & no        \\       
PGC 029301   & PGC  2331  & 9.7477   & -9.0027 & 0.0185 & 80.01       & 6.27$\times 10^{10}$ & 5      & 2         & BGG   & no   & --        \\
  			   & PGC  26059 & 138.6555 & 8.1172  & 0.0335  & 146.35        & 6.89$\times 10^{10}$ & 5      & 1         & isolated   & no   & --        \\
 			   & PGC  43504 & 192.7480 & 47.6715 & 0.0313  & 105.40*  & 6.13$\times 10^{10}$ & 5      & 4        & BGG   & no   & --       \\
  			   & PGC  32078 & 161.2155  & 6.5969  & 0.0307 & 90.68*  & 6.52$\times 10^{10}$ & 5      & 1         & isolated   & no   & --        \\
UGC 12897   & PGC  54861 & 230.5327 & 13.9282 & 0.0320 & 140.05        & 7.52$\times 10^{10}$ & 2      & 1         & isolated   & no   & --        \\
  			   & PGC  69172 & 338.5282 & 5.5703  & 0.0132 & 51.52*  & 7.13$\times 10^{10}$ & 2      & 1         & isolated   & no   & --        \\
  			  & PGC  51167 & 214.8191 & 26.2986 & 0.0390 & 171.31        & 6.80$\times 10^{10}$ & 2      & 4       & satellite   & no   & --        \\
  			  & PGC  49280 & 208.1112 & 14.4909 & 0.0433 & 190.92         & 6.99$\times 10^{10}$ & 2      & 3         & BGG   & no   & --        \\
NGC 550         & PGC  5628  & 22.6696  & -1.9944 & 0.0180  & 77.93        & 6.99$\times 10^{10}$ & 1      & 43       & satellite  & yes   & no        \\
		       & PGC  47131 & 201.8202  & 32.0307 & 0.0265  & 154.12*  & 7.82$\times 10^{10}$ & 1      & 7      & satellite   & yes   & no        \\
  		        & PGC  25875 & 137.6652 & 50.3798 & 0.0184 & 79.39         & 6.78$\times 10^{10}$ & 1      & 3         & satellite   & yes   & yes        \\
  		        & PGC  2279  & 9.5516   & 2.7286  & 0.0156 & 90.98*  & 8.42$\times 10^{10}$ & 1      & 16       & satellite  & no   & --       \\
NGC 2618        & PGC  354   & 1.2760   & 6.9201  & 0.0184 & 79.25*  & 8.47$\times 10^{10}$ & 2      & 3         & BGG   & no   & --   \\
  		        & PGC  41024 & 187.1235 & 17.0850 & 0.0061 & 20.45*  & 8.43$\times 10^{10}$ & 2      & 197      & Satellite & no   & --       \\
  		        & PGC  53508 & 224.6501 & 44.8836 & 0.0374 & 164.15        & 8.45$\times 10^{10}$ & 2      & 1         & isolated   & no   & --        \\
  		        & PGC  916   & 3.4491    & -4.4751   & 0.0346 & 151.44        & 7.80$\times 10^{10}$ & 2      & 1         & isolated   & yes   & no        \\
NGC 3527        & PGC  32872 & 164.0645 & 9.7544  & 0.0362  & 158.84        & 8.81$\times 10^{10}$ & 1      & 17      & satellite  & no   & --        \\
		       & PGC  55817 & 235.4758 & 28.1341 & 0.0347 & 125.28*  & 8.77$\times 10^{10}$  & 1      & 10        & satellite  & yes   & no        \\
  		        & PGC  52171 & 218.9405 & 24.7258 & 0.0381  & 167.05        & 8.96$\times 10^{10}$  & 1      & 2         & BGG   & no   & --        \\
  		        & PGC  36348 & 175.5989 & 10.2641 & 0.0246 & 143.18*  & 8.40$\times 10^{10}$ & 1      & 9        & satellite   & no   & --       \\
 \hline
\hline
\multicolumn{12}{l}{$^{1}$ From \citet{Bonne15}}\\
\multicolumn{12}{l}{$^{2}$ * denotes redshift independent distances from NED, collated by \citet{Bonne15}, otherwise these are flow-corrected distances, calculated by \citet{Bonne15}.}\\
\multicolumn{12}{l}{$^{3}$ From NASA Sloan Atlas}\\
\multicolumn{12}{l}{$^{4}$ Compiled by \citet{Bonne15}, most of which are from \citet{Paturel03}.}\\
\multicolumn{12}{l}{$^{5}$ Environmental properties from \citet{Tully15}.}\\
\multicolumn{12}{l}{$^{6}$ From visual inspection of SDSS images by the authors.}\\
\end{longtable}
\label{Table:comp}
\end{footnotesize}
\end{landscape}
\twocolumn

    \bibliographystyle{mnras}
  \bibliography{thesisbib}

\begin{thebibliography}{}
\makeatletter
\relax
\def\mn@urlcharsother{\let\do\@makeother \do\$\do\&\do\#\do\^\do\_\do\%\do\~}
\def\mn@doi{\begingroup\mn@urlcharsother \@ifnextchar [ {\mn@doi@}
  {\mn@doi@[]}}
\def\mn@doi@[#1]#2{\def\@tempa{#1}\ifx\@tempa\@empty \href
  {http://dx.doi.org/#2} {doi:#2}\else \href {http://dx.doi.org/#2} {#1}\fi
  \endgroup}
\def\mn@eprint#1#2{\mn@eprint@#1:#2::\@nil}
\def\mn@eprint@arXiv#1{\href {http://arxiv.org/abs/#1} {{\tt arXiv:#1}}}
\def\mn@eprint@dblp#1{\href {http://dblp.uni-trier.de/rec/bibtex/#1.xml}
  {dblp:#1}}
\def\mn@eprint@#1:#2:#3:#4\@nil{\def\@tempa {#1}\def\@tempb {#2}\def\@tempc
  {#3}\ifx \@tempc \@empty \let \@tempc \@tempb \let \@tempb \@tempa \fi \ifx
  \@tempb \@empty \def\@tempb {arXiv}\fi \@ifundefined
  {mn@eprint@\@tempb}{\@tempb:\@tempc}{\expandafter \expandafter \csname
  mn@eprint@\@tempb\endcsname \expandafter{\@tempc}}}

\bibitem[\protect\citeauthoryear{{Aguerri}, {M{\'e}ndez-Abreu}  \&
  {Corsini}}{{Aguerri} et~al.}{2009}]{Aguerri09}
{Aguerri} J.~A.~L.,  {M{\'e}ndez-Abreu} J.,   {Corsini} E.~M.,  2009, \mn@doi
  [\aap] {10.1051/0004-6361:200810931}, \href
  {http://adsabs.harvard.edu/abs/2009A%26A...495..491A} {495, 491}

\bibitem[\protect\citeauthoryear{{Athanassoula}}{{Athanassoula}}{2002}]{Athanassoula02}
{Athanassoula} E.,  2002, \mn@doi [\apjl] {10.1086/340784}, \href
  {http://adsabs.harvard.edu/abs/2002ApJ...569L..83A} {569, L83}

\bibitem[\protect\citeauthoryear{{Athanassoula}}{{Athanassoula}}{2013}]{Athanassoula13}
{Athanassoula} E.,  2013, {Bars and secular evolution in disk galaxies:
  Theoretical input}.
p.~305

\bibitem[\protect\citeauthoryear{{Balogh}, {Navarro}  \& {Morris}}{{Balogh}
  et~al.}{2000}]{Balogh00}
{Balogh} M.~L.,  {Navarro} J.~F.,   {Morris} S.~L.,  2000, \mn@doi [\apj]
  {10.1086/309323}, \href {http://adsabs.harvard.edu/abs/2000ApJ...540..113B}
  {540, 113}

\bibitem[\protect\citeauthoryear{{Bamford} et~al.,}{{Bamford}
  et~al.}{2009}]{Bamford09}
{Bamford} S.~P.,  et~al., 2009, \mn@doi [\mnras]
  {10.1111/j.1365-2966.2008.14252.x}, \href
  {http://adsabs.harvard.edu/abs/2009MNRAS.393.1324B} {393, 1324}

\bibitem[\protect\citeauthoryear{{Barazza}, {Jogee}  \& {Marinova}}{{Barazza}
  et~al.}{2008}]{Barazza08}
{Barazza} F.~D.,  {Jogee} S.,   {Marinova} I.,  2008, \mn@doi [\apj]
  {10.1086/526510}, \href {http://adsabs.harvard.edu/abs/2008ApJ...675.1194B}
  {675, 1194}

\bibitem[\protect\citeauthoryear{{Bekki}}{{Bekki}}{2009}]{Bekki09}
{Bekki} K.,  2009, \mn@doi [\mnras] {10.1111/j.1365-2966.2009.15431.x}, \href
  {http://adsabs.harvard.edu/abs/2009MNRAS.399.2221B} {399, 2221}

\bibitem[\protect\citeauthoryear{{Bell} et~al.,}{{Bell} et~al.}{2012}]{Bell12}
{Bell} E.~F.,  et~al., 2012, \mn@doi [\apj] {10.1088/0004-637X/753/2/167},
  \href {http://adsabs.harvard.edu/abs/2012ApJ...753..167B} {753, 167}

\bibitem[\protect\citeauthoryear{{Bonne}, {Brown}, {Jones}  \&
  {Pimbblet}}{{Bonne} et~al.}{2015}]{Bonne15}
{Bonne} N.~J.,  {Brown} M.~J.~I.,  {Jones} H.,   {Pimbblet} K.~A.,  2015,
  \mn@doi [\apj] {10.1088/0004-637X/799/2/160}, \href
  {http://adsabs.harvard.edu/abs/2015ApJ...799..160B} {799, 160}

\bibitem[\protect\citeauthoryear{{Bower}, {Schaye}, {Frenk}, {Theuns},
  {Schaller}, {Crain}  \& {McAlpine}}{{Bower} et~al.}{2017}]{Bower17}
{Bower} R.~G.,  {Schaye} J.,  {Frenk} C.~S.,  {Theuns} T.,  {Schaller} M.,
  {Crain} R.~A.,   {McAlpine} S.,  2017, \mn@doi [\mnras]
  {10.1093/mnras/stw2735}, \href
  {http://adsabs.harvard.edu/abs/2017MNRAS.465...32B} {465, 32}

\bibitem[\protect\citeauthoryear{{Brinchmann}, {Charlot}, {White}, {Tremonti},
  {Kauffmann}, {Heckman}  \& {Brinkmann}}{{Brinchmann}
  et~al.}{2004}]{Brinchmann04}
{Brinchmann} J.,  {Charlot} S.,  {White} S.~D.~M.,  {Tremonti} C.,  {Kauffmann}
  G.,  {Heckman} T.,   {Brinkmann} J.,  2004, \mn@doi [\mnras]
  {10.1111/j.1365-2966.2004.07881.x}, \href
  {http://adsabs.harvard.edu/abs/2004MNRAS.351.1151B} {351, 1151}

\bibitem[\protect\citeauthoryear{{Brown} et~al.,}{{Brown}
  et~al.}{2014}]{Brown14}
{Brown} M.~J.~I.,  et~al., 2014, \mn@doi [\apjs] {10.1088/0067-0049/212/2/18},
  \href {http://adsabs.harvard.edu/abs/2014ApJS..212...18B} {212, 18}

\bibitem[\protect\citeauthoryear{{Bundy} et~al.,}{{Bundy}
  et~al.}{2010}]{Bundy10}
{Bundy} K.,  et~al., 2010, \mn@doi [\apj] {10.1088/0004-637X/719/2/1969}, \href
  {http://adsabs.harvard.edu/abs/2010ApJ...719.1969B} {719, 1969}

\bibitem[\protect\citeauthoryear{{Bundy} et~al.,}{{Bundy}
  et~al.}{2015}]{Bundy15}
{Bundy} K.,  et~al., 2015, \mn@doi [\apj] {10.1088/0004-637X/798/1/7}, \href
  {http://adsabs.harvard.edu/abs/2015ApJ...798....7B} {798, 7}

\bibitem[\protect\citeauthoryear{{Cameron} et~al.,}{{Cameron}
  et~al.}{2010}]{Cameron10}
{Cameron} E.,  et~al., 2010, \mn@doi [\mnras]
  {10.1111/j.1365-2966.2010.17314.x}, \href
  {http://adsabs.harvard.edu/abs/2010MNRAS.409..346C} {409, 346}

\bibitem[\protect\citeauthoryear{{Chabrier}}{{Chabrier}}{2003}]{Chabrier03}
{Chabrier} G.,  2003, \mn@doi [\pasp] {10.1086/376392}, \href
  {http://adsabs.harvard.edu/abs/2003PASP..115..763C} {115, 763}

\bibitem[\protect\citeauthoryear{{Cheung} et~al.,}{{Cheung}
  et~al.}{2013}]{Cheung13}
{Cheung} E.,  et~al., 2013, \mn@doi [\apj] {10.1088/0004-637X/779/2/162}, \href
  {http://adsabs.harvard.edu/abs/2013ApJ...779..162C} {779, 162}

\bibitem[\protect\citeauthoryear{{Combes} \& {Sanders}}{{Combes} \&
  {Sanders}}{1981}]{Combes81}
{Combes} F.,  {Sanders} R.~H.,  1981, \aap, \href
  {http://adsabs.harvard.edu/abs/1981A%26A....96..164C} {96, 164}

\bibitem[\protect\citeauthoryear{{Conroy}, {van Dokkum}  \&
  {Kravtsov}}{{Conroy} et~al.}{2015}]{Conroy15}
{Conroy} C.,  {van Dokkum} P.~G.,   {Kravtsov} A.,  2015, \mn@doi [\apj]
  {10.1088/0004-637X/803/2/77}, \href
  {http://adsabs.harvard.edu/abs/2015ApJ...803...77C} {803, 77}

\bibitem[\protect\citeauthoryear{{Cortese}}{{Cortese}}{2012}]{Cortese12}
{Cortese} L.,  2012, \mn@doi [\aap] {10.1051/0004-6361/201219443}, \href
  {http://adsabs.harvard.edu/abs/2012A%26A...543A.132C} {543, A132}

\bibitem[\protect\citeauthoryear{{Croom} et~al.,}{{Croom}
  et~al.}{2012}]{Croom12}
{Croom} S.~M.,  et~al., 2012, \mn@doi [\mnras]
  {10.1111/j.1365-2966.2011.20365.x}, \href
  {http://adsabs.harvard.edu/abs/2012MNRAS.421..872C} {421, 872}

\bibitem[\protect\citeauthoryear{{Davies} et~al.,}{{Davies}
  et~al.}{2016}]{Davies16}
{Davies} L.~J.~M.,  et~al., 2016, \mn@doi [\mnras] {10.1093/mnras/stv2573},
  \href {http://adsabs.harvard.edu/abs/2016MNRAS.455.4013D} {455, 4013}

\bibitem[\protect\citeauthoryear{{Debattista}, {Mayer}, {Carollo}, {Moore},
  {Wadsley}  \& {Quinn}}{{Debattista} et~al.}{2006}]{Debattista06}
{Debattista} V.~P.,  {Mayer} L.,  {Carollo} C.~M.,  {Moore} B.,  {Wadsley} J.,
   {Quinn} T.,  2006, \mn@doi [\apj] {10.1086/504147}, \href
  {http://adsabs.harvard.edu/abs/2006ApJ...645..209D} {645, 209}

\bibitem[\protect\citeauthoryear{{Dekel} \& {Birnboim}}{{Dekel} \&
  {Birnboim}}{2006}]{Dekel06}
{Dekel} A.,  {Birnboim} Y.,  2006, \mn@doi [\mnras]
  {10.1111/j.1365-2966.2006.10145.x}, \href
  {http://adsabs.harvard.edu/abs/2006MNRAS.368....2D} {368, 2}

\bibitem[\protect\citeauthoryear{{Dekel} \& {Silk}}{{Dekel} \&
  {Silk}}{1986}]{Dekel86}
{Dekel} A.,  {Silk} J.,  1986, \mn@doi [\apj] {10.1086/164050}, \href
  {http://adsabs.harvard.edu/abs/1986ApJ...303...39D} {303, 39}

\bibitem[\protect\citeauthoryear{{Dolley} et~al.,}{{Dolley}
  et~al.}{2014}]{Dolley14}
{Dolley} T.,  et~al., 2014, \mn@doi [\apj] {10.1088/0004-637X/797/2/125}, \href
  {http://adsabs.harvard.edu/abs/2014ApJ...797..125D} {797, 125}

\bibitem[\protect\citeauthoryear{{Eastmond} \& {Abell}}{{Eastmond} \&
  {Abell}}{1978}]{Eastmond78}
{Eastmond} T.~S.,  {Abell} G.~O.,  1978, \mn@doi [\pasp] {10.1086/130341},
  \href {http://adsabs.harvard.edu/abs/1978PASP...90..367E} {90, 367}

\bibitem[\protect\citeauthoryear{{Ellison}, {Nair}, {Patton}, {Scudder},
  {Mendel}  \& {Simard}}{{Ellison} et~al.}{2011}]{Ellison11}
{Ellison} S.~L.,  {Nair} P.,  {Patton} D.~R.,  {Scudder} J.~M.,  {Mendel}
  J.~T.,   {Simard} L.,  2011, \mn@doi [\mnras]
  {10.1111/j.1365-2966.2011.19195.x}, \href
  {http://adsabs.harvard.edu/abs/2011MNRAS.416.2182E} {416, 2182}

\bibitem[\protect\citeauthoryear{{Fabian}, {Arnaud}, {Bautz}  \&
  {Tawara}}{{Fabian} et~al.}{1994}]{Fabian94}
{Fabian} A.~C.,  {Arnaud} K.~A.,  {Bautz} M.~W.,   {Tawara} Y.,  1994, \mn@doi
  [\apjl] {10.1086/187633}, \href
  {http://adsabs.harvard.edu/abs/1994ApJ...436L..63F} {436, L63}

\bibitem[\protect\citeauthoryear{{Fillingham}, {Cooper}, {Wheeler},
  {Garrison-Kimmel}, {Boylan-Kolchin}  \& {Bullock}}{{Fillingham}
  et~al.}{2015}]{Fillingham15}
{Fillingham} S.~P.,  {Cooper} M.~C.,  {Wheeler} C.,  {Garrison-Kimmel} S.,
  {Boylan-Kolchin} M.,   {Bullock} J.~S.,  2015, \mn@doi [\mnras]
  {10.1093/mnras/stv2058}, \href
  {http://adsabs.harvard.edu/abs/2015MNRAS.454.2039F} {454, 2039}

\bibitem[\protect\citeauthoryear{{Fraser-McKelvie}, {Brown}, {Pimbblet},
  {Dolley}, {Crossett}  \& {Bonne}}{{Fraser-McKelvie} et~al.}{2016}]{Fraser16}
{Fraser-McKelvie} A.,  {Brown} M.~J.~I.,  {Pimbblet} K.~A.,  {Dolley} T.,
  {Crossett} J.~P.,   {Bonne} N.~J.,  2016, \mn@doi [\mnras]
  {10.1093/mnrasl/slw117}, \href
  {http://adsabs.harvard.edu/abs/2016MNRAS.462L..11F} {462, L11}

\bibitem[\protect\citeauthoryear{{Friedli} \& {Benz}}{{Friedli} \&
  {Benz}}{1995}]{Friedli95}
{Friedli} D.,  {Benz} W.,  1995, \aap, \href
  {http://adsabs.harvard.edu/abs/1995A%26A...301..649F} {301, 649}

\bibitem[\protect\citeauthoryear{{Gavazzi} et~al.,}{{Gavazzi}
  et~al.}{2015}]{Gavazzi15}
{Gavazzi} G.,  et~al., 2015, \mn@doi [\aap] {10.1051/0004-6361/201425351},
  \href {http://adsabs.harvard.edu/abs/2015A%26A...580A.116G} {580, A116}

\bibitem[\protect\citeauthoryear{{Geha}, {Blanton}, {Yan}  \& {Tinker}}{{Geha}
  et~al.}{2012}]{Geha12}
{Geha} M.,  {Blanton} M.~R.,  {Yan} R.,   {Tinker} J.~L.,  2012, \mn@doi [\apj]
  {10.1088/0004-637X/757/1/85}, \href
  {http://adsabs.harvard.edu/abs/2012ApJ...757...85G} {757, 85}

\bibitem[\protect\citeauthoryear{{Gonz{\'a}lez}}{{Gonz{\'a}lez}}{1993}]{Gonzalez93}
{Gonz{\'a}lez} J.~J.,  1993, PhD thesis, Thesis (PH.D.)--UNIVERSITY OF
  CALIFORNIA, SANTA CRUZ, 1993.Source: Dissertation Abstracts International,
  Volume: 54-05, Section: B, page: 2551.

\bibitem[\protect\citeauthoryear{{Goto} et~al.,}{{Goto} et~al.}{2003}]{Goto03}
{Goto} T.,  et~al., 2003, \mn@doi [\pasj] {10.1093/pasj/55.4.757}, \href
  {http://adsabs.harvard.edu/abs/2003PASJ...55..757G} {55, 757}

\bibitem[\protect\citeauthoryear{{Gunn} \& {Gott}}{{Gunn} \&
  {Gott}}{1972}]{Gunn72}
{Gunn} J.~E.,  {Gott} III J.~R.,  1972, \mn@doi [\apj] {10.1086/151605}, \href
  {http://adsabs.harvard.edu/abs/1972ApJ...176....1G} {176, 1}

\bibitem[\protect\citeauthoryear{{Holmes} et~al.,}{{Holmes}
  et~al.}{2015}]{Holmes15}
{Holmes} L.,  et~al., 2015, \mn@doi [\mnras] {10.1093/mnras/stv1254}, \href
  {http://adsabs.harvard.edu/abs/2015MNRAS.451.4397H} {451, 4397}

\bibitem[\protect\citeauthoryear{{Hopkins}, {Kere{\v s}}, {O{\~n}orbe},
  {Faucher-Gigu{\`e}re}, {Quataert}, {Murray}  \& {Bullock}}{{Hopkins}
  et~al.}{2014}]{PHopkins14}
{Hopkins} P.~F.,  {Kere{\v s}} D.,  {O{\~n}orbe} J.,  {Faucher-Gigu{\`e}re}
  C.-A.,  {Quataert} E.,  {Murray} N.,   {Bullock} J.~S.,  2014, \mn@doi
  [\mnras] {10.1093/mnras/stu1738}, \href
  {http://adsabs.harvard.edu/abs/2014MNRAS.445..581H} {445, 581}

\bibitem[\protect\citeauthoryear{{Ishigaki}, {Goto}  \& {Matsuhara}}{{Ishigaki}
  et~al.}{2007}]{Ishigaki07}
{Ishigaki} M.,  {Goto} T.,   {Matsuhara} H.,  2007, \mn@doi [\mnras]
  {10.1111/j.1365-2966.2007.12356.x}, \href
  {http://adsabs.harvard.edu/abs/2007MNRAS.382..270I} {382, 270}

\bibitem[\protect\citeauthoryear{{Jarrett}, {Chester}, {Cutri}, {Schneider},
  {Skrutskie}  \& {Huchra}}{{Jarrett} et~al.}{2000}]{Jarrett00}
{Jarrett} T.~H.,  {Chester} T.,  {Cutri} R.,  {Schneider} S.,  {Skrutskie} M.,
   {Huchra} J.~P.,  2000, \mn@doi [\aj] {10.1086/301330}, \href
  {http://adsabs.harvard.edu/abs/2000AJ....119.2498J} {119, 2498}

\bibitem[\protect\citeauthoryear{{Jogee}, {Scoville}  \& {Kenney}}{{Jogee}
  et~al.}{2005}]{Jogee05}
{Jogee} S.,  {Scoville} N.,   {Kenney} J.~D.~P.,  2005, \mn@doi [\apj]
  {10.1086/432106}, \href {http://adsabs.harvard.edu/abs/2005ApJ...630..837J}
  {630, 837}

\bibitem[\protect\citeauthoryear{{Kauffmann} et~al.,}{{Kauffmann}
  et~al.}{2003}]{Kauffmann03}
{Kauffmann} G.,  et~al., 2003, \mn@doi [\mnras]
  {10.1046/j.1365-8711.2003.06292.x}, \href
  {http://adsabs.harvard.edu/abs/2003MNRAS.341...54K} {341, 54}

\bibitem[\protect\citeauthoryear{{Kawinwanichakij} et~al.,}{{Kawinwanichakij}
  et~al.}{2017}]{Kawinwanichakij17}
{Kawinwanichakij} L.,  et~al., 2017, preprint, \href
  {http://adsabs.harvard.edu/abs/2017arXiv170603780K} {} (\mn@eprint {arXiv}
  {1706.03780})

\bibitem[\protect\citeauthoryear{{Khoperskov}, {Haywood}, {Di Matteo},
  {Lehnert}  \& {Combes}}{{Khoperskov} et~al.}{2017}]{Khoperskov17}
{Khoperskov} S.,  {Haywood} M.,  {Di Matteo} P.,  {Lehnert} M.~D.,   {Combes}
  F.,  2017, preprint, \href
  {http://adsabs.harvard.edu/abs/2017arXiv170903604K} {} (\mn@eprint {arXiv}
  {1709.03604})

\bibitem[\protect\citeauthoryear{{Knapen}, {P{\'e}rez-Ram{\'{\i}}rez}  \&
  {Laine}}{{Knapen} et~al.}{2002}]{Knapen02}
{Knapen} J.~H.,  {P{\'e}rez-Ram{\'{\i}}rez} D.,   {Laine} S.,  2002, \mn@doi
  [\mnras] {10.1046/j.1365-8711.2002.05840.x}, \href
  {http://adsabs.harvard.edu/abs/2002MNRAS.337..808K} {337, 808}

\bibitem[\protect\citeauthoryear{{Kormendy}}{{Kormendy}}{1979}]{Kormendy79}
{Kormendy} J.,  1979, \mn@doi [\apj] {10.1086/156782}, \href
  {http://adsabs.harvard.edu/abs/1979ApJ...227..714K} {227, 714}

\bibitem[\protect\citeauthoryear{{Kormendy} \& {Ho}}{{Kormendy} \&
  {Ho}}{2013}]{Kormendy13}
{Kormendy} J.,  {Ho} L.~C.,  2013, \mn@doi [\araa]
  {10.1146/annurev-astro-082708-101811}, \href
  {http://adsabs.harvard.edu/abs/2013ARA%26A..51..511K} {51, 511}

\bibitem[\protect\citeauthoryear{{Kormendy} \& {Kennicutt}}{{Kormendy} \&
  {Kennicutt}}{2004}]{Kormendy04}
{Kormendy} J.,  {Kennicutt} Jr. R.~C.,  2004, \mn@doi [\araa]
  {10.1146/annurev.astro.42.053102.134024}, \href
  {http://adsabs.harvard.edu/abs/2004ARA%26A..42..603K} {42, 603}

\bibitem[\protect\citeauthoryear{{Lake}, {Katz}  \& {Moore}}{{Lake}
  et~al.}{1998}]{Lake98}
{Lake} G.,  {Katz} N.,   {Moore} B.,  1998, \mn@doi [\apj] {10.1086/305265},
  \href {http://adsabs.harvard.edu/abs/1998ApJ...495..152L} {495, 152}

\bibitem[\protect\citeauthoryear{{Larson}, {Tinsley}  \& {Caldwell}}{{Larson}
  et~al.}{1980}]{Larson80}
{Larson} R.~B.,  {Tinsley} B.~M.,   {Caldwell} C.~N.,  1980, \mn@doi [\apj]
  {10.1086/157917}, \href {http://adsabs.harvard.edu/abs/1980ApJ...237..692L}
  {237, 692}

\bibitem[\protect\citeauthoryear{{Martig}, {Bournaud}, {Teyssier}  \&
  {Dekel}}{{Martig} et~al.}{2009}]{Martig09}
{Martig} M.,  {Bournaud} F.,  {Teyssier} R.,   {Dekel} A.,  2009, \mn@doi
  [\apj] {10.1088/0004-637X/707/1/250}, \href
  {http://adsabs.harvard.edu/abs/2009ApJ...707..250M} {707, 250}

\bibitem[\protect\citeauthoryear{{Martinet} \& {Friedli}}{{Martinet} \&
  {Friedli}}{1997}]{Martinet97}
{Martinet} L.,  {Friedli} D.,  1997, \aap, \href
  {http://adsabs.harvard.edu/abs/1997A%26A...323..363M} {323, 363}

\bibitem[\protect\citeauthoryear{{Martinez-Valpuesta}, {Shlosman}  \&
  {Heller}}{{Martinez-Valpuesta} et~al.}{2006}]{Martinez06}
{Martinez-Valpuesta} I.,  {Shlosman} I.,   {Heller} C.,  2006, \mn@doi [\apj]
  {10.1086/498338}, \href {http://adsabs.harvard.edu/abs/2006ApJ...637..214M}
  {637, 214}

\bibitem[\protect\citeauthoryear{{Martinez-Valpuesta}, {Knapen}  \&
  {Buta}}{{Martinez-Valpuesta} et~al.}{2007}]{Martinez07}
{Martinez-Valpuesta} I.,  {Knapen} J.~H.,   {Buta} R.,  2007, \mn@doi [\aj]
  {10.1086/522205}, \href {http://adsabs.harvard.edu/abs/2007AJ....134.1863M}
  {134, 1863}

\bibitem[\protect\citeauthoryear{{Martinez-Valpuesta}, {Knapen}  \&
  {Buta}}{{Martinez-Valpuesta} et~al.}{2008}]{Martinez08}
{Martinez-Valpuesta} I.,  {Knapen} J.~H.,   {Buta} R.,  2008, in {Knapen}
  J.~H.,  {Mahoney} T.~J.,   {Vazdekis} A.,  eds,  Astronomical Society of the
  Pacific Conference Series Vol. 390, Pathways Through an Eclectic Universe.
  p.~304

\bibitem[\protect\citeauthoryear{{Masters} et~al.,}{{Masters}
  et~al.}{2010}]{Masters10}
{Masters} K.~L.,  et~al., 2010, \mn@doi [\mnras]
  {10.1111/j.1365-2966.2010.16503.x}, \href
  {http://adsabs.harvard.edu/abs/2010MNRAS.405..783M} {405, 783}

\bibitem[\protect\citeauthoryear{{Masters} et~al.,}{{Masters}
  et~al.}{2011}]{Masters11}
{Masters} K.~L.,  et~al., 2011, \mn@doi [\mnras]
  {10.1111/j.1365-2966.2010.17834.x}, \href
  {http://adsabs.harvard.edu/abs/2011MNRAS.411.2026M} {411, 2026}

\bibitem[\protect\citeauthoryear{{Melvin} et~al.,}{{Melvin}
  et~al.}{2014}]{Melvin14}
{Melvin} T.,  et~al., 2014, \mn@doi [\mnras] {10.1093/mnras/stt2397}, \href
  {http://adsabs.harvard.edu/abs/2014MNRAS.438.2882M} {438, 2882}

\bibitem[\protect\citeauthoryear{{Moore}, {Katz}, {Lake}, {Dressler}  \&
  {Oemler}}{{Moore} et~al.}{1996}]{Moore96}
{Moore} B.,  {Katz} N.,  {Lake} G.,  {Dressler} A.,   {Oemler} A.,  1996,
  \mn@doi [\nat] {10.1038/379613a0}, \href
  {http://adsabs.harvard.edu/abs/1996Natur.379..613M} {379, 613}

\bibitem[\protect\citeauthoryear{{Nair} \& {Abraham}}{{Nair} \&
  {Abraham}}{2010}]{Nair10}
{Nair} P.~B.,  {Abraham} R.~G.,  2010, \mn@doi [\apjs]
  {10.1088/0067-0049/186/2/427}, \href
  {http://adsabs.harvard.edu/abs/2010ApJS..186..427N} {186, 427}

\bibitem[\protect\citeauthoryear{{Ogle}, {Lanz}, {Nader}  \& {Helou}}{{Ogle}
  et~al.}{2016}]{Ogle16}
{Ogle} P.~M.,  {Lanz} L.,  {Nader} C.,   {Helou} G.,  2016, \mn@doi [\apj]
  {10.3847/0004-637X/817/2/109}, \href
  {http://adsabs.harvard.edu/abs/2016ApJ...817..109O} {817, 109}

\bibitem[\protect\citeauthoryear{{Paturel}, {Petit}, {Prugniel}, {Theureau},
  {Rousseau}, {Brouty}, {Dubois}  \& {Cambr{\'e}sy}}{{Paturel}
  et~al.}{2003}]{Paturel03}
{Paturel} G.,  {Petit} C.,  {Prugniel} P.,  {Theureau} G.,  {Rousseau} J.,
  {Brouty} M.,  {Dubois} P.,   {Cambr{\'e}sy} L.,  2003, \mn@doi [\aap]
  {10.1051/0004-6361:20031411}, \href
  {http://adsabs.harvard.edu/abs/2003A%26A...412...45P} {412, 45}

\bibitem[\protect\citeauthoryear{{Peng} et~al.,}{{Peng} et~al.}{2010}]{Peng10}
{Peng} Y.-j.,  et~al., 2010, \mn@doi [\apj] {10.1088/0004-637X/721/1/193},
  \href {http://adsabs.harvard.edu/abs/2010ApJ...721..193P} {721, 193}

\bibitem[\protect\citeauthoryear{{Rowlands} et~al.,}{{Rowlands}
  et~al.}{2012}]{Rowlands12}
{Rowlands} K.,  et~al., 2012, \mn@doi [\mnras]
  {10.1111/j.1365-2966.2011.19905.x}, \href
  {http://adsabs.harvard.edu/abs/2012MNRAS.419.2545R} {419, 2545}

\bibitem[\protect\citeauthoryear{{Scannapieco}, {Tissera}, {White}  \&
  {Springel}}{{Scannapieco} et~al.}{2008}]{Scannapieco08}
{Scannapieco} C.,  {Tissera} P.~B.,  {White} S.~D.~M.,   {Springel} V.,  2008,
  \mn@doi [\mnras] {10.1111/j.1365-2966.2008.13678.x}, \href
  {http://adsabs.harvard.edu/abs/2008MNRAS.389.1137S} {389, 1137}

\bibitem[\protect\citeauthoryear{{Shen} \& {Sellwood}}{{Shen} \&
  {Sellwood}}{2004}]{Shen04}
{Shen} J.,  {Sellwood} J.~A.,  2004, \mn@doi [\apj] {10.1086/382124}, \href
  {http://adsabs.harvard.edu/abs/2004ApJ...604..614S} {604, 614}

\bibitem[\protect\citeauthoryear{{Shlosman}, {Frank}  \& {Begelman}}{{Shlosman}
  et~al.}{1989}]{Shlosman89}
{Shlosman} I.,  {Frank} J.,   {Begelman} M.~C.,  1989, \mn@doi [\nat]
  {10.1038/338045a0}, \href {http://adsabs.harvard.edu/abs/1989Natur.338...45S}
  {338, 45}

\bibitem[\protect\citeauthoryear{{Skibba} et~al.,}{{Skibba}
  et~al.}{2009}]{Skibba09}
{Skibba} R.~A.,  et~al., 2009, \mn@doi [\mnras]
  {10.1111/j.1365-2966.2009.15334.x}, \href
  {http://adsabs.harvard.edu/abs/2009MNRAS.399..966S} {399, 966}

\bibitem[\protect\citeauthoryear{{Spinoso}, {Bonoli}, {Dotti}, {Mayer}, {Madau}
   \& {Bellovary}}{{Spinoso} et~al.}{2017}]{Spinoso17}
{Spinoso} D.,  {Bonoli} S.,  {Dotti} M.,  {Mayer} L.,  {Madau} P.,
  {Bellovary} J.,  2017, \mn@doi [\mnras] {10.1093/mnras/stw2934}, \href
  {http://adsabs.harvard.edu/abs/2017MNRAS.465.3729S} {465, 3729}

\bibitem[\protect\citeauthoryear{{Strateva} et~al.,}{{Strateva}
  et~al.}{2001}]{Strateva01}
{Strateva} I.,  et~al., 2001, \mn@doi [\aj] {10.1086/323301}, \href
  {http://adsabs.harvard.edu/abs/2001AJ....122.1861S} {122, 1861}

\bibitem[\protect\citeauthoryear{{Tabor} \& {Binney}}{{Tabor} \&
  {Binney}}{1993}]{Tabor93}
{Tabor} G.,  {Binney} J.,  1993, \mn@doi [\mnras] {10.1093/mnras/263.2.323},
  \href {http://adsabs.harvard.edu/abs/1993MNRAS.263..323T} {263, 323}

\bibitem[\protect\citeauthoryear{{Thomas}, {Maraston}  \& {Bender}}{{Thomas}
  et~al.}{2003}]{Thomas03}
{Thomas} D.,  {Maraston} C.,   {Bender} R.,  2003, \mn@doi [\mnras]
  {10.1046/j.1365-8711.2003.06248.x}, \href
  {http://adsabs.harvard.edu/abs/2003MNRAS.339..897T} {339, 897}

\bibitem[\protect\citeauthoryear{{Toomre} \& {Toomre}}{{Toomre} \&
  {Toomre}}{1972}]{Toomre72}
{Toomre} A.,  {Toomre} J.,  1972, \mn@doi [\apj] {10.1086/151823}, \href
  {http://adsabs.harvard.edu/abs/1972ApJ...178..623T} {178, 623}

\bibitem[\protect\citeauthoryear{{Tremonti} et~al.,}{{Tremonti}
  et~al.}{2004}]{Tremonti04}
{Tremonti} C.~A.,  et~al., 2004, \mn@doi [\apj] {10.1086/423264}, \href
  {http://adsabs.harvard.edu/abs/2004ApJ...613..898T} {613, 898}

\bibitem[\protect\citeauthoryear{{Tully}}{{Tully}}{2015}]{Tully15}
{Tully} R.~B.,  2015, \mn@doi [\aj] {10.1088/0004-6256/149/5/171}, \href
  {http://adsabs.harvard.edu/abs/2015AJ....149..171T} {149, 171}

\bibitem[\protect\citeauthoryear{{Vazdekis}, {S{\'a}nchez-Bl{\'a}zquez},
  {Falc{\'o}n-Barroso}, {Cenarro}, {Beasley}, {Cardiel}, {Gorgas}  \&
  {Peletier}}{{Vazdekis} et~al.}{2010}]{Vazdekis10}
{Vazdekis} A.,  {S{\'a}nchez-Bl{\'a}zquez} P.,  {Falc{\'o}n-Barroso} J.,
  {Cenarro} A.~J.,  {Beasley} M.~A.,  {Cardiel} N.,  {Gorgas} J.,   {Peletier}
  R.~F.,  2010, \mn@doi [\mnras] {10.1111/j.1365-2966.2010.16407.x}, \href
  {http://adsabs.harvard.edu/abs/2010MNRAS.404.1639V} {404, 1639}

\bibitem[\protect\citeauthoryear{{Weinberg}}{{Weinberg}}{1985}]{Weinberg85}
{Weinberg} M.~D.,  1985, \mn@doi [\mnras] {10.1093/mnras/213.3.451}, \href
  {http://adsabs.harvard.edu/abs/1985MNRAS.213..451W} {213, 451}

\bibitem[\protect\citeauthoryear{{Weinmann}, {van den Bosch}, {Yang}  \&
  {Mo}}{{Weinmann} et~al.}{2006}]{Weinmann06}
{Weinmann} S.~M.,  {van den Bosch} F.~C.,  {Yang} X.,   {Mo} H.~J.,  2006,
  \mn@doi [\mnras] {10.1111/j.1365-2966.2005.09865.x}, \href
  {http://adsabs.harvard.edu/abs/2006MNRAS.366....2W} {366, 2}

\bibitem[\protect\citeauthoryear{{White} \& {Rees}}{{White} \&
  {Rees}}{1978}]{White78}
{White} S.~D.~M.,  {Rees} M.~J.,  1978, \mn@doi [\mnras]
  {10.1093/mnras/183.3.341}, \href
  {http://adsabs.harvard.edu/abs/1978MNRAS.183..341W} {183, 341}

\bibitem[\protect\citeauthoryear{{Wolf} et~al.,}{{Wolf} et~al.}{2009}]{Wolf09}
{Wolf} C.,  et~al., 2009, \mn@doi [\mnras] {10.1111/j.1365-2966.2008.14204.x},
  \href {http://adsabs.harvard.edu/abs/2009MNRAS.393.1302W} {393, 1302}

\bibitem[\protect\citeauthoryear{{Yepes}, {Kates}, {Khokhlov}  \&
  {Klypin}}{{Yepes} et~al.}{1997}]{Yepes97}
{Yepes} G.,  {Kates} R.,  {Khokhlov} A.,   {Klypin} A.,  1997, \mn@doi [\mnras]
  {10.1093/mnras/284.1.235}, \href
  {http://adsabs.harvard.edu/abs/1997MNRAS.284..235Y} {284, 235}

\bibitem[\protect\citeauthoryear{{de Vaucouleurs}}{{de
  Vaucouleurs}}{1961}]{deVauc61}
{de Vaucouleurs} G.,  1961, \mn@doi [\apjs] {10.1086/190064}, \href
  {http://adsabs.harvard.edu/abs/1961ApJS....6..213D} {6, 213}

\bibitem[\protect\citeauthoryear{{de Vaucouleurs}, {de Vaucouleurs}, {Corwin},
  {Buta}, {Paturel}  \& {Fouqu{\'e}}}{{de Vaucouleurs} et~al.}{1991}]{deVac91}
{de Vaucouleurs} G.,  {de Vaucouleurs} A.,  {Corwin} Jr. H.~G.,  {Buta} R.~J.,
  {Paturel} G.,   {Fouqu{\'e}} P.,  1991, {Third Reference Catalogue of Bright
  Galaxies. Volume I: Explanations and references. Volume II: Data for galaxies
  between 0$^{h}$ and 12$^{h}$. Volume III: Data for galaxies between 12$^{h}$
  and 24$^{h}$.}

\bibitem[\protect\citeauthoryear{{van Gorkom}}{{van
  Gorkom}}{2004}]{vanGorkom04}
{van Gorkom} J.~H.,  2004, Clusters of Galaxies: Probes of Cosmological
  Structure and Galaxy Evolution, \href
  {http://adsabs.harvard.edu/abs/2004cgpc.symp..305V} {p.~305}

\bibitem[\protect\citeauthoryear{{van den Bergh}}{{van den
  Bergh}}{1976}]{vandenBergh76}
{van den Bergh} S.,  1976, \mn@doi [\apj] {10.1086/154452}, \href
  {http://adsabs.harvard.edu/abs/1976ApJ...206..883V} {206, 883}

\makeatother
\end{thebibliography}
    \end{document}